\documentclass[aps,prd,showpacs,11pt,tightenlines,preprint,nofootinbib]{revtex4-1} 
\usepackage{url}
\usepackage{bm}
\usepackage{amsmath}
\usepackage{graphicx}
\usepackage{SIunits}
\usepackage{hyperref}
\usepackage{color}

\allowdisplaybreaks

\newcommand*\oline[1]{
  \hspace*{0.2em}
  \vbox{
    \kern-0.35ex
    \hrule height 0.4pt
    \kern0.35ex
    \hbox{
      \kern-0.5em
      \ifmmode#1\else\ensuremath{#1}\fi
      \kern-0.0em
}}}

\newcommand{\CP}{\mathit{CP}}
 
\newcommand{\lam}[1]{\lambda_{#1}}

\hypersetup{pdfstartview=FitV,colorlinks=true,linkcolor=blue,citecolor=red,filecolor=black,urlcolor=blue}

\begin{document} 

\title{Global fits of the two-loop renormalized\\ Two-Higgs-Doublet model with soft $Z_2$ breaking}

\author{Debtosh Chowdhury}
\email{debtosh.chowdhury@roma1.infn.it}

\author{Otto Eberhardt}
\email{otto.eberhardt@roma1.infn.it}

\affiliation{Istituto Nazionale di Fisica Nucleare,
Sezione di Roma, Piazzale Aldo Moro 2, I-00185 Roma, Italy}

\date{\today}

\begin{abstract}

We determine the next-to-leading order renormalization group equations for the Two-Higgs-Doublet model with a softly broken $Z_2$ symmetry and $\CP$ conservation in the scalar potential. We use them to identify the parameter regions which are stable up to the Planck scale and find that in this case the quartic couplings of the Higgs potential cannot be larger than $1$ in magnitude and that the absolute values of the $S$-matrix eigenvalues cannot exceed $2.5$ at the electroweak symmetry breaking scale. Interpreting the \unit{125}{GeV} resonance as the light $\CP$-even Higgs eigenstate, we combine stability constraints, electroweak precision and flavour observables with the latest ATLAS and CMS data on Higgs signal strengths and heavy Higgs searches in global parameter fits to all four types of $Z_2$ symmetry. 
We quantify the maximal deviations from the alignment limit and find that in type II and Y the mass of the heavy $\CP$-even ($\CP$-odd) scalar cannot be smaller than $340$ GeV ($360$ GeV). Also, we pinpoint the physical parameter regions compatible with a stable scalar potential up to the Planck scale.
Motivated by the question how natural a Higgs mass of \unit{125}{GeV} can be in the context of a Two-Higgs-Doublet model, we also address the hierarchy problem and find that the Two-Higgs-Doublet model does not offer a perturbative solution to it beyond \unit{5}{TeV}.

\end{abstract}

\maketitle

\section{Introduction}
\label{sec:intro}

After the discovery of a scalar particle at the LHC \cite{Aad:2012tfa,Chatrchyan:2012ufa}, one of the next questions is whether this is the one Higgs particle predicted by the Standard Model (SM) or whether there are more generations of SU(2) doublets, like it is the case for the fermions. In that sense, the simplest and most straightforward extension of the SM would be the addition of another Higgs doublet, the so-called Two-Higgs-Doublet model (2HDM) \cite{Lee:1973iz,Gunion:2002zf,Branco:2011iw}.
Furthermore, the measured mass of this new scalar \cite{Aad:2015zhl} is a peculiar value for the SM: it tells us that the Higgs potential of this model cannot be stable up to very high energy scales \cite{Degrassi:2012ry,Buttazzo:2013uya}. However, there is the possibility that the electroweak vacuum may just end up being metastable. So either one has to believe that we live in a metastable universe and then there is no need of new physics beyond the SM, or one has to introduce an additional mechanism to stabilize the Higgs potential.
The latter could for instance be achieved by the heavier scalars of the 2HDM. This model might be realized as an intermediate ``effective'' theory which describes physics at energy scales between the electroweak scale $\mu_{\text{ew}}$ of order \unit{10^2}{GeV} and some higher scale $\mu_{\text{high}}$. Beyond the latter, a more comprehensive model would be needed to describe ``physics beyond the 2HDM''. An upper bound on $\mu_{\text{high}}$ is the Planck scale $\mu_{\text{Pl}}\approx 10^{19}$ \unit{}{GeV}, at which gravitational effects become non-negligible in a quantum field theory framework. Large scale differences between $\mu_{\text{ew}}$ and $\mu_{\text{high}}$ bring along hierarchy problems like the fine-tuning of the \unit{125}{GeV} Higgs mass, which could be resolved by mechanisms of the ``complete'' models, but are usually neglected in the effective models. Still one could ask to what extent the 2HDM could possibly mitigate the Higgs mass hierarchy problem and whether it might even be valid up to Planck scale without requiring any other New Physics.

Therefore, we want to analyze the renormalization group evolution behaviour of the 2HDM in this article, focussing on softly-broken $Z_2$ symmetric model realizations, which avoid flavour changing neutral currents at tree-level. Recently these models have attracted a lot of attention. A large number of papers \cite{Cheon:2012rh,Chen:2013kt,Chiang:2013ixa,Grinstein:2013npa,Barroso:2013zxa,Coleppa:2013dya,Eberhardt:2013uba,Belanger:2013xza,Chang:2013ona,Cheung:2013rva,Celis:2013ixa,Wang:2013sha,Baglio:2014nea,Inoue:2014nva,Dumont:2014wha,Kanemura:2014bqa,Ferreira:2014sld,Broggio:2014mna,Dumont:2014kna,Bernon:2014nxa,Chen:2015gaa} have analyzed current data for the \unit{125}{GeV} Higgs-like state within the context of 2HDM, and investigated the phenomenology of the other Higgs states present in the model. Given these results, the prospects for LHC upgrades and for other future colliders were examined in \cite{Chen:2013rba,Craig:2013hca,Barger:2013ofa,Kanemura:2014dea,Wang:2014lta,Barger:2014qva,Gorbahn:2015gxa,Kanemura:2015mxa}.

For renormalization group studies, especially the role of Higgs self-couplings is crucial and has been studied in the literature, in the SM (see for instance \cite{Nierste:1995zx,Degrassi:2012ry}) as well as in the 2HDM \cite{Hill:1985tg,Andrianov:1994za,Branco:2011iw,Bijnens:2011gd,JuarezW.:2012qa,Chakrabarty:2014aya,Chakrabarty:2015yia}, because these quartic couplings tend to destabilize the Higgs potential at some $\mu_{\text{high}}$. Since a break-down of stability would mean that our theory would lose validity beyond a certain scale, we want to impose a stable Higgs potential beyond $\mu_{\text{ew}}$ as a constraint on all couplings. Recently, the impact of stability up to the Planck scale on the parameters in the alignment limit of the 2HDM was discussed in \cite{Das:2015mwa}.
If one wants to solve the Higgs mass fine-tuning problem, one has to guarantee the cancellation of quadratic divergencies of higher order Higgs mass correction terms. The corresponding conditions that need to be fulfilled are called ``Veltman conditions'' \cite{Veltman:1980mj} in general, and in the context of the 2HDM also ``Newton-Wu conditions'' \cite{Newton:1993xc}. They have been analyzed at one-loop level \cite{Ma:2001sj,Jora:2013opa,Biswas:2014uba} and even leading two-loop contributions have been taken into account in type II \cite{Grzadkowski:2009iz,Grzadkowski:2010dn,Grzadkowski:2010se}. A recent idea was to only relax the cancellation of the generically large contributions of quadratic divergencies instead of imposing the strict cancellation using Veltman conditions \cite{Chakraborty:2014oma}.

In this article, we want to improve available results concerning two main aspects: We perform global parameter fits including the most up-to-date ATLAS and CMS results, rather than only using a handful of benchmark scenarios, which might not cover the whole spectrum of interesting features. Secondly, we go beyond leading order precision by employing two-loop renormalization group equations (RGE) in order to analyze vacuum stability of the 2HDM scalar potential. Moreover, we want to make use of the framework of next-to-leading order RGE to find out to what extent Veltman conditions can be fulfilled in the 2HDM.

In the following, we first want to make the reader familiar with the model in Section \ref{sec:model}, and introduce in Section \ref{sec:constraints} its theoretical and experimental constraints and the numerical setup that we use. Then we are ready to compare leading and next-to-leading order renormalization group equations for a benchmark scenario in Section \ref{subsec:benchmark}. We go on examining the quartic couplings varying the stability cut-off scale in global fits without experimental inputs in Section \ref{subsec:withoutexp}. We also address the question of which upper limit to use for the unitarity condition from the perspective of renormalizability. Taking into account experimental data, we analyze the results of global fits to the physical parameters at $\mu_{\text{ew}}$ and $\mu_{\text{Pl}}$ in Section \ref{subsec:withexp}.
The hierarchy problem is discussed in Section \ref{sec:naturalness}, before we conclude in Section \ref{sec:conclusions}. Explicit expressions for the one-loop and two-loop RGE can be found in the \hyperref[sec:appendix]{Appendix}.

\section{Model}
\label{sec:model}

The Two-Higgs-Doublet model with a softly broken $Z_2$ symmetry is characterized by the following scalar potential:
\begin{align}
 V
 &=m_{11}^2\Phi_1^\dagger\Phi_1^{\phantom{\dagger}}
   +m_{22}^2\Phi_2^\dagger\Phi_2^{\phantom{\dagger}}
   -m_{12}^2 ( \Phi_1^\dagger\Phi_2^{\phantom{\dagger}}
              +\Phi_2^\dagger\Phi_1^{\phantom{\dagger}})
   +\tfrac12 \lambda_1(\Phi_1^\dagger\Phi_1^{\phantom{\dagger}})^2
   +\tfrac12 \lambda_2(\Phi_2^\dagger\Phi_2^{\phantom{\dagger}})^2
 \nonumber \\
 &\phantom{{}={}}
  +\lambda_3(\Phi_1^\dagger\Phi_1^{\phantom{\dagger}})
            (\Phi_2^\dagger\Phi_2^{\phantom{\dagger}})
  +\lambda_4(\Phi_1^\dagger\Phi_2^{\phantom{\dagger}})
            (\Phi_2^\dagger\Phi_1^{\phantom{\dagger}})
  +\tfrac12 \lambda_5 \left[ (\Phi_1^\dagger\Phi_2^{\phantom{\dagger}})^2
                      +(\Phi_2^\dagger\Phi_1^{\phantom{\dagger}})^2 \right],\label{eq:pot}
\end{align}
where $\Phi_1$ and $\Phi_2$ are the two Higgs doublets.
In the following, we will use two sets of parameters: the eight potential parameters from Eq.~\eqref{eq:pot}, which we assume to be real, and the physical parameters consisting of the vacuum expectation value $v$, the $\CP$-even Higgs masses $m_h$ and $m_H$, the $\CP$-odd Higgs mass $m_A$, the mass of the charged Higgs, $m_{H^+}$, the two diagonalization angles $\alpha$ and $\beta$, and the soft $Z_2$ breaking parameter $m_{12}^2$. The first two physical parameters can be treated as fixed by measurements, assuming that the \unit{125}{GeV} scalar found at the LHC is the lighter $\CP$-even Higgs. Instead of $\alpha$ and $\beta$ we will use the combinations $\beta-\alpha$ and $\tan \beta$, since they can be directly related to physical observables. The measurements of the light Higgs couplings to fermions and bosons are compatible with the SM, such that the 2HDM is pushed towards the so-called alignment limit \cite{Gunion:2002zf,Delgado:2013zfa,Craig:2013hca,Carena:2013ooa}, in which $\beta-\alpha=\pi/2$. In this limit, it has recently been shown that $\CP$ violation in the scalar potential of $Z_2$ symmetric models with a soft breaking term is strongly suppressed \cite{Grzadkowski:2014ada}, which qualifies our above assumption that the potential parameters are real. The masses of the heavy scalars could in general even be lighter than \unit{125}{GeV}, and are not necessarily in the decoupling limit \cite{Gunion:2002zf} (which itself is a limiting case of the alignment limit). In the following we will consider them to be in the range between \unit{130}{GeV} and \unit{10}{TeV}, that is heavier than the region where the \unit{125}{GeV} scalar was found, yet still in the TeV range, which will be accessible by future colliders.

Neglecting the first two generations of fermions, the Yukawa part of the 2HDM Lagrangian is
\begin{align}
 {\cal L}_Y =& -Y_t \oline Q_{\textit{\tiny{L}}} i\sigma_2 \Phi_2^* t_{\textit{\tiny{R}}} -Y_{b,1} \oline Q_{\textit{\tiny{L}}} \Phi_1 b_{\textit{\tiny{R}}} -Y_{b,2} \oline Q_{\textit{\tiny{L}}} \Phi_2 b_{\textit{\tiny{R}}} -Y_{\tau,1} \oline L_{\textit{\tiny{L}}} \Phi_1 \tau_{\textit{\tiny{R}}} -Y_{\tau,2} \oline L_{\textit{\tiny{L}}} \Phi_2 \tau_{\textit{\tiny{R}}} + {\text{h.c.}}
  \label{eq:yuk}
\end{align}
In the above Lagrangian, the top quark only couples to $\Phi_2$ by convention; its Yukawa coupling is related to the SM value $Y_t^{\rm SM}$ by $Y_t\!=\!Y_t^{\rm SM}/\sin \beta$. Without breaking the $Z_2$ symmetry in the Yukawa sector, there are only four possibilities to couple the Higgs fields to the bottom quark and tau lepton at the tree-level. They are called type I, type II, type X or ``lepton specific'' and type Y or ``flipped''; in Table \ref{tab:types} we show the corresponding Higgs field assignments. Type II is of special interest, as it contains the Higgs part of supersymmetric models.
As soon as we consider any one of the above types, only three Yukawa couplings remain as free parameters, and we can speak of $Y_t$, $Y_b$ and $Y_\tau$ without any ambiguity.
\begin{table}[htb]
  \centering
  \caption{Yukawa assignments in the four possible $Z_2$ symmetric 2HDM types.}\vspace{0.2cm}
  \begin{tabular}{l|l|l|l}
    \hline\hline
      Type I & Type II & Type X (``lepton specific'') & Type Y (``flipped'') \\
    \hline
      $Y_{b,1}=Y_{\tau,1}=0$ & $Y_{b,2}=Y_{\tau,2}=0$ & $Y_{b,1}=Y_{\tau,2}=0$ & $Y_{b,2}=Y_{\tau,1}=0$ \\
      $Y_{b,2}\!=\!Y_b^{\rm SM}/\sin \beta$ & $Y_{b,1}\!=\!Y_b^{\rm SM}/\cos \beta$ & $Y_{b,2}\!=\!Y_b^{\rm SM}/\sin \beta$ & $Y_{b,1}\!=\!Y_b^{\rm SM}/\cos \beta$ \\
      $Y_{\tau,2}\!=\!Y_\tau^{\rm SM}/\sin \beta$ & $Y_{\tau,1}\!=\!Y_\tau^{\rm SM}/\cos \beta$ & $Y_{\tau,1}\!=\!Y_\tau^{\rm SM}/\cos \beta$ & $Y_{\tau,2}\!=\!Y_\tau^{\rm SM}/\sin \beta$ \\
      \hline \hline
  \end{tabular}
  \label{tab:types}
\end{table} 

\section{Constraints and set-up}
\label{sec:constraints}

We will apply the following sets of constraints on the parameter space: On the theoretical side, the positivity of the Higgs potential \cite{Deshpande:1977rw} and the unitarity of the eigenvalues of the $\Phi_i \Phi_j \to \Phi_i \Phi_j$ scattering matrix \cite{Ginzburg:2005dt} are imposed at all scales and vacuum stability \cite{Barroso:2013awa} at the electroweak scale. Moreover, we make sure that the quartic couplings $\lambda_i$ ($i=1,2,3,4,5$) and the Yukawa couplings $Y_i$ ($i=t,b,\tau$) do not run into non-perturbative regions.
On the experimental side, electroweak precision observables, the branching ratio ${\rm Br}(\oline{B}\to X_s\gamma)$, the mass difference in the $B_s$ system and light and heavy Higgs searches constrain the 2HDM parameters at the electroweak scale. For a detailed description on the various constraints, we refer to \cite{Eberhardt:2013uba} and \cite{Baglio:2014nea} except for ${\rm Br}(\oline{B}\to X_s\gamma)$ and the experimental Higgs data: In type II and Y we assume $m_{H^+}>480$\unit{}{GeV} in order to be consistent with the latest bound from ${\rm Br}(\oline{B}\to X_s\gamma)$ \cite{Misiak:2015xwa}. For the light Higgs signal strengths and heavy Higgs searches we use the most up-to-date ATLAS and CMS publications and pre-prints \cite{ATLAS-CONF-2015-007,Aad:2015gra,Khachatryan:2014jba,Khachatryan:2015ila,Aad:2014yja,ATLAS-CONF-2013-067,Aad:2014ioa,ATLAS-CONF-2014-049,Aad:2015wra,Khachatryan:2014wca,Khachatryan:2015cwa,Khachatryan:2015qba,CMS-PAS-HIG-13-032,CMS-PAS-HIG-14-011,Khachatryan:2015yea,Aad:2014kga,CMS-PAS-HIG-13-026,CMS-PAS-HIG-14-020}, applying the narrow width approximation.
We do not make use of (semi-)tauonic $B$ decay observables, which would only be relevant in type II \cite{Mahmoudi:2009zx}, because the existing tension between the measurements can only be explained in a 2HDM with explicitly broken $Z_2$ symmetry \cite{Crivellin:2012ye}.
The SM parameters will be fixed to their best fit values \cite{Eberhardt:2012gv}; for the SM Yukawa couplings in the $\overline{\rm MS}$ renormalization scheme at the scale $m_Z$ we take $Y_t^{\rm SM}=0.961$, $Y_b^{\rm SM}=0.0172$ and $Y_\tau^{\rm SM}=0.0102$.
While variations of the strong coupling $\alpha_s (m_Z)$ within the $3\sigma$ allowed range have no effects on the outcome of our fits, varying the input for $m_t(m_Z)$ can have an impact on a specific parameter region like mentioned in \cite{Chakrabarty:2014aya}. However, we observe that these effects are imperceptible in the results of our global fits.

The two-loop RGE have been obtained with the publicly available package PyR@TE \cite{Lyonnet:2013dna}; we neglect all Yukawa couplings except for the top and bottom quarks and the $\tau$ lepton. The observables have been calculated with the help of Zfitter \cite{Bardin:1989tq, Bardin:1999yd, Arbuzov:2005ma}, FeynArts \cite{Hahn:2000kx}, FormCalc \cite{Hahn:1998yk}, LoopTools \cite{Hahn:2006qw}, HDECAY \cite{Djouadi:1997yw,Butterworth:2010ym,Djouadi:2006bz}, FeynRules \cite{Alloul:2013bka} and MadGraph5 \cite{Alwall:2014hca}. The frequentist fits are performed with the CKMfitter package \cite{Hocker:2001xe}. For the fits involving experimental constraints we use of the naive definition of the $p$-value (Wilks' theorem) \cite{Wilks:1938dza}. If not stated differently, exclusion limits are meant to be at the $2\sigma$ level, which roughly corresponds to the $95\%$ confidence level.

Since we want to discuss various values for the scale $\mu$ in the following, we want to define our notation:
The scale range for the running quantities lies between the electroweak scale $\mu_{\text{ew}}=m_Z$ at the lower end and the Planck scale $\mu_{\text{Pl}}=10^{19}$\unit{}{GeV} at the upper end, as mentioned in the introduction. If -- starting with a given set of parameters at $\mu_{\text{ew}}$ and evolving to higher energy scales -- one of the theoretical constraints is violated, we denote this breakdown of stability as $\mu_{\text{st}}$. When discussing the hierarchy problem, it might be useful to introduce a cut-off scale $\mu_{nat}$, which a priori does not need to be the same as $\mu_{\text{st}}$. Furthermore, we want to introduce the parameter $t_{(X)} = \ln (\mu_{(X)}/$\unit{}{GeV}$)$ as the usual logarithmic scale.

\section{Renormalization at next-to-leading order}
\label{sec:rge}

One can find a plethora of leading order RGE \cite{Inoue:1979nn,Haber:1993an,Grimus:2004yh,Branco:2011iw} and next-to-leading order RGE \cite{Cheung:2012nb,Dev:2014yca} for different realizations of a 2HDM in the literature; however, we failed to find a complete set for a 2HDM with soft $Z_2$ breaking including the mass parameters, so we list the leading order (LO) and next-to-leading order (NLO) expressions in the \hyperref[sec:appendix]{Appendix}. Before scanning over the whole parameter space with our fitting set-up, we want to explain some features of the 2HDM RGE looking at a representative example.

\subsection{A benchmark point}
\label{subsec:benchmark}

\begin{figure}
  \centering
  \resizebox{200pt}{!}{
   \begin{picture}(300,280)(0,0)
    \put(-40,87){\includegraphics[width=340pt]{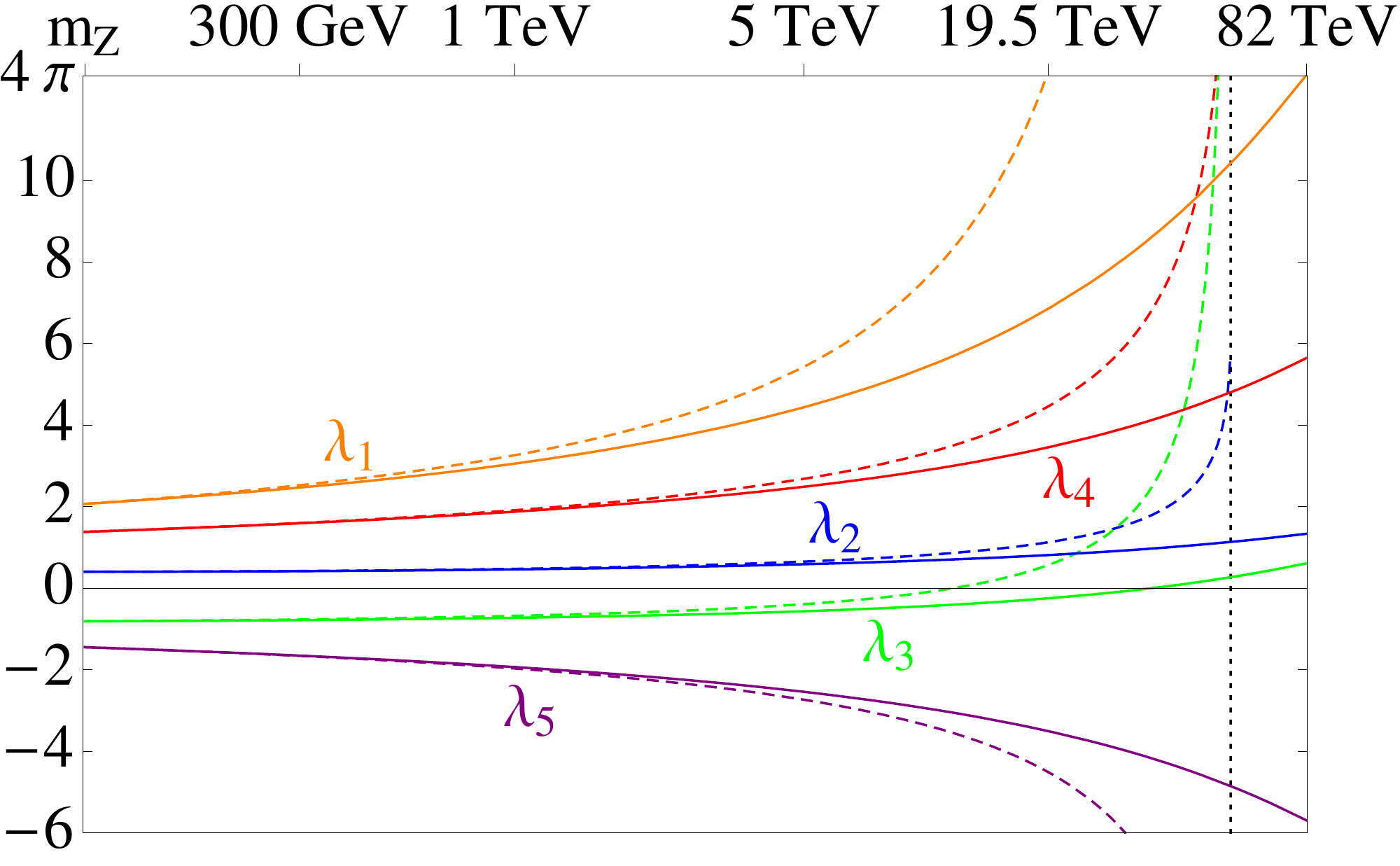}}
    \put(-55,-10){\includegraphics[width=354.5pt]{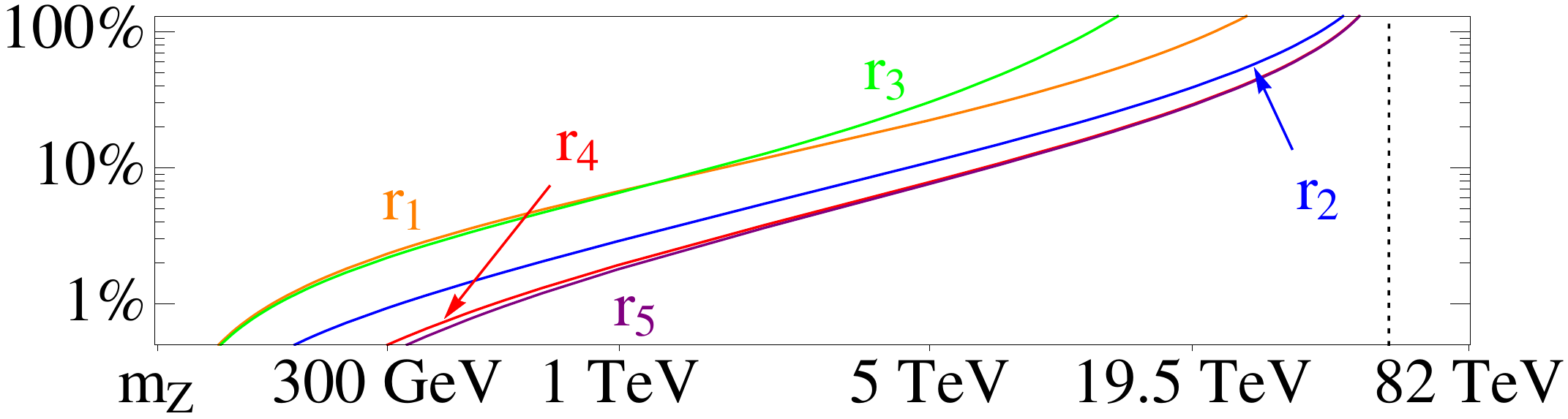}}
   \end{picture}
  }
  \resizebox{200pt}{!}{
   \begin{picture}(300,280)(0,0)
    \put(10,139){\includegraphics[width=346.5pt]{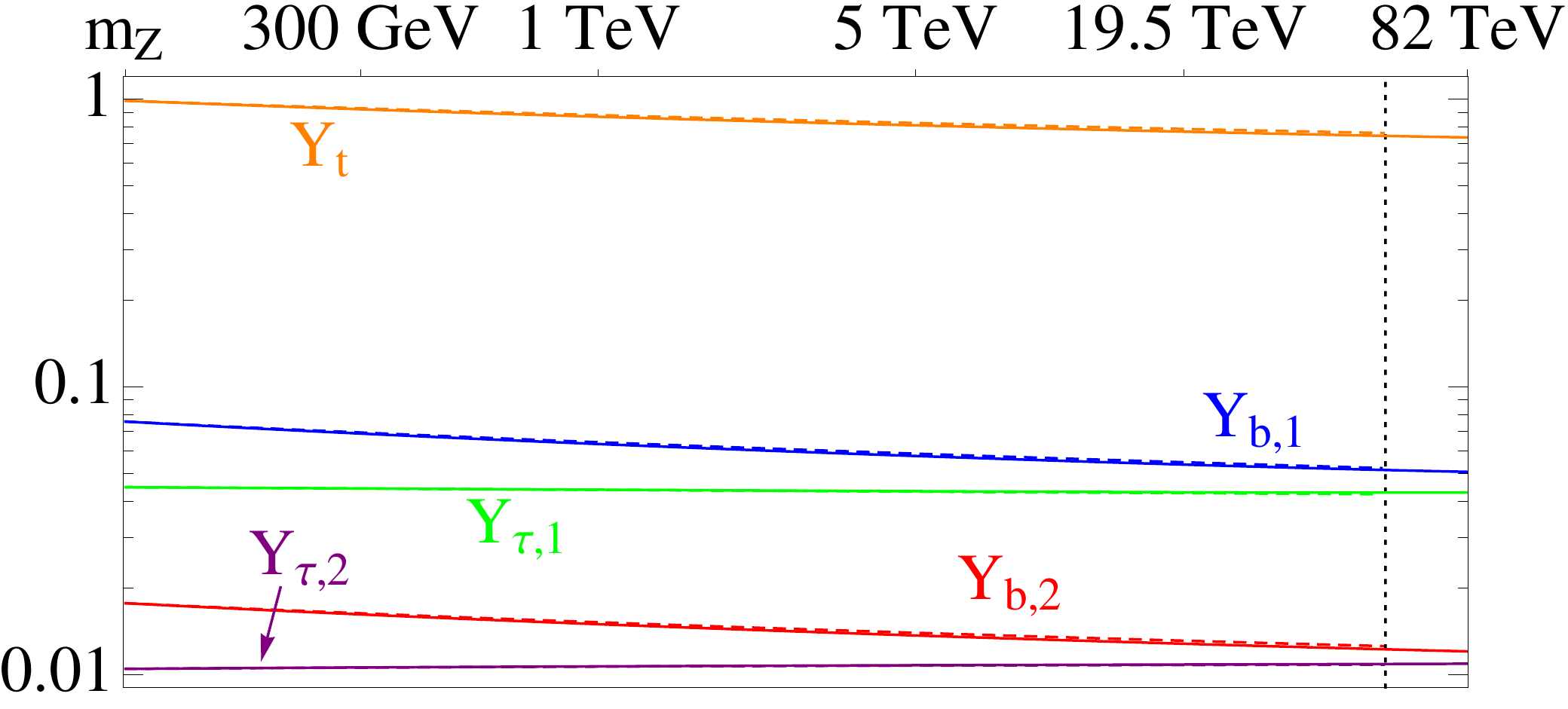}}
    \put(-16.5,-10.5){\includegraphics[width=373pt]{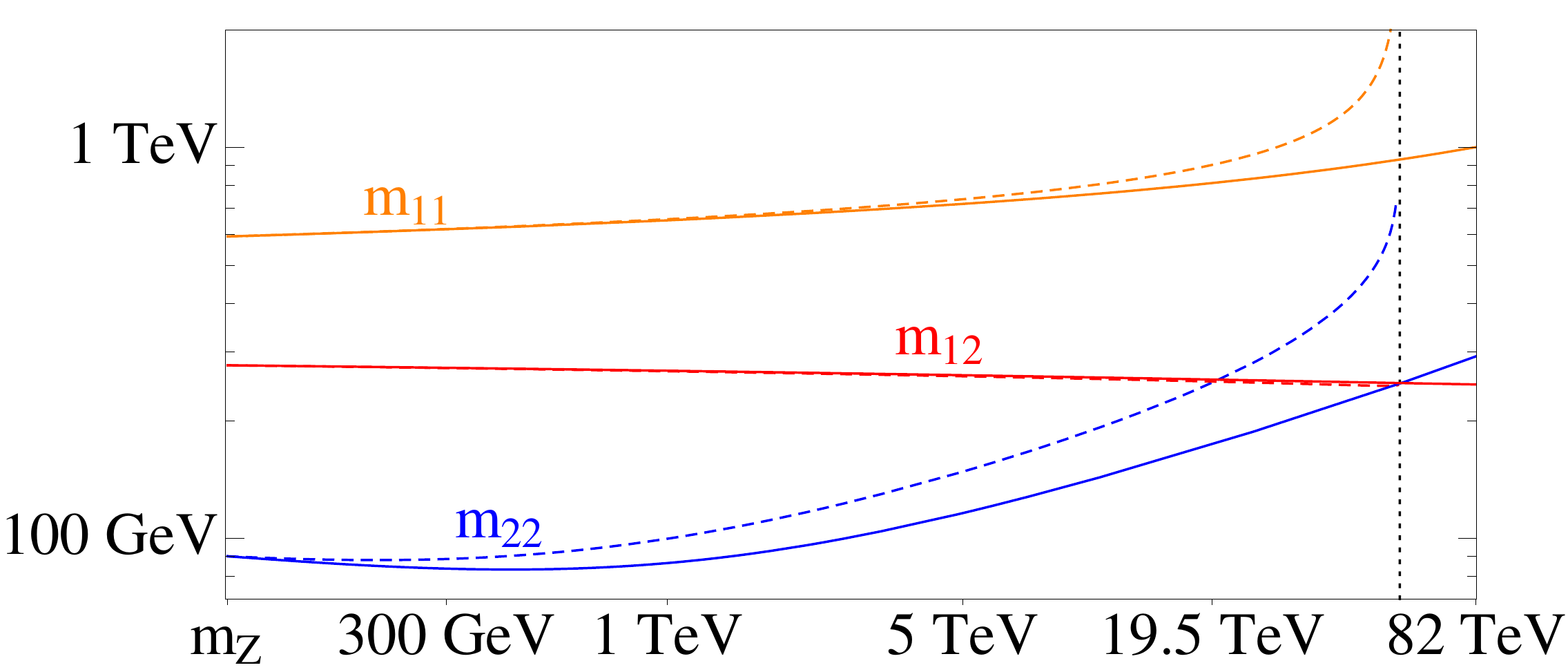}}
   \end{picture}
  }
  \caption{Leading order (dashed) and next-to-leading order (solid) RG running for the benchmark scenario H-4. On the left, we show on top the evolution of the quartic couplings. At LO, $\lambda_1$ hits the perturbativity limit $4\pi$ at \unit{19.5}{GeV}; the Landau pole is at \unit{54}{GeV}, indicated by the vertical dotted line. NLO RGE shift the perturbativity cut-off to \unit{82}{GeV}. In the lower figure on the left, we show the relative error $r_i=\left| (\lambda _i^{\text{\tiny{LO}}}-\lambda _i^{\text{\tiny{NLO}}})/\lambda _i^{\text{\tiny{NLO}}} \right|$ between LO and NLO expressions for the quartic couplings. The Yukawa couplings and the potential mass parameters are shown on the right. All types look the same except for the $b$ and $\tau$ Yukawa couplings.}
  \label{fig:LOvsNLO}
\end{figure}

\begin{figure}
  \centering
  \resizebox{400pt}{!}{
   \begin{picture}(400,130)(0,0)
    \put(-40,0){\includegraphics[width=220pt]{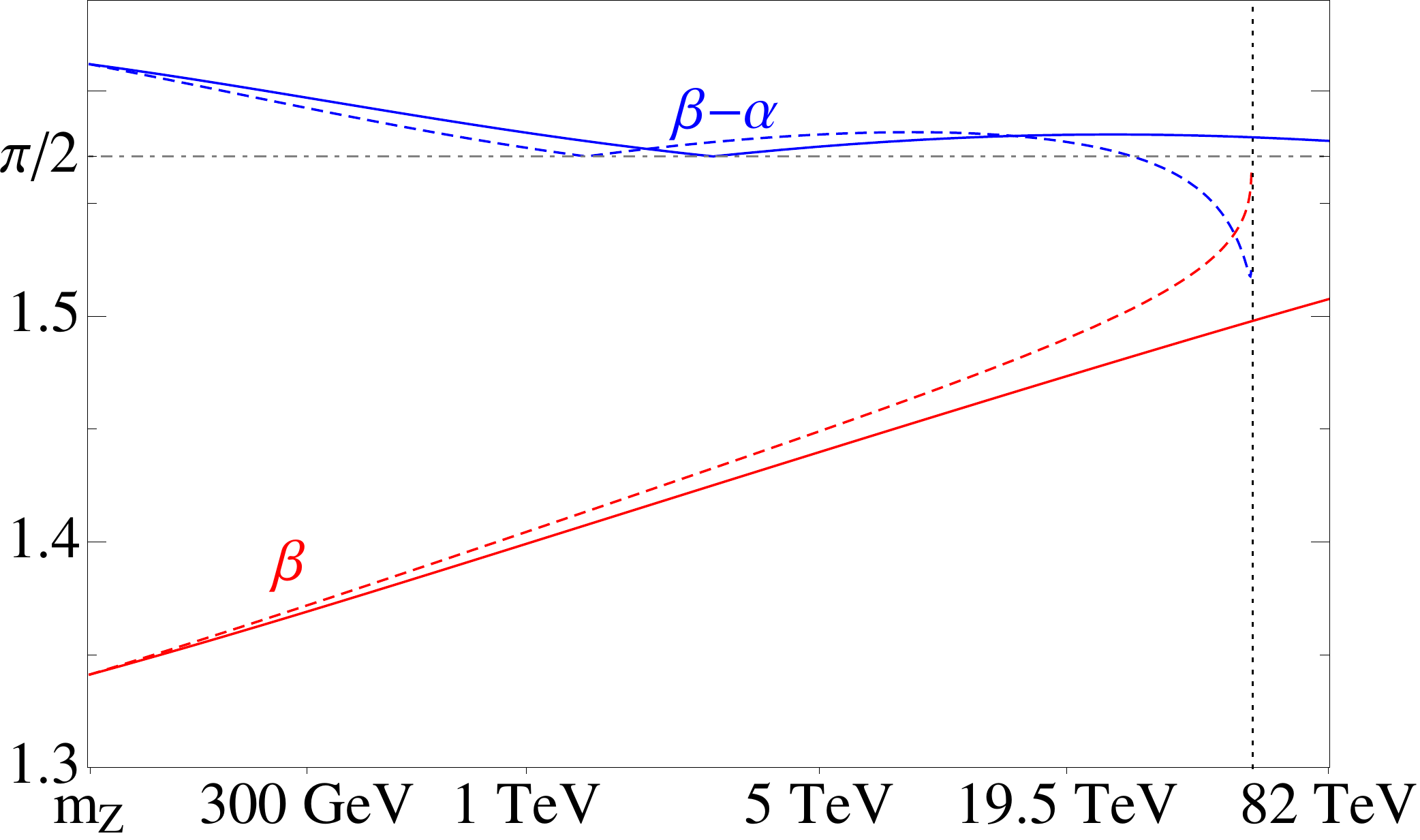}}
    \put(190,0.1){\includegraphics[width=241pt]{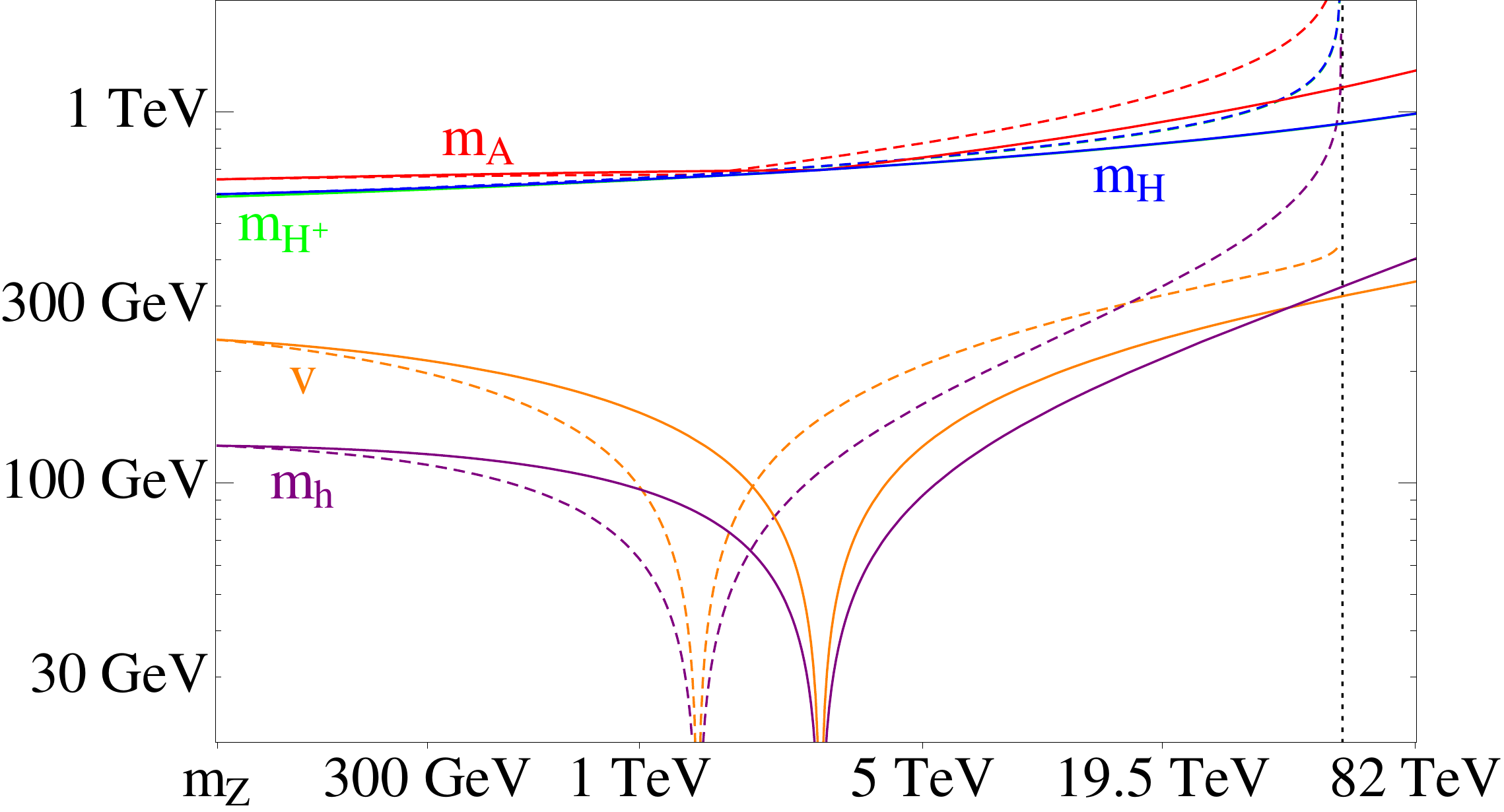}}
   \end{picture}
  }
  \caption{Leading order (dashed) and next-to-leading order (solid) RG running of the physical angles and masses for the benchmark scenario H-4. The lines for $m_{H^+}$ and $m_H$ are almost on top of each other. At a scale of \unit{1.4}{TeV} (\unit{2.8}{TeV}) at LO (NLO) $v$ and $m_h$ are $0$ and $\beta-\alpha$ is in the alignment limit of $\pi/2$.}
  \label{fig:LOvsNLOphys}
\end{figure}

In order to compare the LO and NLO RGE, we choose the scenario H-4 from \cite{Baglio:2014nea} as benchmark scenario, because all quartic couplings are relatively large already at $\mu_{\text{ew}}$. It is defined by $m_H=600$\unit{}{GeV}, $m_A=658$\unit{}{GeV}, $m_{H^+}=591$\unit{}{GeV}, $\beta-\alpha=0.513 \pi$, $\tan \beta=4.28$, and $m_{12}^2=76900$\unit{}{GeV}$^2$, and compatible with all experimental measurements so far.
The cut-off scale, where one of the quartic couplings becomes non-perturbative, is at \unit{19.5}{TeV} at LO (dashed lines), and at \unit{82}{TeV} at NLO (solid lines), see the top left panel of Fig.~\ref{fig:LOvsNLO}. The Landau poles are at \unit{54}{TeV} and \unit{3.2\cdot 10^6}{TeV}, respectively; the former is shown as a vertical dotted line in Figs.~\ref{fig:LOvsNLO} and \ref{fig:LOvsNLOphys}.
The fact that the higher order contributions ``stabilize'' the RG evolution, and thus increase both, cut-off and Landau pole scales, holds for all benchmark points from \cite{Baglio:2014nea} and is a general feature in the 2HDM: All dominant NLO contributions to the RGE of the $\lambda_i$, which are cubic in the quartic couplings, come with a negative coefficient and thus mitigate the positive LO contribution coming from quadratic $\lambda_i$ terms (see \hyperref[sec:appendix]{Appendix}).
Beneath the total values of LO and NLO running we show the relative difference between LO and NLO RGE $r_i=\left| (\lambda _i^{\text{\tiny{LO}}}-\lambda _i^{\text{\tiny{NLO}}})/\lambda _i^{\text{\tiny{NLO}}} \right|$ with respect to the scale. For this benchmark point, the relative change of $\lam{1}$ and $\lam{3}$, is as large as $10\%$ at around \unit{2}{TeV} and the difference increases even at a faster rate at higher scales. This is a first hint that the effect of the NLO contribution to the RGE in the 2HDM is non-negligible. One can see that $r_3$ diverges around \unit{35}{TeV} due to the fact that at this scale $\lambda _3^{\text{\tiny{NLO}}}$ turns to $0$.
A better quantitative measure of the  NLO vs.~LO RGE is the relative distance $\delta_{12}^L$ defined in \cite{Costa:2014qga}: For a dimensionless coupling $L$, we can define the relative distance between the LO and NLO curves  $L^{\phantom{s}\!\!\! ^\text{LO}}(t)$ and $L^{\phantom{s}\!\!\! ^\text{NLO}}(t)$ in the scale range from $t_1$ to $t_2$ as

\begin{align*}
 \delta_{12}^L &=\sqrt{\frac{\int\limits_{t_1}^{t_2}\frac{dt}{t_2-t_1}[L^{\phantom{s}\!\!\! ^\text{LO}}(t)-L^{\phantom{s}\!\!\! ^\text{NLO}}(t)]^2}{\int\limits_{t_1}^{t_2}\frac{dt}{t_2-t_1}L^{\phantom{s}\!\!\! ^\text{NLO}}(t)^2}}.
\end{align*}

For H-4, $\delta_{12}^{\lambda_1}$ is $38\%$, if we integrate from $m_Z$ to the LO cut-off at \unit{19.5}{TeV}. To quantify the typical size of $\delta_{12}^{\lambda_c}$, where $\lambda_c$ is the quartic coupling with the lowest perturbativity violating scale, we checked all benchmark points of \cite{Baglio:2014nea}, and found values between $17\%$ and $45\%$, which indicates that in general the two-loop corrections are not negligible.\footnote{It is important to note that this definition of $\delta_{12}^L$ is only meaningful, if the denominator inside the square root does not become too small.}

In Fig.~\ref{fig:LOvsNLO} we do not show the running of the gauge couplings $g_1$, $g_2$ and $g_3$, since the two-loop corrections are too small to be visible.
Also, the running of the Yukawa couplings is not significantly altered going from LO to NLO. However, due to the different assignment of the Higgs fields in the four types, we start with different values at the low scale (see Table \ref{tab:types}); that is why we denote the Yukawa couplings in the upper right panel of Fig.~\ref{fig:LOvsNLO} as introduced in Eq.~\eqref{eq:yuk}. (Note that only two of them are non-zero, depending on the type of $Z_2$.)
Among the mass parameters, $m_{12}^2$ changes least if we run to higher scales, which we also observe as general feature of all types. $m_{11}^2$ and $m_{22}^2$ can have very different values at different scales, compare the lower right panel of Fig.~\ref{fig:LOvsNLO}. Neither of the mass couplings feeds back to the dimensionless couplings, as the RGE of the latter do not depend on $m_{12}^2$, $m_{11}^2$ or $m_{22}^2$.
Furthermore, we have checked the mentioned benchmark scenarios for fixed point behaviour and do not find any below the perturbativity cut-off.

If we switch to the physical parameter basis, we observe that also the RG running of the mixing angles can be sizable, see left side of Fig.~\ref{fig:LOvsNLOphys}. The scale at which $\beta-\alpha$ hits the alignment limit corresponds to vanishing $v$ and $m_h$, which can be seen in the right panel of Fig.~\ref{fig:LOvsNLOphys}, where we show the running of all physical mass parameters. We find that the breakdown of the vacuum expectation value at some scale above $\mu_{\text{ew}}$ is a general feature and occurs for all benchmark scenarios that we have analyzed; this is also observable in the benchmark points of \cite{JuarezW.:2012qa}.

After scrutinizing one benchmark point, we want to discuss more general features that can be found in comprehensive fits. Especially the general dependence of $\mu_{\text{st}}$ on the value of $\lambda_i (m_Z)$ will be an interesting question in the following.

\subsection{Fits without experimental data}
\label{subsec:withoutexp}

A parametrization independent way of setting upper limits to the quartic couplings of the Higgs potential is the requirement that the scattering matrix of $\Phi_i \Phi_j \to \Phi_i \Phi_j$ processes is unitary. This corresponds to the condition that its absolute eigenvalues should be smaller than $16\pi$. The tree-level expressions \cite{Huffel:1980sk,Maalampi:1991fb,Kanemura:1993hm,Akeroyd:2000wc,Ginzburg:2005dt} are widely used theoretical constraints for the 2HDM; however, it seems that these bounds are very conservative. Studies involving higher order corrections have shown that the eigenvalues cannot be larger than $2\pi$ in the SM \cite{Nierste:1995zx}, and this bound has been adopted for the 2HDM of type II in \cite{Baglio:2014nea}. Analyzing maximally allowed cut-off scales can shed light on how well this bound is motivated from the RGE perspective.

In this section, we only want to impose the Higgs potential bounds, regardless of experimental constraints, in order to show the impact of the former on the 2HDM parameters. Since the assumption of having a stable potential affects the potential parameters, we express our results in terms of the five quartic couplings and $\tan \beta$. (The latter modifies the Yukawa couplings as compared to their SM values, see Table \ref{tab:types}.) Due to the smallness of $Y_b$ and $Y_\tau$, their influence on the RGE is very weak and differences between the four $Z_2$ types are not visible in the $\lambda_i$ planes.

\begin{figure}
  \centering
  \resizebox{350pt}{!}{
   \begin{picture}(300,270)(60,0)
    \put(-10,150){\includegraphics[width=140pt]{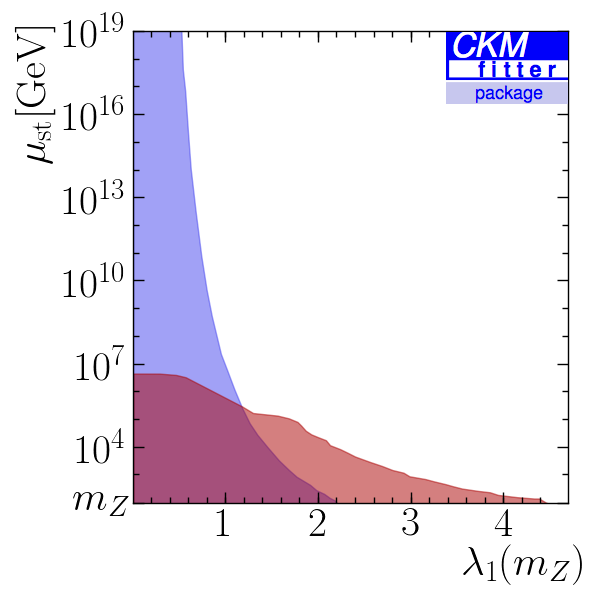}}
    \put(140,150){\includegraphics[width=140pt]{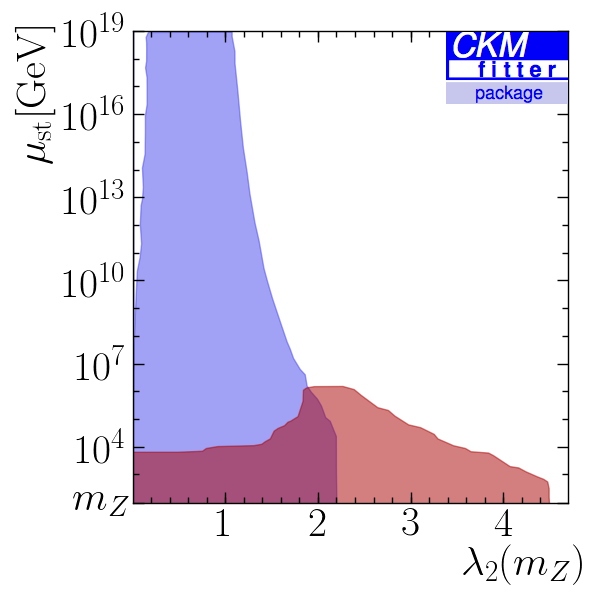}}
    \put(290,150){\includegraphics[width=140pt]{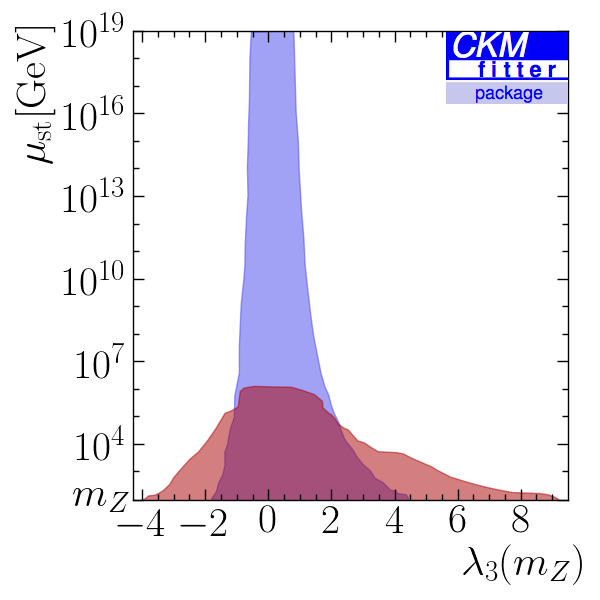}}
    \put(-10,0){\includegraphics[width=140pt]{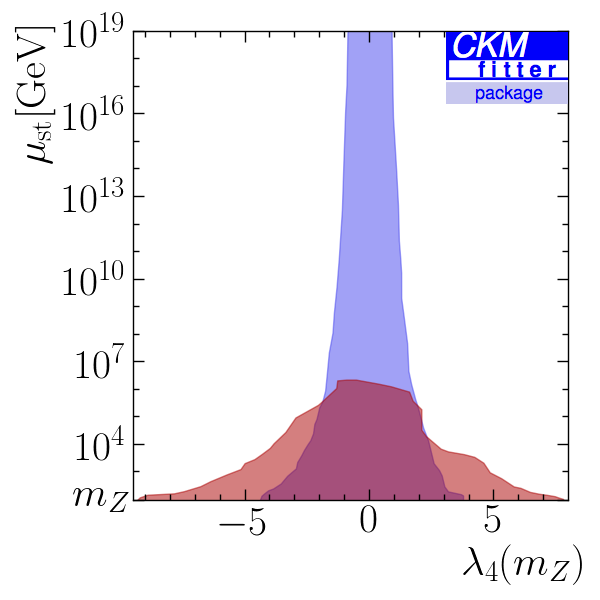}}
    \put(140,0){\includegraphics[width=140pt]{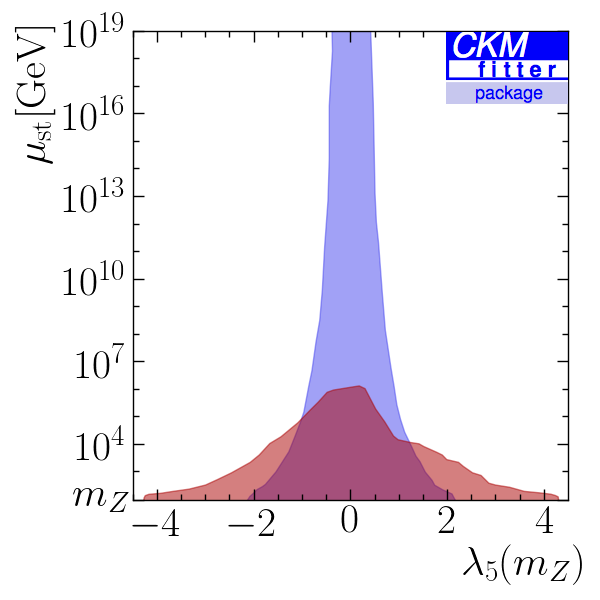}}
    \put(290,0){\includegraphics[width=140pt]{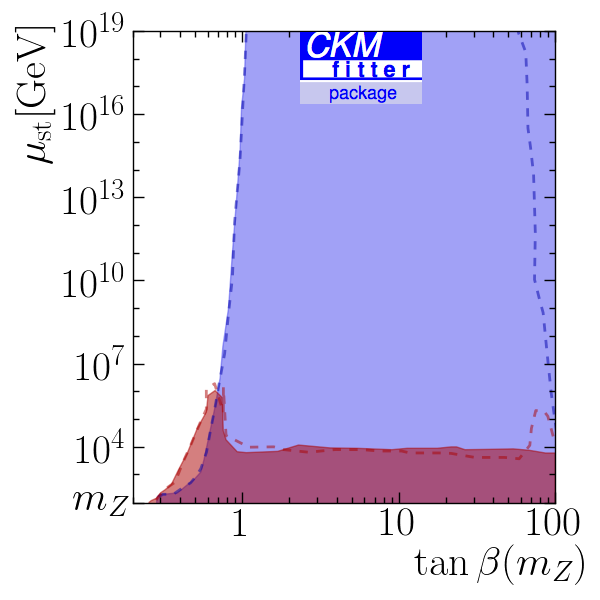}}
   \end{picture}
  }
  \caption{The blue (light) shaded regions show the dependence of $\mu_{\text{st}}$ on the values for the quartic couplings $\lambda_i$ and $\tan \beta$ at $m_Z$ if we choose an upper limit of $2\pi$ for the absolute $S$-matrix eigenvalues. The red (dark) regions illustrate the allowed regions, if we take $4\pi$ instead and force at least one of the eigenvalues to be larger than $2\pi$ in magnitude. All types give the same dependence for the $\lambda_i$. For $\tan \beta$, we show the possible regions in type I and X as shaded, and the areas below the dashed lines correspond to type II and Y.}
  \label{fig:2pivs4pi-I}
\end{figure}

In Fig.~\ref{fig:2pivs4pi-I} we show the dependence of the cut-off scale on the values of quartic couplings and $\tan \beta$ at the electroweak scale, for the two cases that either all eigenvalue moduli are smaller than $2\pi$ or that at least one of them is larger than $2\pi$. Our fits show that forcing at least one eigenvalue of the $S$-matrix to have an absolute value larger than $2\pi$ reduces the maximal cut-off scale $\mu_{\text{st}}$ to be at 5$\cdot$\unit{10^{6}}{GeV} instead of the Planck scale; if we set at least one eigenvalue modulus larger than $4\pi$, the maximal $\mu_{\text{st}}$ is at a few TeV. If we want to maintain a stable Higgs potential up to $\mu_{\text{Pl}}$, the largest eigenvalue can have a magnitude of at most $2.5$ ($\approx 0.8 \pi$).
Naturally, a larger upper bound on the eigenvalues allows for larger quartic couplings. But one can also see that cut-off scales larger than \unit{10}{TeV} are only allowed for a very narrow range of $\tan \beta$ around $0.7$ and -- only in type II and Y -- for an additional narrow range around $80$. While the low $\tan \beta$ scenarios are known to be disfavoured for light 2HDM spectra by flavour observables, we will see in the next section that also the large $\tan \beta$ regions are now excluded in type II. So we can conclude for all types but type Y that assuming $\mu_{\text{st}}>10$\unit{}{TeV} and not too heavy new Higgs states all $S$-matrix eigenvalues need to be smaller than $2\pi$ in magnitude. We will use the upper bound of $2\pi$ in the following.
Fig. \ref{fig:2pivs4pi-I} also shows the allowed $\lambda_i(m_Z)$ and $\tan \beta(m_Z)$ intervals for $\mu_{\text{st}}$ at Planck scale. Roughly speaking, stability up to \unit{10^{19}}{GeV} requires $|\lambda_i (m_Z)|\lesssim1$ and $\tan \beta (m_Z)>1$. In this case, we also observe a lower limit on $\lambda_2(m_Z)$, which cannot be smaller than $0.15$. In type II and Y, $\tan \beta (m_Z)$ is also limited from above and cannot be larger than $60$. We list the precise ranges of the parameters in Table \ref{tab:lambdavalues}.
\begin{table}[htb]
  \centering
  \caption{Allowed intervals for the quartic couplings and $\tan \beta$ at the electroweak scale, if we assume stability at $\mu_{\text{ew}}$ (first line) and up to $\mu_{\text{Pl}}$ (second line).}\vspace{0.2cm}
  \begin{tabular}{l|l|l|l|l|l|l}
    \hline\hline
     & $\lambda_1(m_Z)$ & $\lambda_2(m_Z)$ & $\lambda_3(m_Z)$ & $\lambda_4(m_Z)$ & $\lambda_5(m_Z)$ & $\tan \beta(m_Z)$ \\
    \hline
     $\mu_{\text{st}}=\mu_{\text{ew}}$ & $[0;2.22]$ & $[0;2.20]$ & $[-1.8;4.4]$ & $[-4.4;3.8]$ & $[-2.1;2.1]$ & $>0.3$ \\
    \hline
     $\mu_{\text{st}}=\mu_{\text{Pl}}$ & $[0;0.52]$ & $[0.15;1.06]$ & $[-0.6;0.8]$ & $[-0.9;0.9]$ & $[-0.4;0.4]$ & $>1.0$ in type I and X\\
     &&&&&& $[1.0;60]$ in type II and Y\\
    \hline \hline
  \end{tabular}
  \label{tab:lambdavalues}
\end{table} 

The bounds on $\lambda_5(m_Z)$ give us a handle on the question whether the $Z_2$ symmetry can be exact with stability up to the Planck scale: following \cite{Gunion:2002zf}, we find that the soft breaking parameter can be written as
\begin{align}
 m_{12}^2 &=\frac{\tan\beta}{1+\tan^2\beta}\left( m_A^2 +v^2\lambda_5\right) \label{eq:m12sqformula}.
\end{align}

Increasing $\mu_{\text{st}}$ to higher scales not only gives a stronger lower bound on $\lambda_5$, but simultaneously also excludes low $m_A$ values, such that beyond $\mu_{\text{st}}\approx 10^{10}$\unit{}{GeV} a cancellation between the pseudoscalar and the $\lambda_5$ contribution in \eqref{eq:m12sqformula} is no longer possible. Hence we confirm the LO result of \cite{Das:2015mwa} that a 2HDM with $\mu_{\text{st}}> 10^{10}$\unit{}{GeV} has to be softly broken, which does not change significantly if we use NLO RGE.

The inclusion of experimental bounds has only very little visible impact on the potential parameters, that is why in the following section we switch to the physical basis.

\subsection{Fits with experimental data}
\label{subsec:withexp}

In this section we want to show the impact of the experimental results discussed in Section \ref{sec:constraints} on the physical parameter space at the electroweak scale, once assuming a stable scalar potential at $\mu_{\text{ew}}$ and once for stability up to $\mu_{\text{Pl}}$. We put special emphasis on the dependence of mass parameters on the relevant angles in order to investigate how large deviations from the alignment limit can still be.

\begin{figure}
  \centering
  \resizebox{450pt}{!}{
   \begin{picture}(450,330)(0,0)
    \put(0,180){\includegraphics[width=220pt]{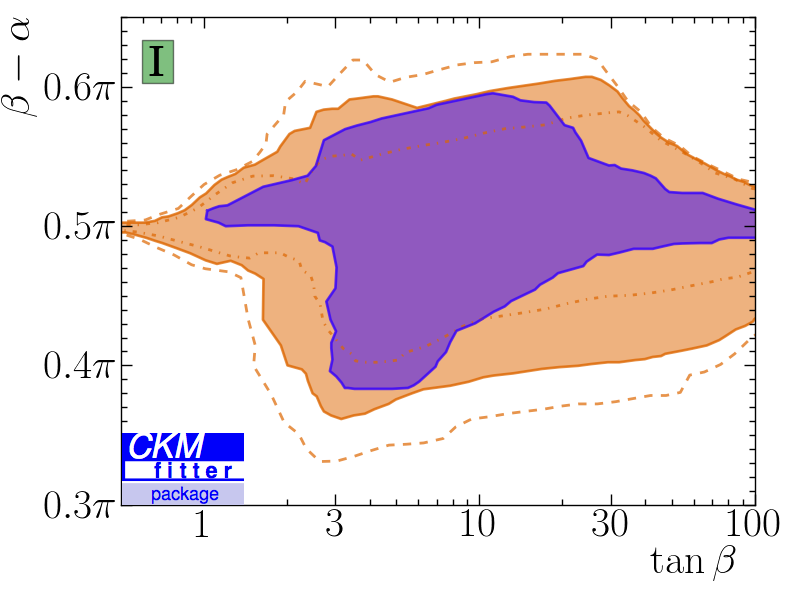}}
    \put(240,180){\includegraphics[width=220pt]{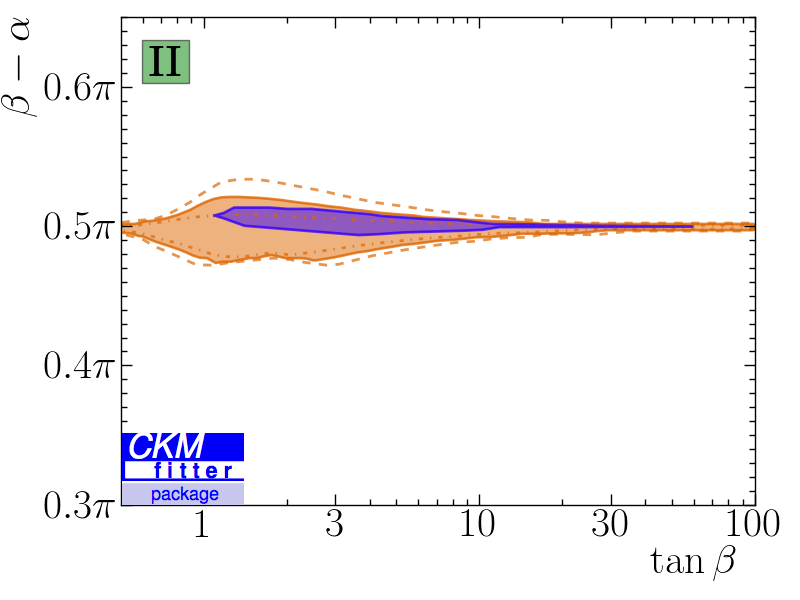}}
    \put(0,0){\includegraphics[width=220pt]{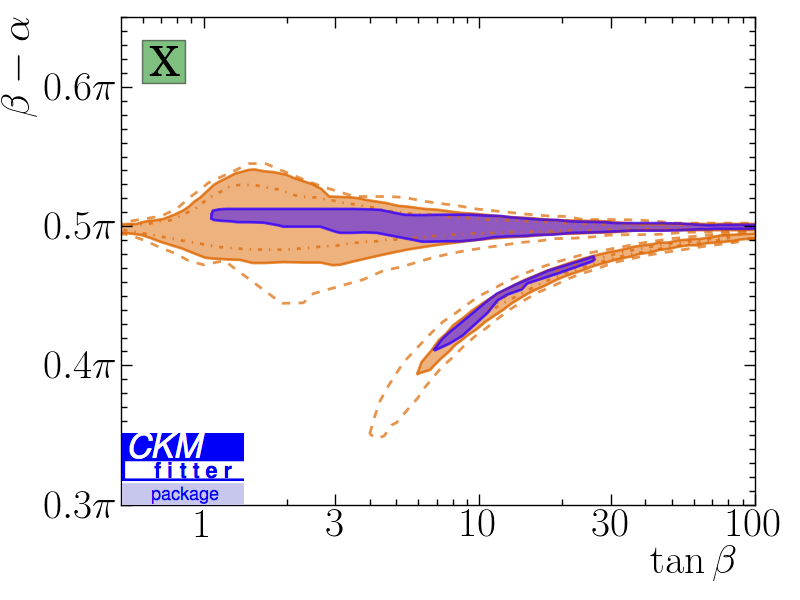}}
    \put(240,0){\includegraphics[width=220pt]{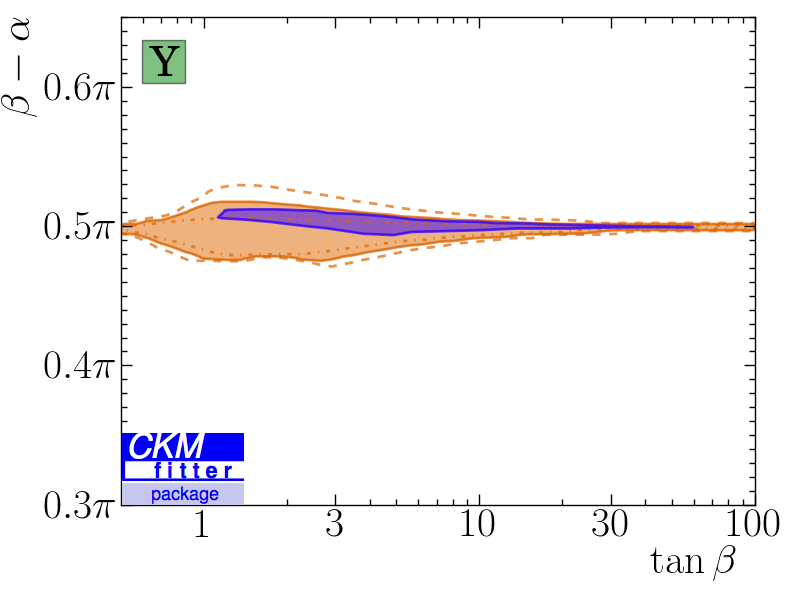}}
   \end{picture}
  }
  \caption{$\tan \beta$--($\beta-\alpha$) plane in type I (top left), type II (top right), type X (bottom left) and type Y (bottom right) at $m_Z$ with stability imposed at $\mu_{\text{ew}}$ in orange (light) and at $\mu_{\text{Pl}}$ in purple (dark). The dash-dotted, continuous and dashed lines border the $1\sigma$, $2\sigma$ and $3\sigma$ allowed regions, respectively; the $2\sigma$ region -- which roughly corresponds to the $95\%$ C.L. area -- is shaded.}
  \label{fig:explogtbvsbma}
\end{figure}

In Fig.~\ref{fig:explogtbvsbma}, we show the $\tan \beta$--$(\beta-\alpha)$ plane for type I on the upper left, for type II on the upper right, for type X on the lower left and for type Y on the lower right. For a stable potential at the electroweak scale (orange) we show the $1\sigma$, $2\sigma$ and $3\sigma$ allowed regions (the $2\sigma$ region is shaded, the $1\sigma$ and $3\sigma$ contours are defined by the dash-dotted and dashed lines, respectively), and for a stable potential at Planck scale (purple shaded) we only present the $2\sigma$ region.
With stability at $\mu_{\text{ew}}$, $\tan \beta$ is not constrained by any observable. For 2HDM masses below \unit{1}{TeV}, however, we find a lower limit of $0.7$ in all types (cf. \cite{Eberhardt:2014kaa}) as well as an upper limit of roughly $60$ in type II.
In contrast, $\beta-\alpha$ is constrained in all types to be fairly close to the alignment limit; the exact limits can be found in Table \ref{tab:bmavalues}. In type I, the deviations from $\beta-\alpha=\pi/2$ can be as large as $0.1\pi$ for a broad range of intermediate values of $\tan \beta$. Only a narrow band which is compatible with all constraints and at the same time allows for deviations from the alignment limit by more than $0.05\pi$ survives the type X fits; within this band $\tan \beta$ is larger than $6$. In type II and Y, this band would in principle also exist, but the new determination of the lower bound on $m_{H^+}$ from ${\rm Br}(\oline{B}\to X_s\gamma)$ excludes scenarios which feature 2HDM heavy scalar masses below \unit{350}{GeV} and cut away the ``lower branches'' in the $\tan \beta$--$(\beta-\alpha)$ plane. This allows us to exclude a deviation by more than $0.03\pi$ from the alignment limit in those two types of $Z_2$ symmetry.
We have seen in Section \ref{subsec:withoutexp} that imposing stability up to $\mu_{\text{Pl}}$ constrains the quartic couplings; at this point, we want to shed light on the effect on the physical parameters. In Fig.~\ref{fig:2pivs4pi-I}, we already observed that $\tan \beta$ is constrained from below in type I and type X and additionally from above in type II and type Y. In type I we can also observe that for $\mu_{\text{st}}=\mu_{\text{Pl}}$, $\beta-\alpha$ has to be closer to $\pi/2$ for low and high values of $\tan \beta$ than in the case of $\mu_{\text{st}}=\mu_{\text{ew}}$. In type X, the ``lower branch'' only occurs at $6.8<\tan \beta<26$ now, and also in type II and Y the allowed region is stronger constrained. Interestingly, the lower bound of $1$ on $\tan \beta$ is not necessarily the same if we impose the alignment limit; in type II and Y we find $\tan \beta \gtrsim 2$ in this case. The reason why this value is smaller than the one found in \cite{Das:2015mwa} is that we use NLO RGE. At leading order, we confirm their result that $\tan \beta<3$ is excluded in the alignment limit.

\begin{table}[htb]
  \centering
  \caption{Allowed intervals for $\beta-\alpha$ at the electroweak scale (and its sine and cosine) for all types of $Z_2$ symmetry, if we assume stability at $\mu_{\text{ew}}$ (first three lines) and up to $\mu_{\text{Pl}}$ (lines four to six).}\vspace{0.2cm}
  \begin{tabular}{l|l|l|l|l|l}
    \hline\hline
    & & Type I & Type II & Type X & Type Y \\
    \hline
     $\mu_{\text{st}}=\mu_{\text{ew}}$ & $\beta-\alpha$ & $[1.14;1.91]$ & $[1.49;1.64]$ & $[1.24;1.70]$ & $[1.50;1.63]$ \\
     & $\cos(\beta-\alpha)$ & $[-0.33;0.42]$ & $[-0.068;0.081]$ & $[-0.13;0.32]$ & $[-0.057;0.076]$ \\
     & $\sin(\beta-\alpha)$ & $[0.908;1]$ & $[0.997;1]$ & $[0.946;1]$ & $[0.997;1]$ \\
    \hline
     $\mu_{\text{st}}=\mu_{\text{Pl}}$ & $\beta-\alpha$ & $[1.21;1.87]$ & $[1.55;1.62]$ & $[1.29;1.61]$ & $[1.55;1.61]$ \\
     & $\cos(\beta-\alpha)$ & $[-0.30;0.36]$ & $[-0.044;0.018]$ & $[-0.04;0.27]$ & $[-0.040;0.018]$ \\
     & $\sin(\beta-\alpha)$ & $[0.934;1]$ & $[0.999;1]$ & $[0.962;1]$ & $[0.999;1]$ \\
    \hline \hline
  \end{tabular}
  \label{tab:bmavalues}
\end{table} 

\begin{figure}
  \centering
  \resizebox{450pt}{!}{
   \begin{picture}(450,330)(0,0)
    \put(0,180){\includegraphics[width=220pt]{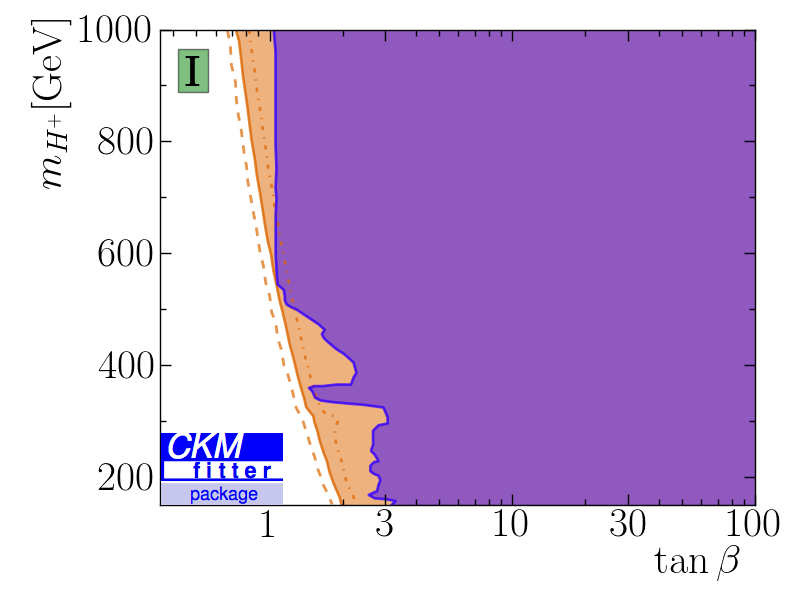}}
    \put(240,180){\includegraphics[width=220pt]{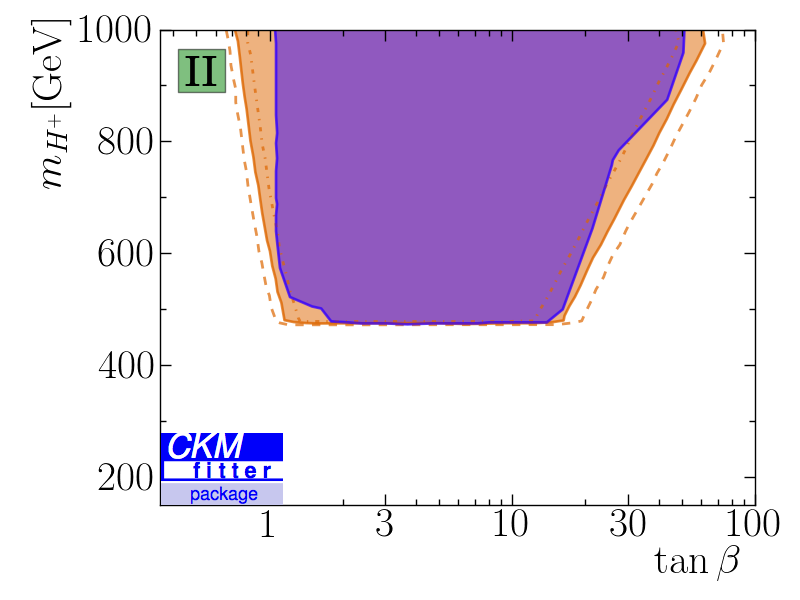}}
    \put(0,0){\includegraphics[width=220pt]{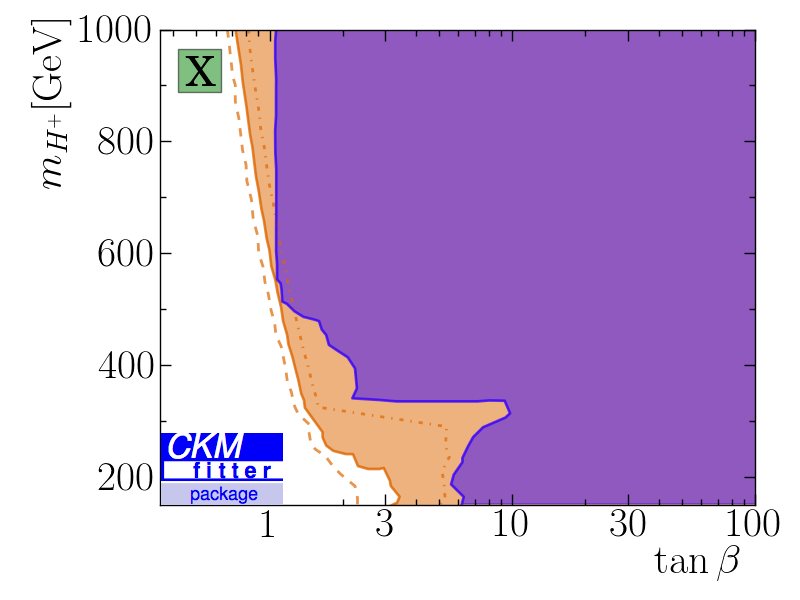}}
    \put(240,0){\includegraphics[width=220pt]{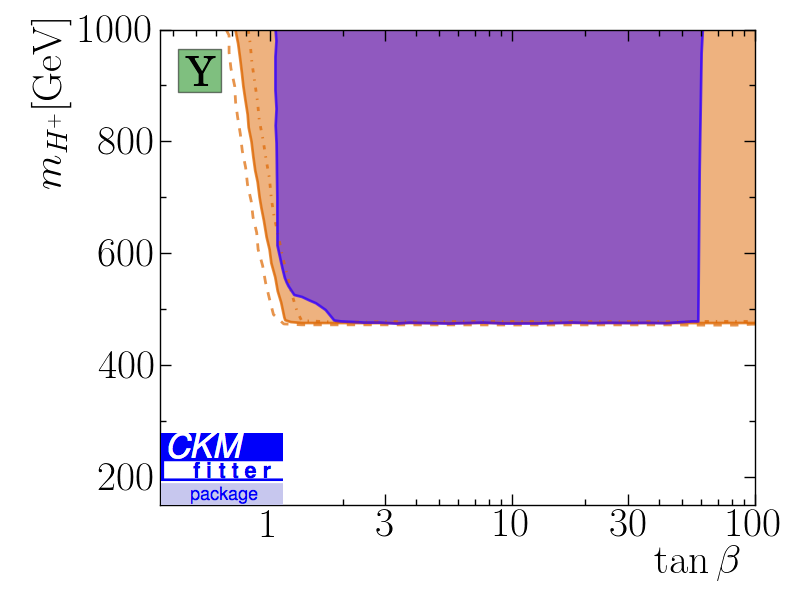}}
   \end{picture}
  }
  \caption{$\tan \beta$--$m_{H^+}$ plane in type I (top left), type II (top right), type X (bottom left) and type Y (bottom right) at $m_Z$ with stability imposed at $\mu_{\text{ew}}$ in orange (light) and at $\mu_{\text{Pl}}$ in purple (dark). The dash-dotted, continuous and dashed lines border the $1\sigma$, $2\sigma$ and $3\sigma$ allowed regions, respectively; the $2\sigma$ region -- which roughly corresponds to the $95\%$ C.L. area -- is shaded.}
  \label{fig:explogtbvsmHp}
\end{figure}

In Fig.~\ref{fig:explogtbvsmHp}, we show the dependence of the charged Higgs mass bounds on $\tan \beta$; let us first discuss the case $\mu_{\text{st}}=\mu_{\text{ew}}$: In type I the strongest constraint for low $\tan \beta$ values comes from the mass difference in the $B_s$ system. The other observables have no visible impact on this plane. The same holds for type X, except for $m_{H^+}<300$\unit{}{GeV}, where direct Higgs searches additionally cut away low $\tan \beta$ values. For type II and type Y, ${\rm Br}(\oline{B}\to X_s\gamma)$ yields a lower limit of \unit{480}{GeV} on $m_{H^+}$; for large masses and low $\tan \beta$, the bound from the mass difference in the $B_s$ system is stronger. In case of the type II we also find that a light charged Higgs is excluded for large $\tan \beta$ values; for instance if $\tan \beta=30$, we obtain $m_{H^+}>700$\unit{}{GeV}. This is an effect only visible in a global fit: for large $\tan \beta$, heavy Higgs searches (mainly the tauonic decays) exclude light $m_H$ and $m_A$. Electroweak precision data, however, are not compatible with too large mass splittings between the heavy neutral and the charged Higgs particles, so also the charged Higgs cannot be too light if $\tan \beta$ is large. This also qualifies that we did not use data from (semi-)tauonic $B$ decays, which would give a weaker bound on the same corner of the type II plane.
Type Y also features this constraint from the neutral Higgs searches, but it is much weaker and would only be visible for $\tan \beta>100$ because the $\tau$ and $b$ couplings to $H$ and $A$ cannot be enhanced simultaneously.
Requiring stability up to $\mu_{\text{Pl}}$ gives almost the same regions as with stability at $\mu_{\text{ew}}$, only that $\tan \beta$ gets constrained at the borders to stay within the limits from Table \ref{tab:lambdavalues}.

\begin{figure}
  \centering
  \resizebox{450pt}{!}{
   \begin{picture}(450,610)(0,0)
    \put(0,495){\includegraphics[width=220pt]{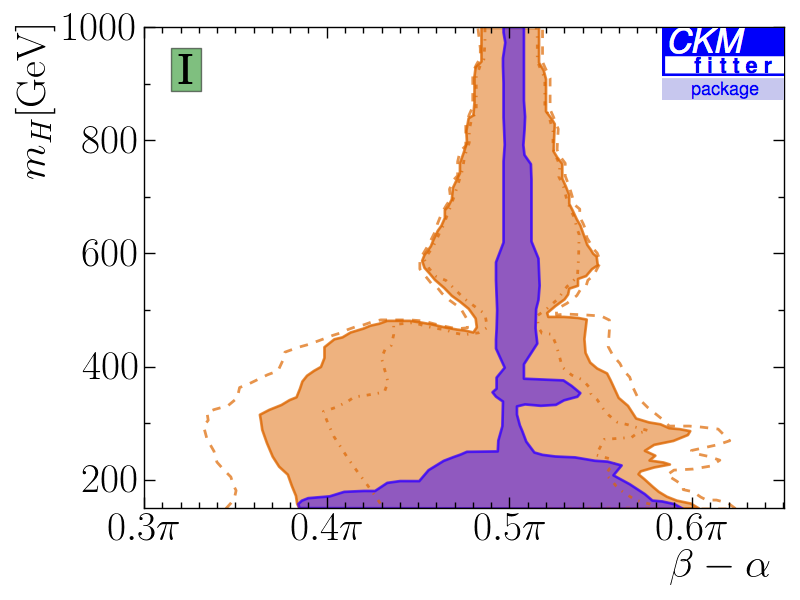}}
    \put(240,495){\includegraphics[width=220pt]{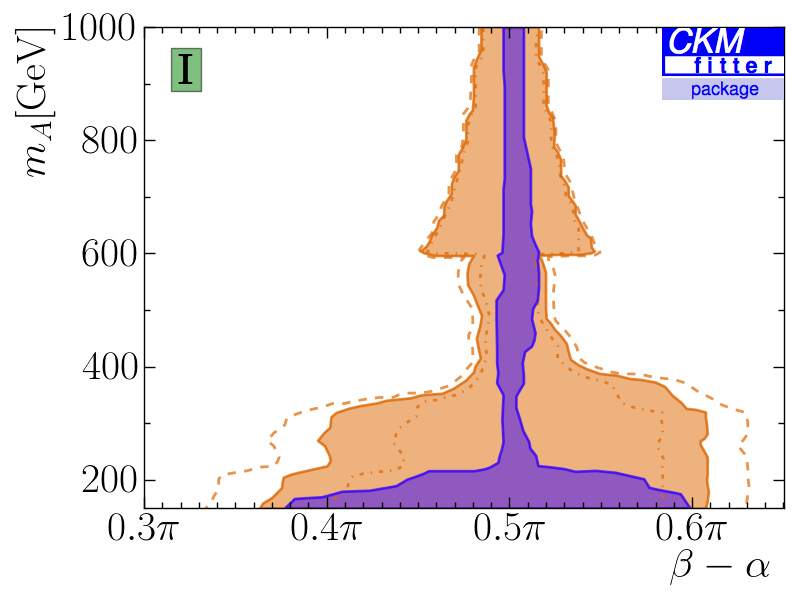}}
    \put(0,330){\includegraphics[width=220pt]{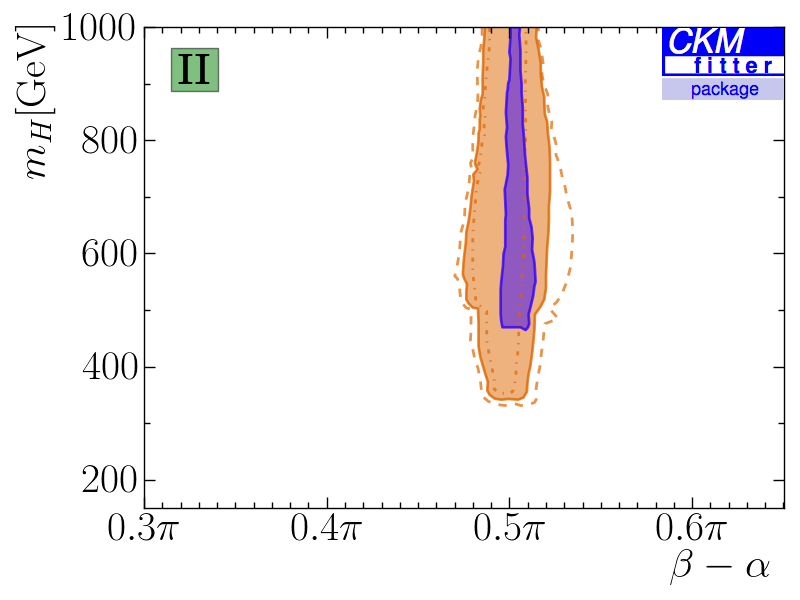}}
    \put(240,330){\includegraphics[width=220pt]{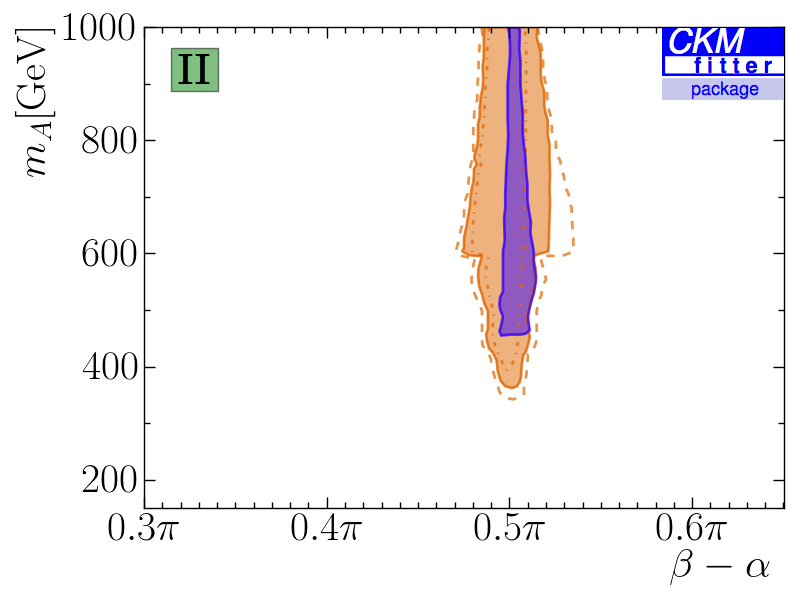}}
    \put(0,165){\includegraphics[width=220pt]{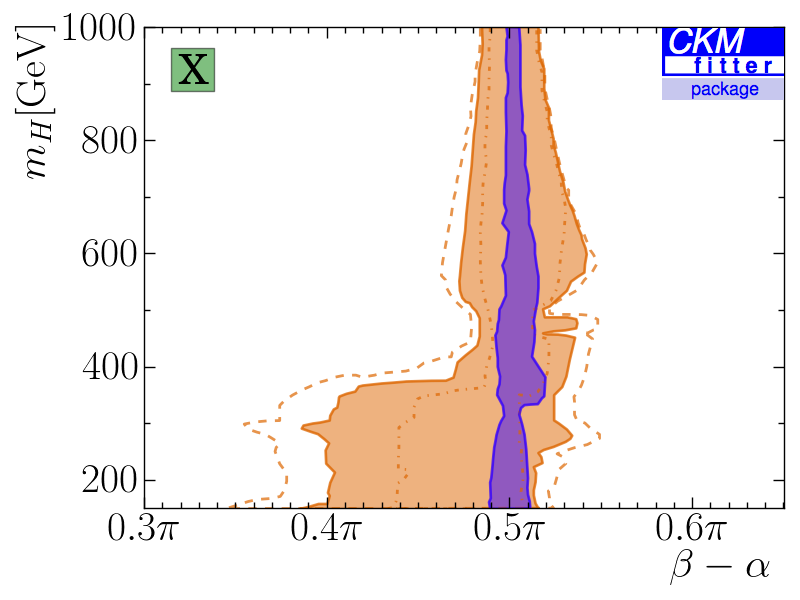}}
    \put(240,165){\includegraphics[width=220pt]{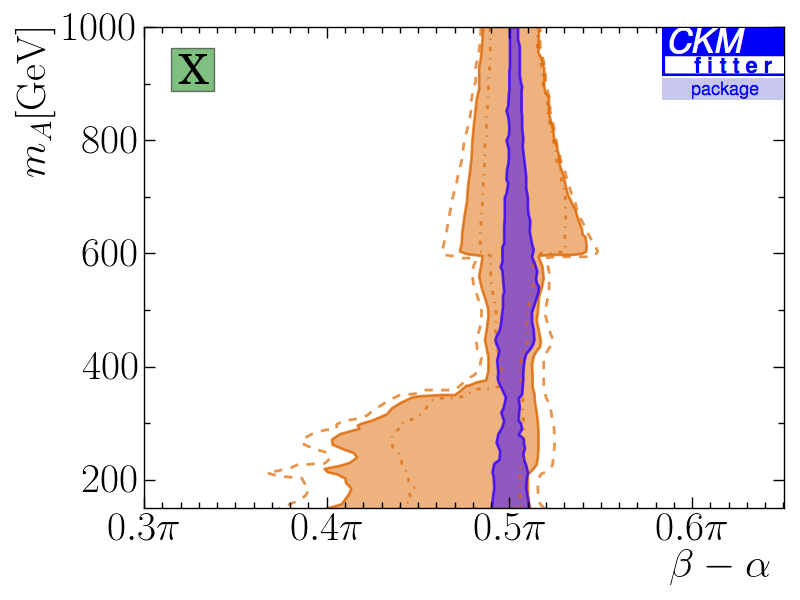}}
    \put(0,0){\includegraphics[width=220pt]{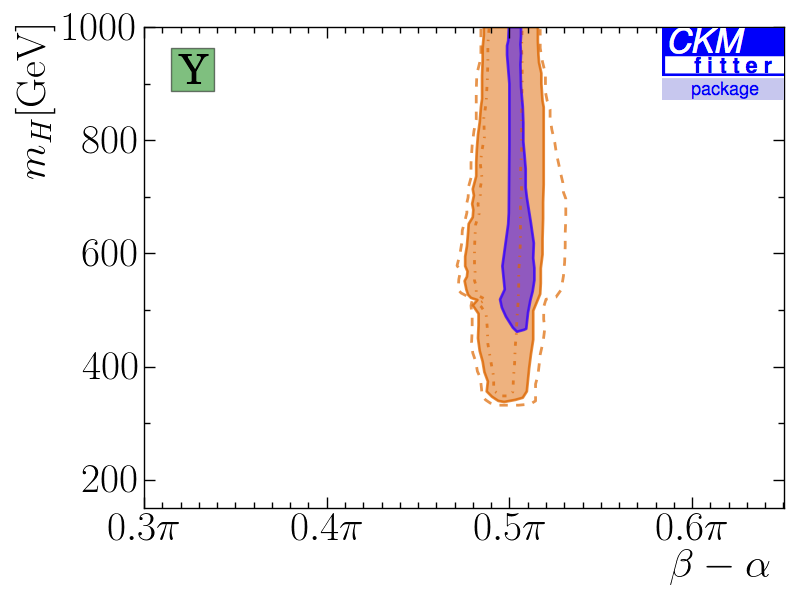}}
    \put(240,0){\includegraphics[width=220pt]{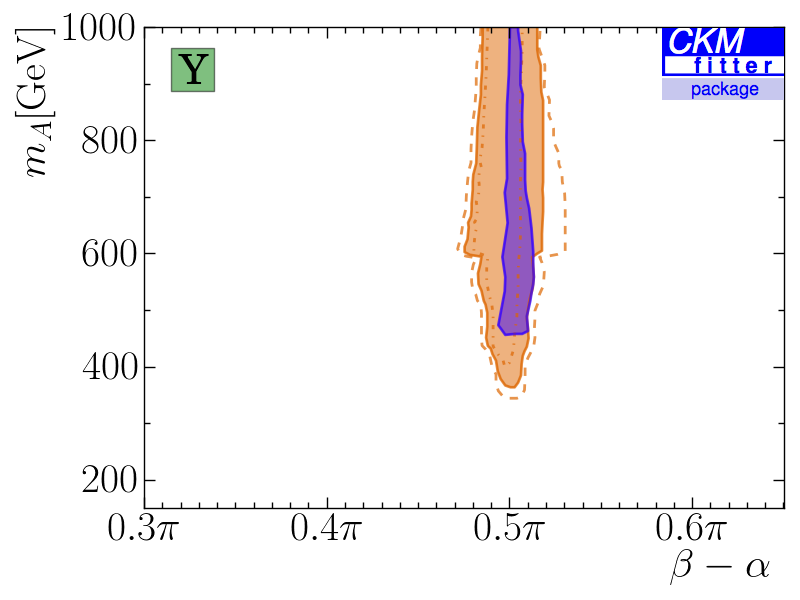}}
   \end{picture}
  }
  \caption{($\beta -\alpha$)--$m_H$ plane (on the left) and ($\beta -\alpha$)--$m_A$ plane (on the right) in type I, type II, type X and type Y (from top to bottom) at $m_Z$ with stability imposed at $\mu_{\text{ew}}$ in orange (light) and at $\mu_{\text{Pl}}$ in purple (dark). The dash-dotted, continuous and dashed lines border the $1\sigma$, $2\sigma$ and $3\sigma$ allowed regions, respectively; the $2\sigma$ region -- which roughly corresponds to the $95\%$ C.L. area -- is shaded.}
  \label{fig:expbmavsmA0}
\end{figure}

While the charged Higgs searches mainly depend on $\tan \beta$, neutral Higgs signals strongly depend on the deviation from the alignment limit, i.e.\ the actual value of $\beta-\alpha$. Therefore, we show in Fig.~\ref{fig:expbmavsmA0} the allowed regions in the $(\beta-\alpha)$--$m_H$ and $(\beta-\alpha)$--$m_A$ planes.
For all types we observe that for neutral masses above \unit{600}{GeV} the deviation of $\beta-\alpha$ from $\pi/2$ can be $0.05\pi$ at most due to the stability bound. The larger deviations in type I and X correspond to neutral masses below \unit{500}{GeV}, where the heavy Higgs searches become relevant constraints. As explained above, these regions are indirectly excluded by $m_{H^+}>480$\unit{}{GeV} in type II and Y and we obtain lower limits of \unit{340}{GeV} and \unit{360}{GeV} for $m_H$ and $m_A$, respectively. This lower bound on the pseudoscalar mass translates directly into a bound on the question whether the $Z_2$ can be exact, and combining Eq.~\eqref{eq:m12sqformula} with the information from the allowed $\lambda_5$ range in Fig.~\ref{fig:2pivs4pi-I}, we can conclude that even with a stability cut-off at the electroweak scale $m_{12}^2=0$ is very hard to achieve in type II and Y.
If we additionally impose stability up to the Planck scale,we can see that sizeable deviations from the alignment limit are only possible for $m_H<250$\unit{}{GeV} and $m_A<230$\unit{}{GeV} in type I. Type X fits do not allow for $\beta-\alpha$ deviations larger than $0.02\pi$ for heavy neutral scalar masses above \unit{150}{GeV}. In type II and Y, the lower bounds on the neutral masses increase to $m_H>460$\unit{}{GeV} and $m_A>455$\unit{}{GeV}, because in general, higher stability cut-off scales allow for less freedom in the mass splittings between $m_H$, $m_A$ and $m_{H^+}$ \cite{Cheon:2012rh,Das:2015mwa}. In our fits we find an upper limit of \unit{45}{GeV} on the absolute mass splittings for all $Z_2$ symmetry types.

\section{The hierarchy problem}
\label{sec:naturalness}

As we have already mentioned, there is a large scale difference between the Planck scale and the scale at which electroweak symmetry breaking occurs. This gap leads to the hierarchy problem of the Higgs mass: if loop corrections can be of order of $\mu_{\text{Pl}}$, why do they cancel each other almost perfectly, such that the Higgs mass is 17 orders of magnitude smaller?
The cancellation of these mass corrections to retain a naturally light $m_h$ was first proposed by Veltman \cite{Veltman:1980mj}, therefore also referred to as ``Veltman conditions'', and was first applied at leading order to the 2HDM by Newton and Wu \cite{Newton:1993xc}. Unlike in supersymmetry, in the 2HDM there is no mechanism which naturally accounts for these cancellations. Still, there might be a hidden symmetry of which we are not aware, so nevertheless it is interesting to address this question. In the framework of the 2HDM, this hierarchy problem does not only affect $m_h$ but in principle also the other scalar masses, if they are not in the decoupling limit. However, since we do not know whether the heavier scalars are decoupled or not, we will only discuss the hierarchy problem of the already discovered \unit{125}{GeV} scalar in the following.\\
The largest one-loop contributions to the Higgs mass come from terms that are quadratic in $\mu_{\text{nat}}$, if we assume that the 2HDM is valid up to a given scale $\mu_{\text{nat}}$ and use this scale as a cut-off. Leading higher order contributions get an additional factor of $[\ln (\mu_{\text{nat}}/\mu_{\text{ew}})]^n$, where $n+1$ is the number of loops. If $\mu_{\text{nat}}$ is large enough, the logarithmic factor might compensate for the loop suppression, and the power series of the higher order corrections no longer converges. So requiring the cancellation of the first order Higgs mass correction -- like often applied in the literature \cite{Andrianov:1994za,Jora:2013opa,Chakraborty:2014oma,Biswas:2014uba} -- is not sufficient if we do not know about the higher order terms. Only if we assume perturbativity of the power series, we can make a valid statement about whether the Higgs mass at the electroweak scale can be natural in the 2HDM or at least whether the hierarchy problem can be mitigated. This assumption of perturbativity is analogous to the one applied above on the Yukawa and quartic Higgs couplings.\\
All higher order leading logarithm mass corrections proportional to $\mu_{\text{nat}}^2$ are given by

\begin{align}
 \delta m_h^2 &= \frac{\mu_{\text{nat}}^2}{16\pi^2} \left[ \sum\limits_{n=0}^\infty f_n(\lambda_i,Y_i,g_i) \left( \ln \frac{\mu_{\text{nat}}}{\mu_{\text{ew}}}\right) ^n\right]. \label{quadraticmasscorrections}
\end{align}

As described in \cite{Kolda:2000wi}, especially for low cut-off scales the power series can be perturbative. However, we need to be careful to keep the leading logarithm sufficiently large with respect to the lower powers in the logarithm assuming that the leading logarithm gives the largest contribution.
The leading coefficient function can be derived from the one-loop Higgs mass corrections and reads as
\begin{align*}
f_0 (\lambda_i,Y_i,g_i)=&-\frac{3}{2}\cos (2 \alpha ) (\lambda_1-\lambda_2) +\frac{3}{2} \lambda_1 +\frac{3}{2} \lambda_2 +2 \lambda_3 + \lambda_4+\frac{3}{4} g_1^2 +\frac{9}{4} g_2^2 \\
& -\cos ^2(\alpha ) \left[6 Y_{b,2}^2+2Y_{\tau,2}^2+6 Y_t^2\right] - \sin ^2(\alpha ) \left[ 6 Y_{b,1}^2 +2Y_{\tau,1}^2\right].
\end{align*}

In order to easily obtain the leading logarithm contributions to higher orders, we use the recursive formula derived by Einhorn and Jones \cite{Einhorn:1992um}, relating the coefficient functions $f_{n+1}$ to $f_n$ and the running of the couplings:

\begin{align*}
 f_{n+1}(\lambda_i,Y_i,g_i) &= \frac{1}{n+1}\sum\limits_{L\in \{\lambda_i,Y_i,g_i\} } \beta_L \frac{\partial }{\partial L} f_n (\lambda_i,Y_i,g_i)
\end{align*}

This recursive relation is based on the following assumptions: the new theory has only one mass scale ($m_h$), and the logarithmic factor has to be large enough to suppress the terms with lower powers of logarithms.
Two-loop effects on the Veltman condition have already been applied to the 2HDM of type II using this approach \cite{Grzadkowski:2009iz,Grzadkowski:2010dn,Grzadkowski:2010se}; the authors found that the Higgs mass hierarchy problem can be ameliorated.\\
An obvious choice of $\mu_{\text{nat}}$ as cut-off would be the breakdown of one of the stability constraints $\mu_{\text{st}}$, so we will use it for the moment.
In order to analyze whether we can make a statement on the Higgs mass naturalness in a 2HDM which is based on a reliable perturbation series, we want to define $k_n$ as the ratio of the $n$-th correction term of Eq.~\eqref{quadraticmasscorrections} to the one of order $n-1$:

\begin{align*}
 k_n &= \frac{f_{n}(\lambda_i,Y_i,g_i)}{f_{n-1} (\lambda_i,Y_i,g_i)} \ln \frac{\mu_{\text{nat}}}{\mu_{\text{ew}}}
\end{align*}

Apart from the obvious logarithmic dependence on $\mu_{\text{nat}}$, the $k_n$ depend on the cut-off scale also indirectly: the latter determines which values for the $\lambda_i$ are allowed, see Fig.~\ref{fig:2pivs4pi-I}.\\
Now we can re-write Eq.~\eqref{quadraticmasscorrections} as

\begin{align}
 \delta m_h^2 &= \frac{\mu_{\text{nat}}^2}{16\pi^2} f_0(\lambda_i,Y_i,g_i) \left[ 1+ \sum\limits_{n=1}^\infty \prod\limits_{\ell=1}^n k_\ell \right] . \label{kformula}
\end{align}
\begin{figure}[!t]
  \centering
   \begin{picture}(320,240)(0,0)
    \put(0,0){\includegraphics[width=320pt]{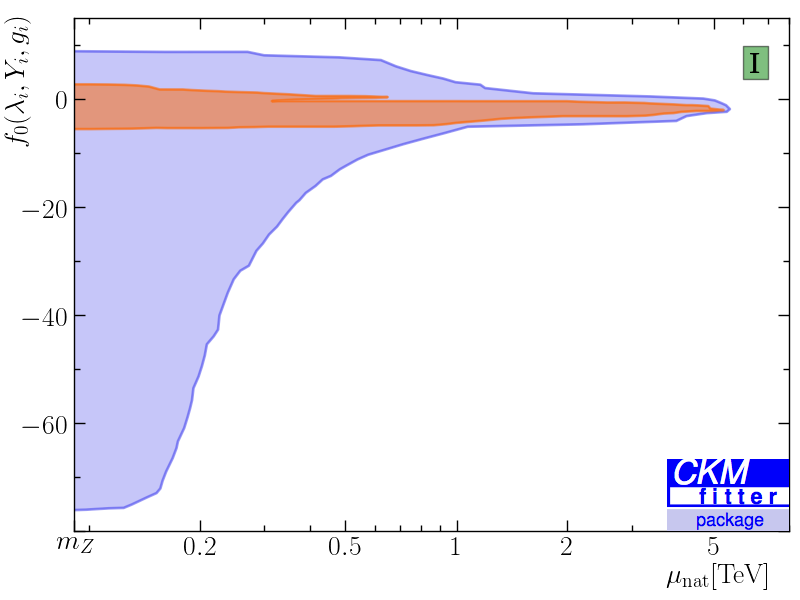}}
   \end{picture}
  \caption{The allowed size of the one-loop coefficient $f_0(\lambda_i,Y_i,g_i)$ of the Veltman condition series depends on the naturalness cut-off $\mu_{\text{nat}}$. Without experimental constraints (light blue shaded), $|f_0(\lambda_i,Y_i,g_i)|$ can be as large as $75$. With the inclusion of the measurements (orange shaded), it gets strongly constrained to be smaller than $6$.
More or less independently of taking into account experimental data, we obtain an upper bound on $\mu_{\text{nat}}$ of \unit{5.3}{TeV}. The plane shows the type I fit, which agrees with the type X fit. In type II and Y, the maximal $\mu_{\text{nat}}$ is already at \unit{3.7}{TeV}.}
  \label{fig:Qvsf0}
\end{figure}

Only if we impose a small value of the leading order coefficient function and a sufficiently small number for $k_1$, we can guarantee a perturbatively stable mitigation of the hierarchy problem of $m_h$, also assuming that the $k_\ell$ for $\ell >1$ are not too large. Note that if we choose $f_0(\lambda_i,Y_i,g_i)$ to be exactly $0$, $k_1$ diverges.\\
If we constrain the first two factors $k_1$ and $k_2$ to be smaller than $1$ in magnitude and that $|\delta m_h^2|<m_h^2$, we observe negative $k_1$ and $k_2$ in most cases, independently of the type of $Z_2$ symmetry. This indicates that the series is alternating, which in turn means that -- except for pathological scenarios -- a suppression of the first two $k_i$ factors should be sufficient to make the series relatively robust with respect to pertubativity.
Cutting the series in Eq.~\eqref{kformula} after the second term (i.e.\ setting $k_3=0$), we find that the maximal $\mu_{\text{nat}}$ is in the TeV range for all types, depending on the value we choose for $f_0(\lambda_i,Y_i,g_i)$. The blue shaded region in Fig.~\ref{fig:Qvsf0} shows this dependence for type I, taking into account only Higgs potential constraints (as in Section \ref{subsec:withoutexp}). There is a lot of freedom for $f_0(\lambda_i,Y_i,g_i)$, which indicate large cancellations between the leading order contribution and higher order terms. This calls into question our assumption that the series can be cut after the second term.
The inclusion of the experimental results (cf. Section \ref{subsec:withexp}, orange shaded in Fig.~\ref{fig:Qvsf0}) limits the choice of $f_0(\lambda_i,Y_i,g_i)$ to be of order $1$, and thus presumably stabilizes the perturbative series. In both cases, however, the maximal $\mu_{\text{nat}}$ is at around \unit{5.3}{TeV} for very small values of $f_0(\lambda_i,Y_i,g_i)$. While we obtain the same results for type X, the maximal $\mu_{\text{nat}}$ is even lower (\unit{3.7}{TeV}) in type II and Y which is a consequence of the much stronger constrained parameter space.

Finally, one could also impose the perturbativity of the power series in Eq.~\eqref{quadraticmasscorrections} as constraint and define $\mu_{\text{nat}}$ as its breakdown scale if it is smaller than $\mu_{\text{st}}$. This, however, would not alter the maximal $\mu_{\text{nat}}$, nor would it constrain the 2HDM parameters stronger than the conventional constraints. It would leave us with the question of what happens beyond the breakdown of perturbative naturalness already at a few TeV.

To put it in a nutshell: softening the Higgs mass hierarchy problem is very difficult in the context of a perturbative 2HDM and can be achieved only for very low cut-off scales $\mu_{\text{nat}}$. 
Nevertheless, this is an improvement of one order of magnitude as compared to the SM hierarchy problem and might hint at a more complete model beyond the 2HDM at TeV scales.

\section{Conclusions}
\label{sec:conclusions}

We obtain the two-loop renormalization group equations for all four $Z_2$ symmetric types of the 2HDM using PyR@TE and show that in general, two-loop corrections to the leading order one-loop expressions should not be neglected. 
We then apply these equations to improve the predictions of renormalization group evolution of the coupling parameters, putting a special emphasis on the quartic couplings $\lambda_i$, which are usually prone to run into non-perturbative regions.
The relative distance between the LO and NLO curves of the $\lambda_i$ can be as large as $45\%$. The quadratic couplings $m_{11}^2$ and $m_{22}^2$ from the Higgs potential can vary by an order of magnitude between the electroweak scale and the perturbativity cut-off, while $m_{12}^2$ is in general found to be rather stable under RG evolution. We do not observe any fixed point behaviour in the regions with a stable Higgs potential.

Imposing positivity and perturbativity bounds at all scales and stability of the vacuum at the electroweak scale, the magnitudes of the $\lambda_i$ at $\mu_{\text{ew}}$ which give a stable Higgs potential up to the Planck scale are found to be typically below $1$; we also find lower limits of $0.15$ for $\lambda_2$ and of $1.0$ for $\tan \beta$. We have checked that these results are the same in all types. In type II and type Y we additionally get an upper limit of $60$ on $\tan \beta$ with stability up to $\mu_{\text{Pl}}$.
Moreover, we address the question of which upper limit for the eigenvalues of the tree-level $\Phi_i \Phi_j \to \Phi_i \Phi_j$ scattering matrix is appropriate and show that as soon as at least one of the eigenvalues exceeds $2\pi$ in magnitude, the maximal scale up to which the Higgs potential can be stable is $5\cdot$\unit{10^{6}}{GeV}. It even reduces to \unit{10}{TeV} in all types if we assume $1<\tan \beta <60$. Imposing stability up to $\mu_{\text{Pl}}$ leads to an upper limit of $2.5$ on the magnitude of the eigenvalues.

Including latest results from the LHC as well as all other relevant experimental data, we show the result of our fits for all four types. We observe that deviations from the alignment limit strongly depend on the value of $\tan \beta$; the maximal deviation of $\beta-\alpha$ from $\pi/2$ is $0.43$, $0.08$, $0.33$ and $0.07$ in type I, II, X and Y, respectively. (This corresponds to deviations of $\sin (\beta-\alpha)$ from $1$ of at most $0.092$, $0.003$, $0.054$ and $0.003$.) Taking the stability constraint up to $\mu_{\text{Pl}}$, the bounds on $\beta-\alpha$ become even stronger and allow for deviations from $\pi/2$ of at most $0.36$, $0.05$, $0.28$ and $0.04$ in the respective types.
The searches for heavy neutral Higgs particles exclude a light charged Higgs boson for large $\tan \beta$ in type II and for very large $\tan \beta$ in type Y. In the $m_{H/A}$--$(\beta-\alpha)$ planes it is visible that deviations from the alignment limit by more than $0.05\pi$ are possible only for $m_H$ and $m_A$ below \unit{500}{GeV} in the types I and X. In type II and Y we obtain lower limits of $340$ GeV and $360$ GeV on $m_H$ and $m_A$, respectively. This makes it very difficult to realize models with an unbroken $Z_2$ symmetry in these two types even if the stability cut-off is only at the electroweak scale. Demanding that the Higgs potential is stable up to the Planck scale, these mass limits are even stronger.

We finally discuss whether a reliable statement on the seemingly fine-tuned Higgs mass $m_h$ can be made in the context of a 2HDM and whether its hierarchy problem can be solved at least partially. Restricting higher order corrections to the perturbative regime, we observe a maximal naturalness cut-off at \unit{5.3}{TeV}.
Our conclusion is that within a perturbative framework a natural cancellation of quadratic divergencies cannot be implemented into a Two-Higgs-Doublet model beyond ${\cal O}$(TeV) scales.

\begin{acknowledgments}
We thank N.~Craig, A.~Kundu, and A.~de la Puente for fruitful
discussions, U.~Nierste for helpful advice and proofreading, and the CKMfitter group for allowing us to use their
statistical analysis framework.
We are very grateful to F.~Lyonnet for support concerning the PyR@TE code and to the MadGraph and FeynRules authors for useful input, and we want to thank H.~Lacker for providing computational power.
The research leading to these results has received funding from the European Research Council under the European Union's Seventh Framework Programme (FP/2007-2013) / ERC Grant Agreement n.~279972.
DC would like to thank the Centro de Ciencias de Benasque Pedro Pascual for its hospitality during the initial stages of this work.
He has also received partial support from the Munich Institute for Astro- and Particle Physics (MIAPP) of the DFG cluster of excellence ``Origin and Structure of the Universe''.

\end{acknowledgments}

\section*{Appendix}
\label{sec:appendix}

Here we list the renormalization group equations for the 2HDM with soft $Z_2$ breaking which we obtained with the PyR@TE code \cite{Lyonnet:2013dna}.\\
For any coupling $L$ the complete $\beta$ functions at NLO can be split into leading and next-to-leading order contributions and further divided into bosonic and fermionic parts as follows:

\begin{align*}
\beta _{L} &\equiv \dfrac{d L}{d t} =\beta _{L}^{LO}+\beta _{L}^{NLO}\\[10pt]
\beta _{L}^{(N)LO} &=\beta _{L}^{(N)LO,b}+\beta _{L}^{(N)LO,f}\\[10pt]
\end{align*}

Except for the Yukawa RGE the bosonic part does not involve fermionic couplings and is type independent, while the fermionic part in general depends on the type of $Z_2$ symmetry. If the expressions for the latter differ for the different types, we will replace the index $f$ by the type label I, II, X or Y, respectively.\\
\newpage

The RGE of the gauge couplings only depend on themselves and on the Yukawa couplings:

\begin{align*}
16 \pi ^2 \beta _{g_1}^{LO} =&7 g_1^3\\[10pt]
(16 \pi ^2)^2 \beta _{g_1}^{NLO,b} =&\left(\frac{104}{9} g_1^2+6 g_2^2+\frac{44}{3} g_3^2\right) g_1^3 \\
(16 \pi ^2)^2 \beta _{g_1}^{NLO,f} =&-\left(\frac{5}{6} Y_b^2+\frac{17}{6} Y_t^2+\frac{5}{2} Y_{\tau}^2\right) g_1^3\\[10pt]
\end{align*}

\begin{align*}
16 \pi ^2 \beta _{g_2}^{LO} =&-3 g_2^3\\[10pt]
(16 \pi ^2)^2 \beta _{g_2}^{NLO,b} =&\left(2 g_1^2+8 g_2^2+12 g_3^2\right) g_2^3 \\
(16 \pi ^2)^2 \beta _{g_2}^{NLO,f} =&-\left(\frac{3}{2} Y_b^2+\frac{3}{2} Y_t^2+\frac{1}{2}Y_{\tau}^2\right) g_2^3\\[10pt]
\end{align*}

\begin{align*}
16 \pi ^2 \beta _{g_3}^{LO} =&-7 g_3^3\\[10pt]
(16 \pi ^2)^2 \beta _{g_3}^{NLO,b} =&\left(\frac{11}{6} g_1^2+\frac{9}{2} g_2^2-26 g_3^2\right) g_3^3 \\
(16 \pi ^2)^2 \beta _{g_3}^{NLO,f} =&-\left(2Y_b^2+2Y_t^2\right) g_3^3\\[10pt]
\end{align*}

As already mentioned, the mass parameters from the Higgs potential do not influence the running of the dimensionless couplings. Their running, however, is not negligible and is given by the following expressions:

\begin{align*}
16 \pi ^2 \beta _{m_{11}^2}^{LO,b} =&\left(-\frac{3}{2} g_1^2-\frac{9}{2}g_2^2 +6 \lambda_1\right) m_{11}^2
+\left(4 \lambda_3 +2\lambda_4 \right) m_{22}^2\\
16 \pi ^2 \beta _{m_{11}^2}^{LO,I} =&0\\
16 \pi ^2 \beta _{m_{11}^2}^{LO,II} =&\left(6 Y_b^2+2Y_{\tau}^2\right) m_{11}^2 \\
16 \pi ^2 \beta _{m_{11}^2}^{LO,X} =&2Y_{\tau}^2 m_{11}^2 \\
16 \pi ^2 \beta _{m_{11}^2}^{LO,Y} =&6 Y_b^2 m_{11}^2 \\[100pt]
(16 \pi ^2)^2 \beta _{m_{11}^2}^{NLO,b} =& \left(\frac{193}{16} g_1^4+\frac{15}{8} g_1^2 g_2^2-\frac{123}{16} g_2^4+12 g_1^2 \lambda_1+36 g_2^2 \lambda_1 \right. \\
&\left.  \hspace*{154pt}  \phantom{\frac{25}{12}} -15 \lambda_1^2-2 \lambda_3^2-2 \lambda_3 \lambda_4-2 \lambda_4^2-3 \lambda_5^2\right) m_{11}^2\\
&+\left(\frac{5}{2} g_1^4+\frac{15}{2} g_2^4 + ( 4 g_1^2 + 12 g_2^2 ) ( 2 \lambda_3 +\lambda_4 ) \right. \\
&\left.  \hspace*{186pt}  \phantom{\frac{25}{12}} -8 \lambda_3^2-8 \lambda_3 \lambda_4-8 \lambda_4^2-12 \lambda_5^2\right) m_{22}^2 \\
(16 \pi ^2)^2 \beta _{m_{11}^2}^{NLO,I} =&- \left(12 Y_b^2 +12 Y_t^2 +4 Y_{\tau}^2\right) (2 \lambda_3+\lambda_4) m_{22}^2 \\
(16 \pi ^2)^2 \beta _{m_{11}^2}^{NLO,II} =& \left(\frac{25}{12} g_1^2 Y_b^2 +\frac{25}{4} g_1^2 Y_{\tau}^2 +\frac{45}{4} g_2^2 Y_b^2 +\frac{15}{4} g_2^2 Y_{\tau}^2 +40 g_3^2 Y_b^2\right. \\
& \hspace*{96pt} \left. \phantom{\frac{25}{12}} -\frac{27}{2} Y_b^4 -\frac{9}{2} Y_b^2 Y_t^2 -\frac{9}{2}Y_{\tau}^4 -36 Y_b^2 \lambda_1 -12 Y_{\tau}^2 \lambda_1\right) m_{11}^2\\
& -12 Y_t^2 (2 \lambda_3+\lambda_4) m_{22}^2\\
(16 \pi ^2)^2 \beta _{m_{11}^2}^{NLO,X} =& \left(\frac{25}{4} g_1^2 +\frac{15}{4} g_2^2 -\frac{9}{2} Y_\tau^2 -12 \lambda_1\right) Y_\tau^2 m_{11}^2 -\left(12Y_b^2+12Y_t^2\right) (2 \lambda_3+\lambda_4) m_{22}^2 \\
(16 \pi ^2)^2 \beta _{m_{11}^2}^{NLO,Y} =& \left(\frac{25}{12} g_1^2 +\frac{45}{4} g_2^2 +40 g_3^2 -\frac{27}{2} Y_b^2 -\frac{9}{2}Y_t^2 -36 \lambda_1\right) Y_b^2 m_{11}^2 \\
& - \left(12 Y_t^2 +4 Y_\tau^2\right) (2 \lambda_3+\lambda_4) m_{22}^2
\\[10pt]
\end{align*}

\begin{align*}
16 \pi ^2 \beta _{m_{22}^2}^{LO,b} =&\left( 4\lambda_3 +2\lambda_4 \right) m_{11}^2 -\left(\frac{3}{2} g_1^2 +\frac{9}{2}g_2^2 -6 \lambda_2 \right) m_{22}^2\\
16 \pi ^2 \beta _{m_{22}^2}^{LO,I} =&\left(6 Y_b^2+6 Y_t^2+2Y_{\tau}^2\right) m_{22}^2\\
16 \pi ^2 \beta _{m_{22}^2}^{LO,II} =&6 Y_t^2 m_{22}^2 \\
16 \pi ^2 \beta _{m_{22}^2}^{LO,X} =&\left(6 Y_b^2 +6 Y_t^2\right) m_{22}^2\\
16 \pi ^2 \beta _{m_{22}^2}^{LO,Y} =&\left(6 Y_t^2 +2Y_\tau^2\right) m_{22}^2\\[10pt]
(16 \pi ^2)^2 \beta _{m_{22}^2}^{NLO,b} =&\left(\frac{5}{2} g_1^4 +8 g_1^2 \lambda_3 +4 g_1^2 \lambda_4 +\frac{15}{2} g_2^4 +24 g_2^2 \lambda_3 +12 g_2^2 \lambda_4 \right. \\
& \hspace*{186pt} \left. \phantom{\frac{25}{12}} -8 \lambda_3^2 -8 \lambda_3 \lambda_4 -8 \lambda_4^2 -12 \lambda_5^2\right) m_{11}^2 \\
&+\left(\frac{193}{16} g_1^4 +\frac{15}{8} g_1^2 g_2^2 +12 g_1^2 \lambda_2 -\frac{123}{16} g_2^4 +36 g_2^2 \lambda_2 \right. \\
& \hspace*{154pt} \left. \phantom{\frac{25}{12}} -15 \lambda_2^2 -2 \lambda_3^2 -2 \lambda_3 \lambda_4 -2 \lambda_4^2 -3 \lambda_5^2\right) m_{22}^2\\[100pt]
(16 \pi ^2)^2 \beta _{m_{22}^2}^{NLO,I} =&\left( g_1^2\left( \frac{25}{12} Y_b^2 +\frac{85}{12} Y_t^2 +\frac{25}{4} Y_{\tau}^2\right) +g_2^2 \left(\frac{45}{4} Y_b^2 +\frac{45}{4} Y_t^2 +\frac{15}{4} Y_{\tau}^2\right) \right. \\
& \hspace*{25pt} +g_3^2 \left(40Y_b^2 +40Y_t^2\right) -\frac{27}{2} Y_b^4 -21 Y_b^2 Y_t^2 -\frac{27}{2} Y_t^4 -\frac{9}{2} Y_{\tau}^4\\
& \hspace*{186pt} \left. \phantom{\frac{25}{12}} -\left( 36 Y_b^2 +36 Y_t^2 +12 Y_{\tau}^2 \right) \lambda_2 \right) m_{22}^2\\
(16 \pi ^2)^2 \beta _{m_{22}^2}^{NLO,II} =&-\left(12 Y_b^2 +4Y_{\tau}^2\right) (2 \lambda_3+\lambda_4) m_{11}^2\\
& \hspace*{92pt} +\left(\frac{85}{12} g_1^2 +\frac{45}{4} g_2^2 +40 g_3^2 -36 \lambda_2 -\frac{9}{2} Y_b^2 -\frac{27}{2} Y_t^2\right) Y_t^2 m_{22}^2 \\
(16 \pi ^2)^2 \beta _{m_{22}^2}^{NLO,X} =& -(8 \lambda_3 +4\lambda_4) Y_\tau^2 m_{11}^2 \\
& +\left( g_1^2 \left(\frac{25}{12} Y_b^2 +\frac{85}{12} Y_t^2\right) +g_2^2 \left( \frac{45}{4} Y_b^2 +\frac{45}{4} Y_t^2\right) + g_3^2 \left( 40Y_b^2 +40Y_t^2\right) \right. \\
& \hspace*{95pt} \left. \phantom{\frac{25}{12}} -\frac{27}{2} Y_b^4 -21 Y_b^2 Y_t^2 -\frac{27}{2} Y_t^4 -36 \left( Y_b^2 +Y_t^2 \right) \lambda_2 \right) m_{22}^2\\
(16 \pi ^2)^2 \beta _{m_{22}^2}^{NLO,Y} =& -(24 \lambda_3 +12\lambda_4) Y_b^2 m_{11}^2 \\
&+\left( g_1^2 \left( \frac{85}{12} Y_t^2 +\frac{25}{4} Y_\tau^2\right) + g_2^2 \left( \frac{45}{4} Y_t^2 +\frac{15}{4}Y_\tau^2 \right) +40 g_3^2 Y_t^2 \right. \\
& \hspace*{114pt} \left. -\frac{9}{2} Y_b^2 Y_t^2 -\frac{27}{2} Y_t^4 -\frac{9}{2} Y_\tau^4 -\left( 36Y_t^2 +12Y_\tau^2\right) \lambda_2 \right) m_{22}^2\\[10pt]
\end{align*}

\begin{align*}
16 \pi ^2 \beta _{m_{12}^2}^{LO,b} =&\left(-\frac{3}{2} g_1^2 -\frac{9}{2} g_2^2 +2\lambda_3 +4 \lambda_4 +6 \lambda_5\right) m_{12}^2 \\
16 \pi ^2 \beta _{m_{12}^2}^{LO,f} =&\left(3 Y_b^2+3 Y_t^2+Y_{\tau}^2\right) m_{12}^2 \\[10pt]
(16 \pi ^2)^2 \beta _{m_{12}^2}^{NLO,b} =&\left(\frac{153}{16} g_1^4 +\frac{15}{8} g_1^2 g_2^2 -\frac{243}{16} g_2^4 +4(g_1^2+3 g_2^2) (\lambda_3+2 \lambda_4+3 \lambda_5) \right.\\
&  \hspace*{25pt} +\frac{3}{2}\lambda_1^2 +\frac{3}{2}\lambda_2^2 -6 (\lambda_1 +\lambda_2) (\lambda_3+\lambda_4+\lambda_5) \\
&  \hspace*{157pt} \left. \phantom{\frac{25}{12}} -6 \lambda_3 \lambda_4 -12 \lambda_3 \lambda_5 -12 \lambda_4 \lambda_5 +3 \lambda_5^2\right) m_{12}^2 \\
(16 \pi ^2)^2 \beta _{m_{12}^2}^{NLO,I} =&\left(g_1^2 \left(\frac{25}{24} Y_b^2+\frac{85}{24} Y_t^2+\frac{25}{8} Y_{\tau}^2\right) + g_2^2 \left(\frac{45}{8} Y_b^2 +\frac{45}{8} Y_t^2 +\frac{15}{8}Y_{\tau}^2\right) \right. \\
& \hspace*{25pt} +g_3^2 \left( 20 Y_b^2 +20 Y_t^2\right) -\frac{27}{4} Y_b^4 +\frac{3}{2} Y_b^2 Y_t^2 -\frac{27}{4} Y_t^4 -\frac{9}{4} Y_{\tau}^4\\
& \hspace*{135pt} \left. \phantom{\frac{25}{12}} -2 (3 Y_b^2 +3 Y_t^2 +Y_{\tau}^2) (\lambda_3+2 \lambda_4+3 \lambda_5) \right) m_{12}^2\\
(16 \pi ^2)^2 \beta _{m_{12}^2}^{NLO,II} =& (16 \pi ^2)^2 \beta _{m_{12}^2}^{NLO,I} -18 Y_b^2 Y_t^2 m_{12}^2\\
(16 \pi ^2)^2 \beta _{m_{12}^2}^{NLO,X} =& (16 \pi ^2)^2 \beta _{m_{12}^2}^{NLO,I}\\
(16 \pi ^2)^2 \beta _{m_{12}^2}^{NLO,Y} =& (16 \pi ^2)^2 \beta _{m_{12}^2}^{NLO,I} -18 Y_b^2 Y_t^2 m_{12}^2\\[10pt]
\end{align*}

Finally, the quartic couplings from the Higgs potential:

\begin{align*}
16 \pi ^2 \beta _{\lambda_1}^{LO,b} =&\frac{3}{4} g_1^4 +\frac{3}{2} g_1^2 g_2^2 +\frac{9}{4} g_2^4 -3g_1^2 \lambda_1 -9 g_2^2 \lambda_1 +12 \lambda_1^2 +4 \lambda_3^2 +4 \lambda_3 \lambda_4 +2\lambda_4^2 +2\lambda_5^2\\
16 \pi ^2 \beta _{\lambda_1}^{LO,I} =&0\\
16 \pi ^2 \beta _{\lambda_1}^{LO,II} =&-12 Y_b^4 -4Y_\tau^4 +12 Y_b^2 \lambda_1 +4 Y_\tau^2 \lambda_1 \\
16 \pi ^2 \beta _{\lambda_1}^{LO,X} =& -4 Y_\tau^4 +4 Y_\tau^2 \lambda_1\\
16 \pi ^2 \beta _{\lambda_1}^{LO,Y} =& -12Y_b^4 +12\lambda_1 Y_b^2\\[10pt]
(16 \pi ^2)^2 \beta _{\lambda_1}^{NLO,b} =& -\frac{131}{8}g_1^6 -\frac{191}{8} g_1^4 g_2^2 -\frac{101}{8}g_1^2 g_2^4 +\frac{291}{8}g_2^6 +g_1^4 \left(\frac{217}{8} \lambda_1 +5 \lambda_3 +\frac{5}{2} \lambda_4\right)\\
& + g_1^2 g_2^2 \left(\frac{39}{4} \lambda_1+5 \lambda_4\right) +g_2^4 \left(-\frac{51}{8}\lambda_1 +15 \lambda_3 +\frac{15}{2} \lambda_4 \right)\\
& +g_1^2 \left(18 \lambda_1^2+8 \lambda_3^2+8 \lambda_3 \lambda_4+4 \lambda_4^2-2 \lambda_5^2\right)
+g_2^2 \left(54 \lambda_1^2+6(2 \lambda_3+\lambda_4)^2\right) \\
& -78 \lambda_1^3 -\lambda_1 \left(20 \lambda_3^2 +20 \lambda_3 \lambda_4 +12 \lambda_4^2 +14 \lambda_5^2\right) \\
& -16 \lambda_3^3 -24 \lambda_3^2 \lambda_4 -32 \lambda_3 \lambda_4^2 -40 \lambda_3 \lambda_5^2 -12 \lambda_4^3 -44 \lambda_4 \lambda_5^2\\
(16 \pi ^2)^2 \beta _{\lambda_1}^{NLO,I} =&-\left(12 Y_b^2 +12 Y_t^2 +4Y_\tau^2\right) \left(2 \lambda_3^2 +2 \lambda_3 \lambda_4 +\lambda_4^2 +\lambda_5^2\right) \\
(16 \pi ^2)^2 \beta _{\lambda_1}^{NLO,II} =& g_1^4 \left( \frac{5}{2} Y_b^2 - \frac{25}{2} Y_\tau^2 \right) +g_1^2 g_2^2 \left(9 Y_b^2 +11 Y_\tau^2\right) -g_2^4 \left(\frac{9}{2} Y_b^2 +\frac{3}{2} Y_\tau^2\right) \\
& +g_1^2 \left(\frac{8}{3} Y_b^4 -8 Y_\tau^4 +\frac{25}{6} Y_b^2 \lambda_1 +\frac{25}{2} Y_\tau^2 \lambda_1 \right) +g_2^2 \left(\frac{45}{2} Y_b^2 \lambda_1 +\frac{15}{2} Y_\tau^2 \lambda_1 \right) \\
& -g_3^2 \left( 64 Y_b^4 -80 Y_b^2 \lambda_1 \right) \\
& +60 Y_b^6 +12 Y_b^4 Y_t^2 +20 Y_\tau^6 -(3 Y_b^4 +9 Y_b^2 Y_t^2 +Y_\tau^4) \lambda_1\\
& -72 Y_b^2 \lambda_1^2 - 12Y_t^2 ( 2 \lambda_3^2 +2 \lambda_3 \lambda_4 + \lambda_4^2 + \lambda_5^2) -24 Y_\tau^2 \lambda_1^2\\
(16 \pi ^2)^2 \beta _{\lambda_1}^{NLO,X} =& -\frac{25}{2} g_1^4 Y_\tau^2 +11g_1^2 g_2^2 Y_\tau^2 -\frac{3}{2} g_2^4 Y_\tau^2 +g_1^2 \left( -8 Y_\tau^4 +\frac{25}{2} Y_\tau^2 \lambda_1\right) +\frac{15}{2} g_2^2 Y_\tau^2 \lambda_1 \\
& +20 Y_\tau^6 -Y_\tau^4 \lambda_1 - \left( 12Y_b^2 +12Y_t^2\right) \left(2 \lambda_3^2+2 \lambda_3 \lambda_4+\lambda_4^2+\lambda_5^2\right) -24 Y_\tau^2 \lambda_1^2 \\
(16 \pi ^2)^2 \beta _{\lambda_1}^{NLO,Y} =&
\frac{5}{2}g_1^4 Y_b^2 +9g_1^2 g_2^2 Y_b^2 -\frac{9}{2}g_2^4 Y_b^2 \\
& +g_1^2\left( \frac{8}{3} Y_b^4 +\frac{25}{6} Y_b^2 \lambda_1 \right) +\frac{45}{2} g_2^2 Y_b^2 \lambda_1 -g_3^2 \left( 64 Y_b^4 -80\lambda_1 Y_b^2 \right) \\
& +60 Y_b^6 +12 Y_b^4 Y_t^2 -\left( 3 Y_b^4 +9 Y_b^2 Y_t^2 \right) \lambda_1 \\
& -72 Y_b^2 \lambda_1^2 -\left( 12 Y_t^2 +4 Y_\tau^2 \right) \left( 2 \lambda_3^2 +2 \lambda_3 \lambda_4 + \lambda_4^2 + \lambda_5^2 \right) \\[10pt]
\end{align*}

\begin{align*}
16 \pi ^2 \beta _{\lambda_2}^{LO,b} =& 16 \pi ^2 \beta _{\lambda_1}^{LO,b} (\lambda1 \leftrightarrow \lambda_2)\\
16 \pi ^2 \beta _{\lambda_2}^{LO,I} =& -12 Y_b^4 -12 Y_t^4 -4Y_\tau^4 +\left( 12 Y_b^2 +12 Y_t^2 +4 Y_\tau^2\right) \lambda_2\\
16 \pi ^2 \beta _{\lambda_2}^{LO,II} =& -12 Y_t^4 +12 Y_t^2 \lambda_2 \\
16 \pi ^2 \beta _{\lambda_2}^{LO,X} =& -12 Y_b^4 -12 Y_t^4 +\left( 12 Y_b^2 +12 Y_t^2 \right) \lambda_2\\
16 \pi ^2 \beta _{\lambda_2}^{LO,Y} =& -12 Y_t^4 -4Y_\tau^4 +\left( 12 Y_t^2 +4 Y_\tau^2\right) \lambda_2 \\[10pt]
(16 \pi ^2)^2 \beta _{\lambda_2}^{NLO,b} =& (16 \pi ^2)^2 \beta _{\lambda_1}^{NLO,b} (\lambda1 \leftrightarrow \lambda_2)\\
(16 \pi ^2)^2 \beta _{\lambda_2}^{NLO,I} =& g_1^4 \left(\frac{5}{2} Y_b^2 -\frac{19}{2} Y_t^2 -\frac{25}{2} Y_\tau^2\right) +g_1^2 g_2^2 \left(9 Y_b^2+21 Y_t^2+11 Y_\tau^2\right) \\
& -g_2^4 \left(\frac{9}{2} Y_b^2 +\frac{9}{2} Y_t^2 +\frac{3}{2}Y_\tau^2\right) \\
& +g_1^2 \left(\frac{8}{3} Y_b^4 -\frac{16}{3} Y_t^4 -8 Y_\tau^4 +\frac{25}{6} Y_b^2 \lambda_2 +\frac{85}{6}  Y_t^2 \lambda_2 +\frac{25}{2} Y_\tau^2 \lambda_2 \right)\\
& +g_2^2 \left(\frac{45}{2} Y_b^2 +\frac{45}{2} Y_t^2 +\frac{15}{2}Y_\tau^2\right) \! \lambda_2 -g_3^2 \left(64 Y_b^4 +64 Y_t^4 -80 Y_b^2 \lambda_2 -80 Y_t^2 \lambda_2 \right) \\
& +60 Y_b^6 -12 Y_b^4 Y_t^2 -12 Y_b^2 Y_t^4 +60 Y_t^6 +20 Y_\tau^6\\
& -\left( 3 Y_b^4 +42 Y_b^2 Y_t^2 +3 Y_t^4 +Y_\tau^4\right) \lambda_2 -(72 Y_b^2 +72 Y_t^2 +24 Y_\tau^2)\lambda_2^2 \\
(16 \pi ^2)^2 \beta _{\lambda_2}^{NLO,II} =& -\frac{19}{2}g_1^4 Y_t^2 +21 g_1^2 g_2^2 Y_t^2 -\frac{9}{2} g_2^4 Y_t^2\\
& +g_1^2 \left( -\frac{16}{3} Y_t^4 +\frac{85}{6} Y_t^2 \lambda_2 \right) +\frac{45}{2} g_2^2 Y_t^2 \lambda_2 -g_3^2 \left( 64 Y_t^4 -80 Y_t^2 \lambda_2 \right) \\
& +12 Y_b^2 Y_t^4 +60 Y_t^6 -\left( 9 Y_b^2 Y_t^2 +3 Y_t^4\right) \lambda_2\\
& -\left( 12 Y_b^2 +4Y_\tau^2 \right) \left( 2\lambda_3^2 +2 \lambda_3 \lambda_4 +\lambda_4^2 +\lambda_5^2 \right) -72 Y_t^2 \lambda_2^2\\
16 \pi ^2 \beta _{\lambda_2}^{NLO,X} =& g_1^4 \left(\frac{5}{2} Y_b^2 -\frac{19}{2} Y_t^2\right) + g_1^2 g_2^2 \left(9 Y_b^2+21 Y_t^2\right) - g_2^4 \left( \frac{9}{2}Y_b^2 +\frac{9}{2}Y_t^2\right) \\
& + g_1^2 \left(\frac{8}{3} Y_b^4 -\frac{16}{3} Y_t^4 +\frac{25}{6} Y_b^2 \lambda_2 +\frac{85}{6} Y_t^2 \lambda_2\right) + g_2^2 \left(\frac{45}{2}Y_b^2+\frac{45}{2}Y_t^2\right) \lambda_2 \\
&-g_3^2 \left( 64 Y_b^4 +64 Y_t^4 -80 Y_b^2 \lambda_2 -80 Y_t^2 \lambda_2 \right) \\
&+60 Y_b^6 -12 Y_b^4 Y_t^2 -12 Y_b^2 Y_t^4 +60 Y_t^6\\
& -\left( 3 Y_b^4 +42 Y_b^2 Y_t^2 +3 Y_t^4 \right) \lambda_2\\
& -72 Y_b^2 \lambda_2^2 -72 Y_t^2 \lambda_2^2 -4Y_\tau^2 \left( 2\lambda_3^2 +2 \lambda_3 \lambda_4 +\lambda_4^2 +\lambda_5^2\right) \\
16 \pi ^2 \beta _{\lambda_2}^{NLO,Y} =& g_1^4 \left(-\frac{19}{2} Y_t^2 -\frac{25}{2}  Y_\tau^2\right)  +g_1^2 g_2^2 \left(21 Y_t^2+11 Y_\tau^2\right)  -g_2^4 \left( \frac{9}{2} Y_t^2 +\frac{3}{2}Y_\tau^2\right) \\
& +g_1^2 \left( -\frac{16}{3} Y_t^4 -8 Y_\tau^4 +\frac{85}{6} Y_t^2 \lambda_2 +\frac{25}{2} Y_\tau^2 \lambda_2\right) +g_2^2 \left(\frac{45}{2} Y_t^2 +\frac{15}{2}Y_\tau^2\right) \lambda_2 \\
& -g_3^2 \left( 64 Y_t^4 -80 Y_t^2 \lambda_2\right) \\
& +12 Y_b^2 Y_t^4 +60 Y_t^6 +20 Y_\tau^6 -\left( 9 Y_b^2 Y_t^2 +3 Y_t^4 +Y_\tau^4 \right) \lambda_2 \\
& -12 Y_b^2 \left(2 \lambda_3^2+2 \lambda_3 \lambda_4+\lambda_4^2+\lambda_5^2\right) -72 Y_t^2  \lambda_2^2 -24 Y_\tau^2 \lambda_2^2 \\[10pt]
\end{align*}

\begin{align*}
16 \pi ^2 \beta _{\lambda_3}^{LO,b} =& \frac{3}{4} g_1^4 -\frac{3}{2} g_1^2 g_2^2 +\frac{9}{4} g_2^4 -3g_1^2 \lambda_3 -9 g_2^2 \lambda_3 \\
& +(\lambda_1 +\lambda_2 ) \left( 6  \lambda_3 +2\lambda_4\right) +4 \lambda_3^2 +2\lambda_4^2 +2\lambda_5^2\\
16 \pi ^2 \beta _{\lambda_3}^{LO,I} =&\left(6 Y_b^2 +6 Y_t^2 +2 Y_\tau^2\right) \lambda_3\\
16 \pi ^2 \beta _{\lambda_3}^{LO,II} =& -12 Y_b^2 Y_t^2 + \left( 6 Y_b^2 +6 Y_t^2 +2 Y_\tau^2\right) \lambda_3\\
16 \pi ^2 \beta _{\lambda_3}^{LO,X} =& 16 \pi ^2 \beta _{\lambda_3}^{LO,I}\\
16 \pi ^2 \beta _{\lambda_3}^{LO,Y} =& 16 \pi ^2 \beta _{\lambda_3}^{LO,II}\\[10pt]
(16 \pi ^2)^2 \beta _{\lambda_3}^{NLO,b} =& -\frac{131}{8} g_1^6 +\frac{101}{8} g_1^4 g_2^2 +\frac{11}{8} g_1^2 g_2^4 +\frac{291}{8} g_2^6\\
&
+g_1^4 \left( \frac{15}{4} \lambda_1 +\frac{15}{4} \lambda_2 +\frac{197}{8} \lambda_3 +\frac{5}{2} \lambda_4\right) - g_1^2 g_2^2 \left( \frac{5}{2}\lambda_1 +\frac{5}{2}\lambda_2 -\frac{11}{4}\lambda_3 +3\lambda_4\right) \\
& + g_2^4 \left( \frac{45}{4}\lambda_1 +\frac{45}{4}\lambda_2 -\frac{111}{8} \lambda_3 +\frac{15}{2} \lambda_4\right) \\
& +g_1^2 \left( (\lambda_1 +\lambda_2) (12 \lambda_3 +4\lambda_4) +2\lambda_3^2 -2\lambda_4^2 +4 \lambda_5^2\right) \\
& +g_2^2 \left( (\lambda_1 +\lambda_2) (36 \lambda_3 +18\lambda_4) +6(\lambda_3-\lambda_4)^2\right) \\
& -\left( \lambda_1^2 +\lambda_2^2\right) \left( 15 \lambda_3 +4 \lambda_4 \right) -\left( \lambda_1 +\lambda_2\right) \left( 36 \lambda_3^2 +16 \lambda_3 \lambda_4 +14 \lambda_4^2 +18 \lambda_5^2\right) \\
& -12 \lambda_3^3 -4 \lambda_3^2 \lambda_4 -16 \lambda_3 \lambda_4^2 -18 \lambda_3 \lambda_5^2 -12 \lambda_4^3 -44 \lambda_4 \lambda_5^2 \\
(16 \pi ^2)^2 \beta _{\lambda_3}^{NLO,I} =& g_1^4 \left( \frac{5}{4} Y_b^2 -\frac{19}{4} Y_t^2 -\frac{25}{4} Y_\tau^2\right) -g_1^2 g_2^2 \left( \frac{9}{2} Y_b^2 +\frac{21}{2} Y_t^2 +\frac{11}{2} Y_\tau^2\right) \\
& -g_2^4 \left( \frac{9}{4} Y_b^2 +\frac{9}{4} Y_t^2 +\frac{3}{4}Y_\tau^2\right) +g_1^2 \left( \frac{25}{12} Y_b^2 +\frac{85}{12} Y_t^2 +\frac{25}{4} Y_\tau^2\right) \lambda_3 \\
& + g_2^2 \left( \frac{45}{4} Y_b^2 +\frac{45}{4} Y_t^2 +\frac{15}{4} Y_\tau^2\right) \lambda_3 +g_3^2 \left( 40 Y_b^2 +40 Y_t^2\right) \lambda_3 \\
& -\frac{27}{2} Y_b^4 \lambda_3 -Y_b^2 Y_t^2 (21 \lambda_3 +24 \lambda_4) -\frac{27}{2} Y_t^4 \lambda_3 -\frac{9}{2} Y_\tau^4 \lambda_3 \\
& -\left( 3 Y_b^2 +3 Y_t^2 +Y_\tau^2\right) \left( 12 \lambda_2 \lambda_3 +4 \lambda_2 \lambda_4 +4 \lambda_3^2 +2\lambda_4^2 +2\lambda_5^2\right) \\
(16 \pi ^2)^2 \beta _{\lambda_3}^{NLO,II} =& (16 \pi ^2)^2 \beta _{\lambda_3}^{NLO,I} -\frac{4}{3}g_1^2 Y_b^2 Y_t^2 -64 g_3^2 Y_b^2 Y_t^2 +36 Y_b^4 Y_t^2 +36 Y_b^2 Y_t^4\\
& +Y_b^2 Y_t^2 (36 \lambda_3+24 \lambda_4) - \left( 3 Y_b^2 +Y_\tau^2\right) (\lambda_1-\lambda_2) (12 \lambda_3+4\lambda_4) \\
16 \pi ^2 \beta _{\lambda_3}^{NLO,X} =& (16 \pi ^2)^2 \beta _{\lambda_3}^{NLO,I}+ Y_\tau^2 (\lambda_1-\lambda_2) (12 \lambda_3+4\lambda_4) \\
16 \pi ^2 \beta _{\lambda_3}^{NLO,Y} =& (16 \pi ^2)^2 \beta _{\lambda_3}^{NLO,II}+ Y_\tau^2 (\lambda_1-\lambda_2) (12 \lambda_3+4\lambda_4) \\[10pt]
\end{align*}

\begin{align*}
16 \pi ^2 \beta _{\lambda_4}^{LO,b} =& 3 g_1^2 g_2^2 -\left( 3 g_1^2 +9 g_2^2 \right) \lambda_4 +2 \lambda_1 \lambda_4+2 \lambda_2 \lambda_4+8 \lambda_3 \lambda_4+4 \lambda_4^2+8 \lambda_5^2\\
16 \pi ^2 \beta _{\lambda_4}^{LO,I} =&\left( 6 Y_b^2 +6 Y_t^2 +2 Y_\tau^2\right) \lambda_4\\
16 \pi ^2 \beta _{\lambda_4}^{LO,II} =&12 Y_b^2 Y_t^2 + \left( 6 Y_b^2 +6 Y_t^2 +2 Y_\tau^2\right) \lambda_4\\
16 \pi ^2 \beta _{\lambda_4}^{LO,X} =& 16 \pi ^2 \beta _{\lambda_4}^{LO,I}\\
16 \pi ^2 \beta _{\lambda_4}^{LO,Y} =& 16 \pi ^2 \beta _{\lambda_4}^{LO,II}\\[10pt]
(16 \pi ^2)^2 \beta _{\lambda_4}^{NLO,b} =& -\frac{73}{2} g_1^4 g_2^2 -14 g_1^2 g_2^4 \\
& +\frac{157 }{8} g_1^4 \lambda_4 +g_1^2 g_2^2 \left(5 \lambda_1 +5 \lambda_2 +2 \lambda_3 +\frac{51}{4} \lambda_4\right) -\frac{231}{8}g_2^4 \lambda_4 \\
& +g_1^2 \left( 4\lambda_1 \lambda_4 +4\lambda_2 \lambda_4 +4\lambda_3 \lambda_4 +8 \lambda_4^2 +16 \lambda_5^2\right) +g_2^2 \left( 36 \lambda_3 \lambda_4 +18\lambda_4^2 +54 \lambda_5^2\right) \\
& -\left( 7\lambda_1^2 +7\lambda_2^2 \right) \lambda_4 -(\lambda_1 +\lambda_2) \left( 40 \lambda_3 \lambda_4 +20 \lambda_4^2 +24 \lambda_5^2 \right) \\
& -28 \lambda_3^2 \lambda_4 -28 \lambda_3 \lambda_4^2 -48 \lambda_3 \lambda_5^2 -26 \lambda_4 \lambda_5^2 \\
(16 \pi ^2)^2 \beta _{\lambda_4}^{NLO,I} =& g_1^2 g_2^2 \left(9 Y_b^2+21 Y_t^2+11 Y_\tau^2\right) +g_1^2 \left( \frac{25}{12} Y_b^2 +\frac{85}{12} Y_t^2 +\frac{25}{4} Y_\tau^2\right) \lambda_4 \\
& +g_2^2 \left( \frac{45}{4} Y_b^2 +\frac{45}{4} Y_t^2 +\frac{15}{4} Y_\tau^2\right) \lambda_4 +40 g_3^2 \left(Y_b^2+Y_t^2\right) \lambda_4 \\
& -\frac{27}{2} \left( Y_b^2-Y_t^2\right) ^2 \lambda_4 -\frac{9}{2} Y_\tau^4 \lambda_4 \\
& - \left( 12Y_b^2 +12Y_t^2 +4Y_\tau^2 \right) \left( \lambda_2 \lambda_4 +2 \lambda_3 \lambda_4 +\lambda_4^2 +2 \lambda_5^2\right) \\
(16 \pi ^2)^2 \beta _{\lambda_4}^{NLO,II} =& (16 \pi ^2)^2 \beta _{\lambda_4}^{NLO,I} +\frac{4}{3}g_1^2 Y_b^2 Y_t^2 +64 g_3^2 Y_b^2 Y_t^2 -24 Y_b^4 Y_t^2 -24 Y_b^2 Y_t^4 \\
&-Y_b^2Y_t^2 (24 \lambda_3 +60 \lambda_4) -(12 Y_b^2 +4 Y_\tau^2) (\lambda_1-\lambda_2) \lambda_4 \\
16 \pi ^2 \beta _{\lambda_4}^{NLO,X} =& (16 \pi ^2)^2 \beta _{\lambda_4}^{NLO,I} -4 Y_\tau^2 (\lambda_1-\lambda_2) \lambda_4\\
16 \pi ^2 \beta _{\lambda_4}^{NLO,Y} =& (16 \pi ^2)^2 \beta _{\lambda_4}^{NLO,II} +4 Y_\tau^2 (\lambda_1-\lambda_2) \lambda_4\\[10pt]
\end{align*}

\begin{align*}
16 \pi ^2 \beta _{\lambda_5}^{LO,b} =& \left(-3 g_1^2 -9 g_2^2 +2\lambda_1 +2\lambda_2 +8 \lambda_3 +12 \lambda_4\right) \lambda_5 \\
16 \pi ^2 \beta _{\lambda_5}^{LO,f} = & \left(6 Y_b^2 +6 Y_t^2 +2Y_\tau^2\right) \lambda_5\\[10pt]
(16 \pi ^2)^2 \beta _{\lambda_5}^{NLO,b} =& \left( \frac{157}{8} g_1^4 +\frac{19}{4} g_1^2 g_2^2 -\frac{231}{8} g_2^4 -g_1^2 \left( 2\lambda_1 +2\lambda_2 -16 \lambda_3 -24 \lambda_4\right) \right. \\
& \hspace*{25pt} +g_2^2 \left( 36\lambda_3 +72 \lambda_4 \right) -7 \lambda_1^2 -7 \lambda_2^2 -\left( \lambda_1 +\lambda_2\right) \left( 40 \lambda_3+44 \lambda_4 \right) \\
& \hspace*{149pt} \left. \phantom{\frac{231}{8}} \hspace*{27pt} -28 \lambda_3^2 -76 \lambda_3 \lambda_4 -32 \lambda_4^2 +6 \lambda_5^2\right) \lambda_5\\
(16 \pi ^2)^2 \beta _{\lambda_5}^{NLO,I} =& \left( g_1^2 \left( \frac{25}{12} Y_b^2 +\frac{85}{12} Y_t^2 +\frac{25}{4} Y_\tau^2\right) +g_2^2 \left( \frac{45}{4} Y_b^2 +\frac{45}{4} Y_t^2 +\frac{15}{4}Y_\tau^2\right) \right. \\
& \hspace*{25pt} +g_3^2 \left(40Y_b^2 +40Y_t^2\right) -\frac{3}{2} \left( Y_b^2 -Y_t^2\right) ^2 -\frac{1}{2}Y_\tau^4 \\
& \left. \phantom{\frac{85}{12}} \hspace*{127pt} -\left( 12 Y_b^2 +12 Y_t^2 +4 Y_\tau^2 \right) \left( \lambda_2+2 \lambda_3+3 \lambda_4\right)\right) \lambda_5\\
(16 \pi ^2)^2 \beta _{\lambda_5}^{NLO,II} =& (16 \pi ^2)^2 \beta _{\lambda_5}^{NLO,I} -\left( 36 Y_b^2 Y_t^2 +12 Y_b^2 \left( \lambda_1-\lambda_2\right) +4 Y_\tau^2 (\lambda_1-\lambda_2)\right) \lambda_5\\
16 \pi ^2 \beta _{\lambda_5}^{NLO,X} =& (16 \pi ^2)^2 \beta _{\lambda_5}^{NLO,I} - 4 Y_\tau^2 (\lambda_1-\lambda_2) \lambda_5 \\
16 \pi ^2 \beta _{\lambda_5}^{NLO,Y} =& (16 \pi ^2)^2 \beta _{\lambda_5}^{NLO,II} +4 Y_\tau^2 (\lambda_1-\lambda_2) \lambda_5\\[10pt]
\end{align*}

Strictly speaking (and according to the introduced notation), there are only fermonic contributions to the $\beta$ functions of the Yukawa couplings. However, we will denote the bosonic loop contributions which are the same in all types as $\beta _{Y_i}^{(N)LO,b}$. Note that one bosonic contribution of $\beta _{Y_b}^{NLO}$ and $\beta _{Y_\tau}^{NLO}$ also depends on the type; thus we add the term to the corresponding $\beta _{Y_i}^{NLO,f}$.

\begin{align*}
16 \pi ^2 \beta _{Y_t}^{LO,b} &=-\left( \frac{17}{12} g_1^2 +\frac{9}{4} g_2^2 +8 g_3^2\right) Y_t\\
16 \pi ^2 \beta _{Y_t}^{LO,I} &=\left( \frac{3}{2} Y_b^2 +\frac{9}{2} Y_t^2 +Y_\tau^2\right) Y_t\\
16 \pi ^2 \beta _{Y_t}^{LO,II} &=16 \pi ^2 \beta _{Y_t}^{LO,I} -\left( Y_b^2 + Y_\tau^2 \right) Y_t\\
16 \pi ^2 \beta _{Y_t}^{LO,X} &=16 \pi ^2 \beta _{Y_t}^{LO,I} - Y_\tau^2 Y_t\\
16 \pi ^2 \beta _{Y_t}^{LO,Y} &=16 \pi ^2 \beta _{Y_t}^{LO,I} -Y_b^2 Y_t \\[10pt]
(16 \pi ^2)^2 \beta _{Y_t}^{NLO,b} &=\left( \frac{1267}{216} g_1^4 -\frac{3}{4}g_1^2 g_2^2 +\frac{19}{9}g_1^2 g_3^2 -\frac{21}{4} g_2^4 +9 g_2^2 g_3^2 -108 g_3^4 \right. \\
& \left. \hspace*{214pt} +\frac{3}{2} \lambda_2^2 +\lambda_3^2 +\lambda_3 \lambda_4 + \lambda_4^2 +\frac{3}{2} \lambda_5^2 \right) Y_t\\
(16 \pi ^2)^2 \beta _{Y_t}^{NLO,I} &= \left( g_1^2 \left( \frac{7}{48} Y_b^2 +\frac{131}{16} Y_t^2 +\frac{25}{8} Y_\tau^2\right) +g_2^2 \left( \frac{99}{16} Y_b^2 +\frac{225}{16} Y_t^2 +\frac{15}{8} Y_\tau^2\right) \right. \\
& \hspace*{25pt} +g_3^2 \left( 4  Y_b^2 +36 Y_t^2\right) \\
&\left. \hspace*{52pt} -\frac{1}{4}Y_b^4 -\frac{11}{4} Y_b^2 Y_t^2 +\frac{5}{4} Y_b^2 Y_\tau^2 -12 Y_t^4 -\frac{9}{4} Y_t^2 Y_\tau^2 -\frac{9}{4} Y_\tau^4 -6 Y_t^2 \lambda_2 \right) Y_t \\
(16 \pi ^2)^2 \beta _{Y_t}^{NLO,II} &= \left( g_1^2 \left( -\frac{41}{144} Y_b^2 +\frac{131}{16} Y_t^2\right) + g_2^2 \left( \frac{33}{16} Y_b^2 +\frac{225}{16} Y_t^2\right) +g_3^2 \left( \frac{16}{3} Y_b^2 + 36Y_t^2\right) \right. \\
&\left. \hspace*{54pt} -\frac{5}{2} Y_b^4 -\frac{5}{2}Y_b^2 Y_t^2 -\frac{3}{4}Y_b^2 Y_\tau^2 -12 Y_t^4 -2Y_b^2 \lambda_3 +2Y_b^2 \lambda_4 -6 Y_t^2 \lambda_2 \right) Y_t \\
16 \pi ^2 \beta _{Y_t}^{NLO,X} &= (16 \pi ^2)^2 \beta _{Y_t}^{NLO,I} -\left( \frac{25}{8} g_1^2 +\frac{15}{8} g_2^2 +\frac{5}{4} Y_b^2 -\frac{9}{4} Y_t^2 -\frac{9}{4} Y_\tau^2\right) Y_\tau^2 Y_t\\
16 \pi ^2 \beta _{Y_t}^{NLO,Y} &= (16 \pi ^2)^2 \beta _{Y_t}^{NLO,II} + \left( \frac{25}{8} g_1^2 + \frac{15}{8} g_2^2 +\frac{3}{4} Y_b^2 -\frac{9}{4} Y_t^2 -\frac{9}{4} Y_\tau^2\right) Y_\tau^2 Y_t \\[10pt]
\end{align*}

\begin{align*}
16 \pi ^2 \beta _{Y_b}^{LO,b} &=-\left( \frac{5}{12} g_1^2 +\frac{9}{4} g_2^2 +8 g_3^2\right) Y_b\\
16 \pi ^2 \beta _{Y_b}^{LO,I} &=\left( \frac{9}{2} Y_b^2 +\frac{3}{2} Y_t^2 +Y_\tau^2\right) Y_b\\
16 \pi ^2 \beta _{Y_b}^{LO,II} &=16 \pi ^2 \beta _{Y_b}^{LO,I} -Y_t^2 Y_b \\
16 \pi ^2 \beta _{Y_b}^{LO,X} &=16 \pi ^2 \beta _{Y_b}^{LO,I} - Y_\tau^2 Y_b\\
16 \pi ^2 \beta _{Y_b}^{LO,Y} &=16 \pi ^2 \beta _{Y_b}^{LO,I} - \left( Y_t^2 +Y_\tau^2\right) Y_b\\[10pt]
(16 \pi ^2)^2 \beta _{Y_b}^{NLO,b} &=\left( -\frac{113}{216} g_1^4 -\frac{9}{4} g_1^2 g_2^2 +\frac{31}{9} g_1^2 g_3^2 -\frac{21}{4} g_2^4 +9 g_2^2 g_3^2 -108 g_3^4 \right. \\
& \left. \hspace*{248pt} +\lambda_3^2 +\lambda_3 \lambda_4 +\lambda_4^2 +\frac{3}{2} \lambda_5^2 \right) Y_b \\
(16 \pi ^2)^2 \beta _{Y_b}^{NLO,I} &= \left( g_1^2 \left( \frac{79}{16} Y_b^2 +\frac{91}{48} Y_t^2 +\frac{25}{8} Y_\tau^2\right) +g_2^2 \left( \frac{225}{16} Y_b^2 +\frac{99}{16} Y_t^2 +\frac{15}{8} Y_\tau^2\right) \right. \\
& \hspace*{25pt} +g_3^2 \left(36 Y_b^2 +4Y_t^2\right) -12 Y_b^4 -\frac{11}{4} Y_b^2Y_t^2 -\frac{9}{4} Y_b^2Y_\tau^2 \\
& \left. \hspace*{166pt} -\frac{1}{4}Y_t^4 +\frac{5}{4} Y_t^2 Y_\tau^2 -\frac{9}{4} Y_\tau^4 - 6 Y_b^2 \lambda_2 +\frac{3}{2} \lambda_2^2\right) Y_b \\
(16 \pi ^2)^2 \beta _{Y_b}^{NLO,II} &= \left( g_1^2 \left( \frac{79}{16} Y_b^2 -\frac{53}{144} Y_t^2 +\frac{25}{8} Y_\tau^2\right) +g_2^2 \left(\frac{225}{16} Y_b^2 +\frac{33}{16} Y_t^2 +\frac{15}{8} Y_\tau^2\right) \right. \\
& \hspace*{25pt} +g_3^2 \left( 36Y_b^2 +\frac{16}{3} Y_t^2\right) -12 Y_b^4 -\frac{5}{2} Y_b^2 Y_t^2 -\frac{9}{4} Y_b^2 Y_\tau^2\\
& \left. \hspace*{125pt} -\frac{5}{2} Y_t^4 -\frac{9}{4} Y_\tau^4 -6 Y_b^2 \lambda_1 -2 Y_t^2 \lambda_3 +2 Y_t^2\lambda_4 +\frac{3}{2} \lambda_1^2 \right) Y_b \\
16 \pi ^2 \beta _{Y_b}^{NLO,X} &=(16 \pi ^2)^2 \beta _{Y_b}^{NLO,I} - \left( \frac{25}{8} g_1^2 +\frac{15}{8} g_2^2 -\frac{9}{4} Y_b^2 +\frac{5}{4} Y_t^2 -\frac{9}{4} Y_\tau^2\right) Y_\tau^2 Y_b\\
16 \pi ^2 \beta _{Y_b}^{NLO,Y} &=(16 \pi ^2)^2 \beta _{Y_b}^{NLO,II} - \left( \frac{25}{8} g_1^2 +\frac{15}{8} g_2^2 -\frac{9}{4} Y_b^2 +\frac{3}{4} Y_t^2 -\frac{9}{4} Y_\tau^2 \right) Y_\tau^2 Y_b\\[10pt]
\end{align*}

\begin{align*}
16 \pi ^2 \beta _{Y_\tau}^{LO,b} &=-\left( \frac{15}{4}g_1^2 +\frac{9}{4}g_2^2\right) Y_\tau\\
16 \pi ^2 \beta _{Y_\tau}^{LO,I} &=\left( 3 Y_b^2 +3 Y_t^2 +\frac{5}{2}Y_\tau^2\right) Y_\tau\\
16 \pi ^2 \beta _{Y_\tau}^{LO,II} &=16 \pi ^2 \beta _{Y_\tau}^{LO,I} -3 Y_t^2 Y_\tau \\
16 \pi ^2 \beta _{Y_\tau}^{LO,X} &=16 \pi ^2 \beta _{Y_\tau}^{LO,I} -\left( 3 Y_t^2 +3 Y_b^2\right) Y_\tau \\
16 \pi ^2 \beta _{Y_\tau}^{LO,Y} &=16 \pi ^2 \beta _{Y_\tau}^{LO,I} -3 Y_b^2 Y_\tau \\[10pt]
(16 \pi ^2)^2 \beta _{Y_\tau}^{NLO,b} &=\left( \frac{161}{8} g_1^4 +\frac{9}{4}g_1^2 g_2^2 -\frac{21}{4} g_2^4 +\lambda_3^2 +\lambda_3 \lambda_4 +\lambda_4^2 +\frac{3}{2} \lambda_5^2 \right) Y_\tau \\[100pt]
(16 \pi ^2)^2 \beta _{Y_\tau}^{NLO,I} &= \left( g_1^2 \left( \frac{25}{24} Y_b^2 +\frac{85}{24} Y_t^2 +\frac{179}{16} Y_\tau^2\right) +g_2^2 \left( \frac{45}{8} Y_b^2 +\frac{45}{8} Y_t^2 +\frac{165}{16} Y_\tau^2\right) \right. \\
& \hspace*{25pt} +g_3^2 \left( 20Y_b^2 +20Y_t^2\right) -\frac{27}{4} Y_b^4 +\frac{3}{2} Y_b^2 Y_t^2 -\frac{27}{4} Y_b^2 Y_\tau^2 \\
& \left. \hspace*{156pt} -\frac{27}{4} Y_t^4 -\frac{27}{4} Y_t^2 Y_\tau^2 -3 Y_\tau^4 -6 Y_\tau^2 \lambda_2 +\frac{3}{2} \lambda_2^2\right) Y_\tau \\
(16 \pi ^2)^2 \beta _{Y_\tau}^{NLO,II} &=
\left( g_1^2 \left( \frac{25}{24} Y_b^2 +\frac{179}{16} Y_\tau^2\right) +g_2^2 \left( \frac{45}{8} Y_b^2 +\frac{165}{16} Y_\tau^2\right) +20 g_3^2 Y_b^2 \right. \\
& \left. \hspace*{104pt} -\frac{27}{4} Y_b^4 -\frac{9}{4} Y_b^2 Y_t^2 -\frac{27}{4} Y_b^2 Y_\tau^2 -3 Y_\tau^4 -6 Y_\tau^2 \lambda_1 +\frac{3}{2} \lambda_1^2\right) Y_\tau \\
16 \pi ^2 \beta _{Y_\tau}^{NLO,X} &=(16 \pi ^2)^2 \beta _{Y_\tau}^{NLO,II} -\left( \frac{25}{24} g_1^2 +\frac{45}{8} g_2^2 +20 g_3^2 -\frac{27}{4} Y_b^2 -\frac{9}{4} Y_t^2 -\frac{27}{4} Y_\tau^2 \right) Y_b^2 Y_\tau \\
16 \pi ^2 \beta _{Y_\tau}^{NLO,Y} &=(16 \pi ^2)^2 \beta _{Y_\tau}^{NLO,I} 
-\left( \frac{25}{24} g_1^2 +\frac{45}{8} g_2^2 +20 g_3^2 -\frac{27}{4} Y_b^2 +\frac{15}{4} Y_t^2 -\frac{27}{4} Y_\tau^2\right) Y_b^2 Y_\tau \\[10pt]
\end{align*}

\bibliographystyle{apsrev4-1}
\bibliography{lambdathdm}

\begin{thebibliography}{109}%
\makeatletter
\providecommand \@ifxundefined [1]{%
 \@ifx{#1\undefined}
}%
\providecommand \@ifnum [1]{%
 \ifnum #1\expandafter \@firstoftwo
 \else \expandafter \@secondoftwo
 \fi
}%
\providecommand \@ifx [1]{%
 \ifx #1\expandafter \@firstoftwo
 \else \expandafter \@secondoftwo
 \fi
}%
\providecommand \natexlab [1]{#1}%
\providecommand \enquote  [1]{``#1''}%
\providecommand \bibnamefont  [1]{#1}%
\providecommand \bibfnamefont [1]{#1}%
\providecommand \citenamefont [1]{#1}%
\providecommand \href@noop [0]{\@secondoftwo}%
\providecommand \href [0]{\begingroup \@sanitize@url \@href}%
\providecommand \@href[1]{\@@startlink{#1}\@@href}%
\providecommand \@@href[1]{\endgroup#1\@@endlink}%
\providecommand \@sanitize@url [0]{\catcode `\\12\catcode `\$12\catcode
  `\&12\catcode `\#12\catcode `\^12\catcode `\_12\catcode `\%12\relax}%
\providecommand \@@startlink[1]{}%
\providecommand \@@endlink[0]{}%
\providecommand \url  [0]{\begingroup\@sanitize@url \@url }%
\providecommand \@url [1]{\endgroup\@href {#1}{\urlprefix }}%
\providecommand \urlprefix  [0]{URL }%
\providecommand \Eprint [0]{\href }%
\providecommand \doibase [0]{http://dx.doi.org/}%
\providecommand \selectlanguage [0]{\@gobble}%
\providecommand \bibinfo  [0]{\@secondoftwo}%
\providecommand \bibfield  [0]{\@secondoftwo}%
\providecommand \translation [1]{[#1]}%
\providecommand \BibitemOpen [0]{}%
\providecommand \bibitemStop [0]{}%
\providecommand \bibitemNoStop [0]{.\EOS\space}%
\providecommand \EOS [0]{\spacefactor3000\relax}%
\providecommand \BibitemShut  [1]{\csname bibitem#1\endcsname}%
\let\auto@bib@innerbib\@empty
\bibitem [{\citenamefont {Aad}\ \emph {et~al.}(2012)\citenamefont {Aad} \emph
  {et~al.}}]{Aad:2012tfa}%
  \BibitemOpen
  \bibfield  {author} {\bibinfo {author} {\bibfnamefont {G.}~\bibnamefont
  {Aad}} \emph {et~al.} (\bibinfo {collaboration} {{ATLAS} Collaboration}),\
  }\href {\doibase 10.1016/j.physletb.2012.08.020} {\bibfield  {journal}
  {\bibinfo  {journal} {Phys.Lett.}\ }\textbf {\bibinfo {volume} {B716}},\
  \bibinfo {pages} {1} (\bibinfo {year} {2012})},\ \Eprint
  {http://arxiv.org/abs/1207.7214} {arXiv:1207.7214 [hep-ex]} \BibitemShut
  {NoStop}%
\bibitem [{\citenamefont {Chatrchyan}\ \emph {et~al.}(2012)\citenamefont
  {Chatrchyan} \emph {et~al.}}]{Chatrchyan:2012ufa}%
  \BibitemOpen
  \bibfield  {author} {\bibinfo {author} {\bibfnamefont {S.}~\bibnamefont
  {Chatrchyan}} \emph {et~al.} (\bibinfo {collaboration} {{CMS}
  Collaboration}),\ }\href {\doibase 10.1016/j.physletb.2012.08.021} {\bibfield
   {journal} {\bibinfo  {journal} {Phys.Lett.}\ }\textbf {\bibinfo {volume}
  {B716}},\ \bibinfo {pages} {30} (\bibinfo {year} {2012})},\ \Eprint
  {http://arxiv.org/abs/1207.7235} {arXiv:1207.7235 [hep-ex]} \BibitemShut
  {NoStop}%
\bibitem [{\citenamefont {Lee}(1973)}]{Lee:1973iz}%
  \BibitemOpen
  \bibfield  {author} {\bibinfo {author} {\bibfnamefont {T.}~\bibnamefont
  {Lee}},\ }\href {\doibase 10.1103/PhysRevD.8.1226} {\bibfield  {journal}
  {\bibinfo  {journal} {Phys.Rev.}\ }\textbf {\bibinfo {volume} {D8}},\
  \bibinfo {pages} {1226} (\bibinfo {year} {1973})}\BibitemShut {NoStop}%
\bibitem [{\citenamefont {Gunion}\ and\ \citenamefont
  {Haber}(2003)}]{Gunion:2002zf}%
  \BibitemOpen
  \bibfield  {author} {\bibinfo {author} {\bibfnamefont {J.~F.}\ \bibnamefont
  {Gunion}}\ and\ \bibinfo {author} {\bibfnamefont {H.~E.}\ \bibnamefont
  {Haber}},\ }\href {\doibase 10.1103/PhysRevD.67.075019} {\bibfield  {journal}
  {\bibinfo  {journal} {Phys.Rev.}\ }\textbf {\bibinfo {volume} {D67}},\
  \bibinfo {pages} {075019} (\bibinfo {year} {2003})},\ \Eprint
  {http://arxiv.org/abs/hep-ph/0207010} {arXiv:hep-ph/0207010 [hep-ph]}
  \BibitemShut {NoStop}%
\bibitem [{\citenamefont {Branco}\ \emph {et~al.}(2012)\citenamefont {Branco},
  \citenamefont {Ferreira}, \citenamefont {Lavoura}, \citenamefont {Rebelo},
  \citenamefont {Sher} \emph {et~al.}}]{Branco:2011iw}%
  \BibitemOpen
  \bibfield  {author} {\bibinfo {author} {\bibfnamefont {G.}~\bibnamefont
  {Branco}}, \bibinfo {author} {\bibfnamefont {P.}~\bibnamefont {Ferreira}},
  \bibinfo {author} {\bibfnamefont {L.}~\bibnamefont {Lavoura}}, \bibinfo
  {author} {\bibfnamefont {M.}~\bibnamefont {Rebelo}}, \bibinfo {author}
  {\bibfnamefont {M.}~\bibnamefont {Sher}},  \emph {et~al.},\ }\href {\doibase
  10.1016/j.physrep.2012.02.002} {\bibfield  {journal} {\bibinfo  {journal}
  {Phys.Rept.}\ }\textbf {\bibinfo {volume} {516}},\ \bibinfo {pages} {1}
  (\bibinfo {year} {2012})},\ \Eprint {http://arxiv.org/abs/1106.0034}
  {arXiv:1106.0034 [hep-ph]} \BibitemShut {NoStop}%
\bibitem [{\citenamefont {Aad}\ \emph {et~al.}(2015{\natexlab{a}})\citenamefont
  {Aad} \emph {et~al.}}]{Aad:2015zhl}%
  \BibitemOpen
  \bibfield  {author} {\bibinfo {author} {\bibfnamefont {G.}~\bibnamefont
  {Aad}} \emph {et~al.} (\bibinfo {collaboration} {CMS s}),\ }\href@noop {} {\
  (\bibinfo {year} {2015}{\natexlab{a}})},\ \Eprint
  {http://arxiv.org/abs/1503.07589} {arXiv:1503.07589 [hep-ex]} \BibitemShut
  {NoStop}%
\bibitem [{\citenamefont {Degrassi}\ \emph {et~al.}(2012)\citenamefont
  {Degrassi}, \citenamefont {Di~Vita}, \citenamefont {Elias-Miro},
  \citenamefont {Espinosa}, \citenamefont {Giudice} \emph
  {et~al.}}]{Degrassi:2012ry}%
  \BibitemOpen
  \bibfield  {author} {\bibinfo {author} {\bibfnamefont {G.}~\bibnamefont
  {Degrassi}}, \bibinfo {author} {\bibfnamefont {S.}~\bibnamefont {Di~Vita}},
  \bibinfo {author} {\bibfnamefont {J.}~\bibnamefont {Elias-Miro}}, \bibinfo
  {author} {\bibfnamefont {J.~R.}\ \bibnamefont {Espinosa}}, \bibinfo {author}
  {\bibfnamefont {G.~F.}\ \bibnamefont {Giudice}},  \emph {et~al.},\ }\href
  {\doibase 10.1007/JHEP08(2012)098} {\bibfield  {journal} {\bibinfo  {journal}
  {JHEP}\ }\textbf {\bibinfo {volume} {1208}},\ \bibinfo {pages} {098}
  (\bibinfo {year} {2012})},\ \Eprint {http://arxiv.org/abs/1205.6497}
  {arXiv:1205.6497 [hep-ph]} \BibitemShut {NoStop}%
\bibitem [{\citenamefont {Buttazzo}\ \emph {et~al.}(2013)\citenamefont
  {Buttazzo}, \citenamefont {Degrassi}, \citenamefont {Giardino}, \citenamefont
  {Giudice}, \citenamefont {Sala} \emph {et~al.}}]{Buttazzo:2013uya}%
  \BibitemOpen
  \bibfield  {author} {\bibinfo {author} {\bibfnamefont {D.}~\bibnamefont
  {Buttazzo}}, \bibinfo {author} {\bibfnamefont {G.}~\bibnamefont {Degrassi}},
  \bibinfo {author} {\bibfnamefont {P.~P.}\ \bibnamefont {Giardino}}, \bibinfo
  {author} {\bibfnamefont {G.~F.}\ \bibnamefont {Giudice}}, \bibinfo {author}
  {\bibfnamefont {F.}~\bibnamefont {Sala}},  \emph {et~al.},\ }\href {\doibase
  10.1007/JHEP12(2013)089} {\bibfield  {journal} {\bibinfo  {journal} {JHEP}\
  }\textbf {\bibinfo {volume} {1312}},\ \bibinfo {pages} {089} (\bibinfo {year}
  {2013})},\ \Eprint {http://arxiv.org/abs/1307.3536} {arXiv:1307.3536
  [hep-ph]} \BibitemShut {NoStop}%
\bibitem [{\citenamefont {Cheon}\ and\ \citenamefont
  {Kang}(2013)}]{Cheon:2012rh}%
  \BibitemOpen
  \bibfield  {author} {\bibinfo {author} {\bibfnamefont {H.}~\bibnamefont
  {Cheon}}\ and\ \bibinfo {author} {\bibfnamefont {S.~K.}\ \bibnamefont
  {Kang}},\ }\href {\doibase 10.1007/JHEP09(2013)085} {\bibfield  {journal}
  {\bibinfo  {journal} {JHEP}\ }\textbf {\bibinfo {volume} {1309}},\ \bibinfo
  {pages} {085} (\bibinfo {year} {2013})},\ \Eprint
  {http://arxiv.org/abs/1207.1083} {arXiv:1207.1083 [hep-ph]} \BibitemShut
  {NoStop}%
\bibitem [{\citenamefont {Chen}\ and\ \citenamefont
  {Dawson}(2013)}]{Chen:2013kt}%
  \BibitemOpen
  \bibfield  {author} {\bibinfo {author} {\bibfnamefont {C.-Y.}\ \bibnamefont
  {Chen}}\ and\ \bibinfo {author} {\bibfnamefont {S.}~\bibnamefont {Dawson}},\
  }\href {\doibase 10.1103/PhysRevD.87.055016} {\bibfield  {journal} {\bibinfo
  {journal} {Phys.Rev.}\ }\textbf {\bibinfo {volume} {D87}},\ \bibinfo {pages}
  {055016} (\bibinfo {year} {2013})},\ \Eprint {http://arxiv.org/abs/1301.0309}
  {arXiv:1301.0309 [hep-ph]} \BibitemShut {NoStop}%
\bibitem [{\citenamefont {Chiang}\ and\ \citenamefont
  {Yagyu}(2013)}]{Chiang:2013ixa}%
  \BibitemOpen
  \bibfield  {author} {\bibinfo {author} {\bibfnamefont {C.-W.}\ \bibnamefont
  {Chiang}}\ and\ \bibinfo {author} {\bibfnamefont {K.}~\bibnamefont {Yagyu}},\
  }\href {\doibase 10.1007/JHEP07(2013)160} {\bibfield  {journal} {\bibinfo
  {journal} {JHEP}\ }\textbf {\bibinfo {volume} {1307}},\ \bibinfo {pages}
  {160} (\bibinfo {year} {2013})},\ \Eprint {http://arxiv.org/abs/1303.0168}
  {arXiv:1303.0168 [hep-ph]} \BibitemShut {NoStop}%
\bibitem [{\citenamefont {Grinstein}\ and\ \citenamefont
  {Uttayarat}(2013)}]{Grinstein:2013npa}%
  \BibitemOpen
  \bibfield  {author} {\bibinfo {author} {\bibfnamefont {B.}~\bibnamefont
  {Grinstein}}\ and\ \bibinfo {author} {\bibfnamefont {P.}~\bibnamefont
  {Uttayarat}},\ }\href {\doibase 10.1007/JHEP09(2013)110,
  10.1007/JHEP06(2013)094} {\bibfield  {journal} {\bibinfo  {journal} {JHEP}\
  }\textbf {\bibinfo {volume} {1306}},\ \bibinfo {pages} {094} (\bibinfo {year}
  {2013})},\ \Eprint {http://arxiv.org/abs/1304.0028} {arXiv:1304.0028
  [hep-ph]} \BibitemShut {NoStop}%
\bibitem [{\citenamefont {Barroso}\ \emph
  {et~al.}(2013{\natexlab{a}})\citenamefont {Barroso}, \citenamefont
  {Ferreira}, \citenamefont {Santos}, \citenamefont {Sher},\ and\ \citenamefont
  {Silva}}]{Barroso:2013zxa}%
  \BibitemOpen
  \bibfield  {author} {\bibinfo {author} {\bibfnamefont {A.}~\bibnamefont
  {Barroso}}, \bibinfo {author} {\bibfnamefont {P.}~\bibnamefont {Ferreira}},
  \bibinfo {author} {\bibfnamefont {R.}~\bibnamefont {Santos}}, \bibinfo
  {author} {\bibfnamefont {M.}~\bibnamefont {Sher}}, \ and\ \bibinfo {author}
  {\bibfnamefont {J.~P.}\ \bibnamefont {Silva}},\ }\href@noop {} {\  (\bibinfo
  {year} {2013}{\natexlab{a}})},\ \Eprint {http://arxiv.org/abs/1304.5225}
  {arXiv:1304.5225 [hep-ph]} \BibitemShut {NoStop}%
\bibitem [{\citenamefont {Coleppa}\ \emph {et~al.}(2014)\citenamefont
  {Coleppa}, \citenamefont {Kling},\ and\ \citenamefont
  {Su}}]{Coleppa:2013dya}%
  \BibitemOpen
  \bibfield  {author} {\bibinfo {author} {\bibfnamefont {B.}~\bibnamefont
  {Coleppa}}, \bibinfo {author} {\bibfnamefont {F.}~\bibnamefont {Kling}}, \
  and\ \bibinfo {author} {\bibfnamefont {S.}~\bibnamefont {Su}},\ }\href
  {\doibase 10.1007/JHEP01(2014)161} {\bibfield  {journal} {\bibinfo  {journal}
  {JHEP}\ }\textbf {\bibinfo {volume} {1401}},\ \bibinfo {pages} {161}
  (\bibinfo {year} {2014})},\ \Eprint {http://arxiv.org/abs/1305.0002}
  {arXiv:1305.0002 [hep-ph]} \BibitemShut {NoStop}%
\bibitem [{\citenamefont {Eberhardt}\ \emph {et~al.}(2013)\citenamefont
  {Eberhardt}, \citenamefont {Nierste},\ and\ \citenamefont
  {Wiebusch}}]{Eberhardt:2013uba}%
  \BibitemOpen
  \bibfield  {author} {\bibinfo {author} {\bibfnamefont {O.}~\bibnamefont
  {Eberhardt}}, \bibinfo {author} {\bibfnamefont {U.}~\bibnamefont {Nierste}},
  \ and\ \bibinfo {author} {\bibfnamefont {M.}~\bibnamefont {Wiebusch}},\
  }\href {\doibase 10.1007/JHEP07(2013)118} {\bibfield  {journal} {\bibinfo
  {journal} {JHEP}\ }\textbf {\bibinfo {volume} {1307}},\ \bibinfo {pages}
  {118} (\bibinfo {year} {2013})},\ \Eprint {http://arxiv.org/abs/1305.1649}
  {arXiv:1305.1649 [hep-ph]} \BibitemShut {NoStop}%
\bibitem [{\citenamefont {Belanger}\ \emph {et~al.}(2013)\citenamefont
  {Belanger}, \citenamefont {Dumont}, \citenamefont {Ellwanger}, \citenamefont
  {Gunion},\ and\ \citenamefont {Kraml}}]{Belanger:2013xza}%
  \BibitemOpen
  \bibfield  {author} {\bibinfo {author} {\bibfnamefont {G.}~\bibnamefont
  {Belanger}}, \bibinfo {author} {\bibfnamefont {B.}~\bibnamefont {Dumont}},
  \bibinfo {author} {\bibfnamefont {U.}~\bibnamefont {Ellwanger}}, \bibinfo
  {author} {\bibfnamefont {J.}~\bibnamefont {Gunion}}, \ and\ \bibinfo {author}
  {\bibfnamefont {S.}~\bibnamefont {Kraml}},\ }\href {\doibase
  10.1103/PhysRevD.88.075008} {\bibfield  {journal} {\bibinfo  {journal}
  {Phys.Rev.}\ }\textbf {\bibinfo {volume} {D88}},\ \bibinfo {pages} {075008}
  (\bibinfo {year} {2013})},\ \Eprint {http://arxiv.org/abs/1306.2941}
  {arXiv:1306.2941 [hep-ph]} \BibitemShut {NoStop}%
\bibitem [{\citenamefont {Chang}\ \emph {et~al.}(2014)\citenamefont {Chang},
  \citenamefont {Kang}, \citenamefont {Lee}, \citenamefont {Lee}, \citenamefont
  {Park} \emph {et~al.}}]{Chang:2013ona}%
  \BibitemOpen
  \bibfield  {author} {\bibinfo {author} {\bibfnamefont {S.}~\bibnamefont
  {Chang}}, \bibinfo {author} {\bibfnamefont {S.~K.}\ \bibnamefont {Kang}},
  \bibinfo {author} {\bibfnamefont {J.-P.}\ \bibnamefont {Lee}}, \bibinfo
  {author} {\bibfnamefont {K.~Y.}\ \bibnamefont {Lee}}, \bibinfo {author}
  {\bibfnamefont {S.~C.}\ \bibnamefont {Park}},  \emph {et~al.},\ }\href
  {\doibase 10.1007/JHEP09(2014)101} {\bibfield  {journal} {\bibinfo  {journal}
  {JHEP}\ }\textbf {\bibinfo {volume} {1409}},\ \bibinfo {pages} {101}
  (\bibinfo {year} {2014})},\ \Eprint {http://arxiv.org/abs/1310.3374}
  {arXiv:1310.3374 [hep-ph]} \BibitemShut {NoStop}%
\bibitem [{\citenamefont {Cheung}\ \emph {et~al.}(2014)\citenamefont {Cheung},
  \citenamefont {Lee},\ and\ \citenamefont {Tseng}}]{Cheung:2013rva}%
  \BibitemOpen
  \bibfield  {author} {\bibinfo {author} {\bibfnamefont {K.}~\bibnamefont
  {Cheung}}, \bibinfo {author} {\bibfnamefont {J.~S.}\ \bibnamefont {Lee}}, \
  and\ \bibinfo {author} {\bibfnamefont {P.-Y.}\ \bibnamefont {Tseng}},\ }\href
  {\doibase 10.1007/JHEP01(2014)085} {\bibfield  {journal} {\bibinfo  {journal}
  {JHEP}\ }\textbf {\bibinfo {volume} {1401}},\ \bibinfo {pages} {085}
  (\bibinfo {year} {2014})},\ \Eprint {http://arxiv.org/abs/1310.3937}
  {arXiv:1310.3937 [hep-ph]} \BibitemShut {NoStop}%
\bibitem [{\citenamefont {Celis}\ \emph {et~al.}(2013)\citenamefont {Celis},
  \citenamefont {Ilisie},\ and\ \citenamefont {Pich}}]{Celis:2013ixa}%
  \BibitemOpen
  \bibfield  {author} {\bibinfo {author} {\bibfnamefont {A.}~\bibnamefont
  {Celis}}, \bibinfo {author} {\bibfnamefont {V.}~\bibnamefont {Ilisie}}, \
  and\ \bibinfo {author} {\bibfnamefont {A.}~\bibnamefont {Pich}},\ }\href
  {\doibase 10.1007/JHEP12(2013)095} {\bibfield  {journal} {\bibinfo  {journal}
  {JHEP}\ }\textbf {\bibinfo {volume} {1312}},\ \bibinfo {pages} {095}
  (\bibinfo {year} {2013})},\ \Eprint {http://arxiv.org/abs/1310.7941}
  {arXiv:1310.7941 [hep-ph]} \BibitemShut {NoStop}%
\bibitem [{\citenamefont {Wang}\ and\ \citenamefont
  {Han}(2014{\natexlab{a}})}]{Wang:2013sha}%
  \BibitemOpen
  \bibfield  {author} {\bibinfo {author} {\bibfnamefont {L.}~\bibnamefont
  {Wang}}\ and\ \bibinfo {author} {\bibfnamefont {X.-F.}\ \bibnamefont {Han}},\
  }\href {\doibase 10.1007/JHEP04(2014)128} {\bibfield  {journal} {\bibinfo
  {journal} {JHEP}\ }\textbf {\bibinfo {volume} {1404}},\ \bibinfo {pages}
  {128} (\bibinfo {year} {2014}{\natexlab{a}})},\ \Eprint
  {http://arxiv.org/abs/1312.4759} {arXiv:1312.4759 [hep-ph]} \BibitemShut
  {NoStop}%
\bibitem [{\citenamefont {Baglio}\ \emph {et~al.}(2014)\citenamefont {Baglio},
  \citenamefont {Eberhardt}, \citenamefont {Nierste},\ and\ \citenamefont
  {Wiebusch}}]{Baglio:2014nea}%
  \BibitemOpen
  \bibfield  {author} {\bibinfo {author} {\bibfnamefont {J.}~\bibnamefont
  {Baglio}}, \bibinfo {author} {\bibfnamefont {O.}~\bibnamefont {Eberhardt}},
  \bibinfo {author} {\bibfnamefont {U.}~\bibnamefont {Nierste}}, \ and\
  \bibinfo {author} {\bibfnamefont {M.}~\bibnamefont {Wiebusch}},\ }\href@noop
  {} {\bibfield  {journal} {\bibinfo  {journal} {arXiv:1403.1264, accepted by
  Phys.Rev.~D}\ } (\bibinfo {year} {2014})},\ \Eprint
  {http://arxiv.org/abs/1403.1264} {arXiv:1403.1264 [hep-ph]} \BibitemShut
  {NoStop}%
\bibitem [{\citenamefont {Inoue}\ \emph {et~al.}(2014)\citenamefont {Inoue},
  \citenamefont {Ramsey-Musolf},\ and\ \citenamefont {Zhang}}]{Inoue:2014nva}%
  \BibitemOpen
  \bibfield  {author} {\bibinfo {author} {\bibfnamefont {S.}~\bibnamefont
  {Inoue}}, \bibinfo {author} {\bibfnamefont {M.~J.}\ \bibnamefont
  {Ramsey-Musolf}}, \ and\ \bibinfo {author} {\bibfnamefont {Y.}~\bibnamefont
  {Zhang}},\ }\href {\doibase 10.1103/PhysRevD.89.115023} {\bibfield  {journal}
  {\bibinfo  {journal} {Phys.Rev.}\ }\textbf {\bibinfo {volume} {D89}},\
  \bibinfo {pages} {115023} (\bibinfo {year} {2014})},\ \Eprint
  {http://arxiv.org/abs/1403.4257} {arXiv:1403.4257 [hep-ph]} \BibitemShut
  {NoStop}%
\bibitem [{\citenamefont {Dumont}\ \emph
  {et~al.}(2014{\natexlab{a}})\citenamefont {Dumont}, \citenamefont {Gunion},
  \citenamefont {Jiang},\ and\ \citenamefont {Kraml}}]{Dumont:2014wha}%
  \BibitemOpen
  \bibfield  {author} {\bibinfo {author} {\bibfnamefont {B.}~\bibnamefont
  {Dumont}}, \bibinfo {author} {\bibfnamefont {J.~F.}\ \bibnamefont {Gunion}},
  \bibinfo {author} {\bibfnamefont {Y.}~\bibnamefont {Jiang}}, \ and\ \bibinfo
  {author} {\bibfnamefont {S.}~\bibnamefont {Kraml}},\ }\href {\doibase
  10.1103/PhysRevD.90.035021} {\bibfield  {journal} {\bibinfo  {journal}
  {Phys.Rev.}\ }\textbf {\bibinfo {volume} {D90}},\ \bibinfo {pages} {035021}
  (\bibinfo {year} {2014}{\natexlab{a}})},\ \Eprint
  {http://arxiv.org/abs/1405.3584} {arXiv:1405.3584 [hep-ph]} \BibitemShut
  {NoStop}%
\bibitem [{\citenamefont {Kanemura}\ \emph
  {et~al.}(2014{\natexlab{a}})\citenamefont {Kanemura}, \citenamefont
  {Tsumura}, \citenamefont {Yagyu},\ and\ \citenamefont
  {Yokoya}}]{Kanemura:2014bqa}%
  \BibitemOpen
  \bibfield  {author} {\bibinfo {author} {\bibfnamefont {S.}~\bibnamefont
  {Kanemura}}, \bibinfo {author} {\bibfnamefont {K.}~\bibnamefont {Tsumura}},
  \bibinfo {author} {\bibfnamefont {K.}~\bibnamefont {Yagyu}}, \ and\ \bibinfo
  {author} {\bibfnamefont {H.}~\bibnamefont {Yokoya}},\ }\href {\doibase
  10.1103/PhysRevD.90.075001} {\bibfield  {journal} {\bibinfo  {journal}
  {Phys.Rev.}\ }\textbf {\bibinfo {volume} {D90}},\ \bibinfo {pages} {075001}
  (\bibinfo {year} {2014}{\natexlab{a}})},\ \Eprint
  {http://arxiv.org/abs/1406.3294} {arXiv:1406.3294 [hep-ph]} \BibitemShut
  {NoStop}%
\bibitem [{\citenamefont {Ferreira}\ \emph {et~al.}(2014)\citenamefont
  {Ferreira}, \citenamefont {Guedes}, \citenamefont {Gunion}, \citenamefont
  {Haber}, \citenamefont {Sampaio} \emph {et~al.}}]{Ferreira:2014sld}%
  \BibitemOpen
  \bibfield  {author} {\bibinfo {author} {\bibfnamefont {P.}~\bibnamefont
  {Ferreira}}, \bibinfo {author} {\bibfnamefont {R.}~\bibnamefont {Guedes}},
  \bibinfo {author} {\bibfnamefont {J.~F.}\ \bibnamefont {Gunion}}, \bibinfo
  {author} {\bibfnamefont {H.~E.}\ \bibnamefont {Haber}}, \bibinfo {author}
  {\bibfnamefont {M.~O.~P.}\ \bibnamefont {Sampaio}},  \emph {et~al.},\
  }\href@noop {} {\  (\bibinfo {year} {2014})},\ \Eprint
  {http://arxiv.org/abs/1407.4396} {arXiv:1407.4396 [hep-ph]} \BibitemShut
  {NoStop}%
\bibitem [{\citenamefont {Broggio}\ \emph {et~al.}(2014)\citenamefont
  {Broggio}, \citenamefont {Chun}, \citenamefont {Passera}, \citenamefont
  {Patel},\ and\ \citenamefont {Vempati}}]{Broggio:2014mna}%
  \BibitemOpen
  \bibfield  {author} {\bibinfo {author} {\bibfnamefont {A.}~\bibnamefont
  {Broggio}}, \bibinfo {author} {\bibfnamefont {E.~J.}\ \bibnamefont {Chun}},
  \bibinfo {author} {\bibfnamefont {M.}~\bibnamefont {Passera}}, \bibinfo
  {author} {\bibfnamefont {K.~M.}\ \bibnamefont {Patel}}, \ and\ \bibinfo
  {author} {\bibfnamefont {S.~K.}\ \bibnamefont {Vempati}},\ }\href {\doibase
  10.1007/JHEP11(2014)058} {\bibfield  {journal} {\bibinfo  {journal} {JHEP}\
  }\textbf {\bibinfo {volume} {1411}},\ \bibinfo {pages} {058} (\bibinfo {year}
  {2014})},\ \Eprint {http://arxiv.org/abs/1409.3199} {arXiv:1409.3199
  [hep-ph]} \BibitemShut {NoStop}%
\bibitem [{\citenamefont {Dumont}\ \emph
  {et~al.}(2014{\natexlab{b}})\citenamefont {Dumont}, \citenamefont {Gunion},
  \citenamefont {Jiang},\ and\ \citenamefont {Kraml}}]{Dumont:2014kna}%
  \BibitemOpen
  \bibfield  {author} {\bibinfo {author} {\bibfnamefont {B.}~\bibnamefont
  {Dumont}}, \bibinfo {author} {\bibfnamefont {J.~F.}\ \bibnamefont {Gunion}},
  \bibinfo {author} {\bibfnamefont {Y.}~\bibnamefont {Jiang}}, \ and\ \bibinfo
  {author} {\bibfnamefont {S.}~\bibnamefont {Kraml}},\ }\href@noop {} {\
  (\bibinfo {year} {2014}{\natexlab{b}})},\ \Eprint
  {http://arxiv.org/abs/1409.4088} {arXiv:1409.4088 [hep-ph]} \BibitemShut
  {NoStop}%
\bibitem [{\citenamefont {Bernon}\ \emph {et~al.}(2014)\citenamefont {Bernon},
  \citenamefont {Gunion}, \citenamefont {Jiang},\ and\ \citenamefont
  {Kraml}}]{Bernon:2014nxa}%
  \BibitemOpen
  \bibfield  {author} {\bibinfo {author} {\bibfnamefont {J.}~\bibnamefont
  {Bernon}}, \bibinfo {author} {\bibfnamefont {J.~F.}\ \bibnamefont {Gunion}},
  \bibinfo {author} {\bibfnamefont {Y.}~\bibnamefont {Jiang}}, \ and\ \bibinfo
  {author} {\bibfnamefont {S.}~\bibnamefont {Kraml}},\ }\href@noop {} {\
  (\bibinfo {year} {2014})},\ \Eprint {http://arxiv.org/abs/1412.3385}
  {arXiv:1412.3385 [hep-ph]} \BibitemShut {NoStop}%
\bibitem [{\citenamefont {Chen}\ \emph {et~al.}(2015)\citenamefont {Chen},
  \citenamefont {Dawson},\ and\ \citenamefont {Zhang}}]{Chen:2015gaa}%
  \BibitemOpen
  \bibfield  {author} {\bibinfo {author} {\bibfnamefont {C.-Y.}\ \bibnamefont
  {Chen}}, \bibinfo {author} {\bibfnamefont {S.}~\bibnamefont {Dawson}}, \ and\
  \bibinfo {author} {\bibfnamefont {Y.}~\bibnamefont {Zhang}},\ }\href@noop {}
  {\  (\bibinfo {year} {2015})},\ \Eprint {http://arxiv.org/abs/1503.01114}
  {arXiv:1503.01114 [hep-ph]} \BibitemShut {NoStop}%
\bibitem [{\citenamefont {Chen}\ \emph {et~al.}(2013)\citenamefont {Chen},
  \citenamefont {Dawson},\ and\ \citenamefont {Sher}}]{Chen:2013rba}%
  \BibitemOpen
  \bibfield  {author} {\bibinfo {author} {\bibfnamefont {C.-Y.}\ \bibnamefont
  {Chen}}, \bibinfo {author} {\bibfnamefont {S.}~\bibnamefont {Dawson}}, \ and\
  \bibinfo {author} {\bibfnamefont {M.}~\bibnamefont {Sher}},\ }\href {\doibase
  10.1103/PhysRevD.88.015018, 10.1103/PhysRevD.88.039901} {\bibfield  {journal}
  {\bibinfo  {journal} {Phys.Rev.}\ }\textbf {\bibinfo {volume} {D88}},\
  \bibinfo {pages} {015018} (\bibinfo {year} {2013})},\ \Eprint
  {http://arxiv.org/abs/1305.1624} {arXiv:1305.1624 [hep-ph]} \BibitemShut
  {NoStop}%
\bibitem [{\citenamefont {Craig}\ \emph {et~al.}(2013)\citenamefont {Craig},
  \citenamefont {Galloway},\ and\ \citenamefont {Thomas}}]{Craig:2013hca}%
  \BibitemOpen
  \bibfield  {author} {\bibinfo {author} {\bibfnamefont {N.}~\bibnamefont
  {Craig}}, \bibinfo {author} {\bibfnamefont {J.}~\bibnamefont {Galloway}}, \
  and\ \bibinfo {author} {\bibfnamefont {S.}~\bibnamefont {Thomas}},\
  }\href@noop {} {\  (\bibinfo {year} {2013})},\ \Eprint
  {http://arxiv.org/abs/1305.2424} {arXiv:1305.2424 [hep-ph]} \BibitemShut
  {NoStop}%
\bibitem [{\citenamefont {Barger}\ \emph {et~al.}(2013)\citenamefont {Barger},
  \citenamefont {Everett}, \citenamefont {Logan},\ and\ \citenamefont
  {Shaughnessy}}]{Barger:2013ofa}%
  \BibitemOpen
  \bibfield  {author} {\bibinfo {author} {\bibfnamefont {V.}~\bibnamefont
  {Barger}}, \bibinfo {author} {\bibfnamefont {L.~L.}\ \bibnamefont {Everett}},
  \bibinfo {author} {\bibfnamefont {H.~E.}\ \bibnamefont {Logan}}, \ and\
  \bibinfo {author} {\bibfnamefont {G.}~\bibnamefont {Shaughnessy}},\ }\href
  {\doibase 10.1103/PhysRevD.88.115003} {\bibfield  {journal} {\bibinfo
  {journal} {Phys.Rev.}\ }\textbf {\bibinfo {volume} {D88}},\ \bibinfo {pages}
  {115003} (\bibinfo {year} {2013})},\ \Eprint {http://arxiv.org/abs/1308.0052}
  {arXiv:1308.0052 [hep-ph]} \BibitemShut {NoStop}%
\bibitem [{\citenamefont {Kanemura}\ \emph
  {et~al.}(2014{\natexlab{b}})\citenamefont {Kanemura}, \citenamefont
  {Yokoya},\ and\ \citenamefont {Zheng}}]{Kanemura:2014dea}%
  \BibitemOpen
  \bibfield  {author} {\bibinfo {author} {\bibfnamefont {S.}~\bibnamefont
  {Kanemura}}, \bibinfo {author} {\bibfnamefont {H.}~\bibnamefont {Yokoya}}, \
  and\ \bibinfo {author} {\bibfnamefont {Y.-J.}\ \bibnamefont {Zheng}},\ }\href
  {\doibase 10.1016/j.nuclphysb.2014.07.00, 10.1016/j.nuclphysb.2014.07.007}
  {\bibfield  {journal} {\bibinfo  {journal} {Nucl.Phys.}\ }\textbf {\bibinfo
  {volume} {B886}},\ \bibinfo {pages} {524} (\bibinfo {year}
  {2014}{\natexlab{b}})},\ \Eprint {http://arxiv.org/abs/1404.5835}
  {arXiv:1404.5835 [hep-ph]} \BibitemShut {NoStop}%
\bibitem [{\citenamefont {Wang}\ and\ \citenamefont
  {Han}(2014{\natexlab{b}})}]{Wang:2014lta}%
  \BibitemOpen
  \bibfield  {author} {\bibinfo {author} {\bibfnamefont {L.}~\bibnamefont
  {Wang}}\ and\ \bibinfo {author} {\bibfnamefont {X.-F.}\ \bibnamefont {Han}},\
  }\href {\doibase 10.1007/JHEP11(2014)085} {\bibfield  {journal} {\bibinfo
  {journal} {JHEP}\ }\textbf {\bibinfo {volume} {1411}},\ \bibinfo {pages}
  {085} (\bibinfo {year} {2014}{\natexlab{b}})},\ \Eprint
  {http://arxiv.org/abs/1404.7437} {arXiv:1404.7437 [hep-ph]} \BibitemShut
  {NoStop}%
\bibitem [{\citenamefont {Barger}\ \emph {et~al.}(2014)\citenamefont {Barger},
  \citenamefont {Everett}, \citenamefont {Jackson}, \citenamefont {Peterson},\
  and\ \citenamefont {Shaughnessy}}]{Barger:2014qva}%
  \BibitemOpen
  \bibfield  {author} {\bibinfo {author} {\bibfnamefont {V.}~\bibnamefont
  {Barger}}, \bibinfo {author} {\bibfnamefont {L.~L.}\ \bibnamefont {Everett}},
  \bibinfo {author} {\bibfnamefont {C.~B.}\ \bibnamefont {Jackson}}, \bibinfo
  {author} {\bibfnamefont {A.~D.}\ \bibnamefont {Peterson}}, \ and\ \bibinfo
  {author} {\bibfnamefont {G.}~\bibnamefont {Shaughnessy}},\ }\href {\doibase
  10.1103/PhysRevD.90.095006} {\bibfield  {journal} {\bibinfo  {journal}
  {Phys.Rev.}\ }\textbf {\bibinfo {volume} {D90}},\ \bibinfo {pages} {095006}
  (\bibinfo {year} {2014})},\ \Eprint {http://arxiv.org/abs/1408.2525}
  {arXiv:1408.2525 [hep-ph]} \BibitemShut {NoStop}%
\bibitem [{\citenamefont {Gorbahn}\ \emph {et~al.}(2015)\citenamefont
  {Gorbahn}, \citenamefont {No},\ and\ \citenamefont {Sanz}}]{Gorbahn:2015gxa}%
  \BibitemOpen
  \bibfield  {author} {\bibinfo {author} {\bibfnamefont {M.}~\bibnamefont
  {Gorbahn}}, \bibinfo {author} {\bibfnamefont {J.~M.}\ \bibnamefont {No}}, \
  and\ \bibinfo {author} {\bibfnamefont {V.}~\bibnamefont {Sanz}},\ }\href@noop
  {} {\  (\bibinfo {year} {2015})},\ \Eprint {http://arxiv.org/abs/1502.07352}
  {arXiv:1502.07352 [hep-ph]} \BibitemShut {NoStop}%
\bibitem [{\citenamefont {Kanemura}\ \emph {et~al.}(2015)\citenamefont
  {Kanemura}, \citenamefont {Kikuchi},\ and\ \citenamefont
  {Yagyu}}]{Kanemura:2015mxa}%
  \BibitemOpen
  \bibfield  {author} {\bibinfo {author} {\bibfnamefont {S.}~\bibnamefont
  {Kanemura}}, \bibinfo {author} {\bibfnamefont {M.}~\bibnamefont {Kikuchi}}, \
  and\ \bibinfo {author} {\bibfnamefont {K.}~\bibnamefont {Yagyu}},\
  }\href@noop {} {\  (\bibinfo {year} {2015})},\ \Eprint
  {http://arxiv.org/abs/1502.07716} {arXiv:1502.07716 [hep-ph]} \BibitemShut
  {NoStop}%
\bibitem [{\citenamefont {Nierste}\ and\ \citenamefont
  {Riesselmann}(1996)}]{Nierste:1995zx}%
  \BibitemOpen
  \bibfield  {author} {\bibinfo {author} {\bibfnamefont {U.}~\bibnamefont
  {Nierste}}\ and\ \bibinfo {author} {\bibfnamefont {K.}~\bibnamefont
  {Riesselmann}},\ }\href {\doibase 10.1103/PhysRevD.53.6638} {\bibfield
  {journal} {\bibinfo  {journal} {Phys.Rev.}\ }\textbf {\bibinfo {volume}
  {D53}},\ \bibinfo {pages} {6638} (\bibinfo {year} {1996})},\ \Eprint
  {http://arxiv.org/abs/hep-ph/9511407} {arXiv:hep-ph/9511407 [hep-ph]}
  \BibitemShut {NoStop}%
\bibitem [{\citenamefont {Hill}\ \emph {et~al.}(1985)\citenamefont {Hill},
  \citenamefont {Leung},\ and\ \citenamefont {Rao}}]{Hill:1985tg}%
  \BibitemOpen
  \bibfield  {author} {\bibinfo {author} {\bibfnamefont {C.~T.}\ \bibnamefont
  {Hill}}, \bibinfo {author} {\bibfnamefont {C.~N.}\ \bibnamefont {Leung}}, \
  and\ \bibinfo {author} {\bibfnamefont {S.}~\bibnamefont {Rao}},\ }\href
  {\doibase 10.1016/0550-3213(85)90328-1} {\bibfield  {journal} {\bibinfo
  {journal} {Nucl.Phys.}\ }\textbf {\bibinfo {volume} {B262}},\ \bibinfo
  {pages} {517} (\bibinfo {year} {1985})}\BibitemShut {NoStop}%
\bibitem [{\citenamefont {Andrianov}\ \emph {et~al.}(1995)\citenamefont
  {Andrianov}, \citenamefont {Rodenberg},\ and\ \citenamefont
  {Romanenko}}]{Andrianov:1994za}%
  \BibitemOpen
  \bibfield  {author} {\bibinfo {author} {\bibfnamefont {A.~A.}\ \bibnamefont
  {Andrianov}}, \bibinfo {author} {\bibfnamefont {R.}~\bibnamefont
  {Rodenberg}}, \ and\ \bibinfo {author} {\bibfnamefont {N.}~\bibnamefont
  {Romanenko}},\ }\href {\doibase 10.1007/BF02816853} {\bibfield  {journal}
  {\bibinfo  {journal} {Nuovo Cim.}\ }\textbf {\bibinfo {volume} {A108}},\
  \bibinfo {pages} {577} (\bibinfo {year} {1995})},\ \Eprint
  {http://arxiv.org/abs/hep-ph/9408301} {arXiv:hep-ph/9408301 [hep-ph]}
  \BibitemShut {NoStop}%
\bibitem [{\citenamefont {Bijnens}\ \emph {et~al.}(2012)\citenamefont
  {Bijnens}, \citenamefont {Lu},\ and\ \citenamefont
  {Rathsman}}]{Bijnens:2011gd}%
  \BibitemOpen
  \bibfield  {author} {\bibinfo {author} {\bibfnamefont {J.}~\bibnamefont
  {Bijnens}}, \bibinfo {author} {\bibfnamefont {J.}~\bibnamefont {Lu}}, \ and\
  \bibinfo {author} {\bibfnamefont {J.}~\bibnamefont {Rathsman}},\ }\href
  {\doibase 10.1007/JHEP05(2012)118} {\bibfield  {journal} {\bibinfo  {journal}
  {JHEP}\ }\textbf {\bibinfo {volume} {1205}},\ \bibinfo {pages} {118}
  (\bibinfo {year} {2012})},\ \Eprint {http://arxiv.org/abs/1111.5760}
  {arXiv:1111.5760 [hep-ph]} \BibitemShut {NoStop}%
\bibitem [{\citenamefont {Juarez~W.}\ \emph {et~al.}(2012)\citenamefont
  {Juarez~W.}, \citenamefont {Morales~C.},\ and\ \citenamefont
  {Kielanowski}}]{JuarezW.:2012qa}%
  \BibitemOpen
  \bibfield  {author} {\bibinfo {author} {\bibfnamefont {S.}~\bibnamefont
  {Juarez~W.}}, \bibinfo {author} {\bibfnamefont {D.}~\bibnamefont
  {Morales~C.}}, \ and\ \bibinfo {author} {\bibfnamefont {P.}~\bibnamefont
  {Kielanowski}},\ }\href@noop {} {\  (\bibinfo {year} {2012})},\ \Eprint
  {http://arxiv.org/abs/1201.1876} {arXiv:1201.1876 [hep-ph]} \BibitemShut
  {NoStop}%
\bibitem [{\citenamefont {Chakrabarty}\ \emph {et~al.}(2014)\citenamefont
  {Chakrabarty}, \citenamefont {Dey},\ and\ \citenamefont
  {Mukhopadhyaya}}]{Chakrabarty:2014aya}%
  \BibitemOpen
  \bibfield  {author} {\bibinfo {author} {\bibfnamefont {N.}~\bibnamefont
  {Chakrabarty}}, \bibinfo {author} {\bibfnamefont {U.~K.}\ \bibnamefont
  {Dey}}, \ and\ \bibinfo {author} {\bibfnamefont {B.}~\bibnamefont
  {Mukhopadhyaya}},\ }\href@noop {} {\  (\bibinfo {year} {2014})},\ \Eprint
  {http://arxiv.org/abs/1407.2145} {arXiv:1407.2145 [hep-ph]} \BibitemShut
  {NoStop}%
\bibitem [{\citenamefont {Chakrabarty}\ \emph {et~al.}(2015)\citenamefont
  {Chakrabarty}, \citenamefont {Ghosh}, \citenamefont {Mukhopadhyaya},\ and\
  \citenamefont {Saha}}]{Chakrabarty:2015yia}%
  \BibitemOpen
  \bibfield  {author} {\bibinfo {author} {\bibfnamefont {N.}~\bibnamefont
  {Chakrabarty}}, \bibinfo {author} {\bibfnamefont {D.~K.}\ \bibnamefont
  {Ghosh}}, \bibinfo {author} {\bibfnamefont {B.}~\bibnamefont
  {Mukhopadhyaya}}, \ and\ \bibinfo {author} {\bibfnamefont {I.}~\bibnamefont
  {Saha}},\ }\href@noop {} {\  (\bibinfo {year} {2015})},\ \Eprint
  {http://arxiv.org/abs/1501.03700} {arXiv:1501.03700 [hep-ph]} \BibitemShut
  {NoStop}%
\bibitem [{\citenamefont {Das}\ and\ \citenamefont {Saha}(2015)}]{Das:2015mwa}%
  \BibitemOpen
  \bibfield  {author} {\bibinfo {author} {\bibfnamefont {D.}~\bibnamefont
  {Das}}\ and\ \bibinfo {author} {\bibfnamefont {I.}~\bibnamefont {Saha}},\
  }\href@noop {} {\  (\bibinfo {year} {2015})},\ \Eprint
  {http://arxiv.org/abs/1503.02135} {arXiv:1503.02135 [hep-ph]} \BibitemShut
  {NoStop}%
\bibitem [{\citenamefont {Veltman}(1981)}]{Veltman:1980mj}%
  \BibitemOpen
  \bibfield  {author} {\bibinfo {author} {\bibfnamefont {M.}~\bibnamefont
  {Veltman}},\ }\href@noop {} {\bibfield  {journal} {\bibinfo  {journal} {Acta
  Phys.Polon.}\ }\textbf {\bibinfo {volume} {B12}},\ \bibinfo {pages} {437}
  (\bibinfo {year} {1981})}\BibitemShut {NoStop}%
\bibitem [{\citenamefont {Newton}\ and\ \citenamefont
  {Wu}(1994)}]{Newton:1993xc}%
  \BibitemOpen
  \bibfield  {author} {\bibinfo {author} {\bibfnamefont {C.}~\bibnamefont
  {Newton}}\ and\ \bibinfo {author} {\bibfnamefont {T.~T.}\ \bibnamefont
  {Wu}},\ }\href {\doibase 10.1007/BF01560241} {\bibfield  {journal} {\bibinfo
  {journal} {Z.Phys.}\ }\textbf {\bibinfo {volume} {C62}},\ \bibinfo {pages}
  {253} (\bibinfo {year} {1994})}\BibitemShut {NoStop}%
\bibitem [{\citenamefont {Ma}(2001)}]{Ma:2001sj}%
  \BibitemOpen
  \bibfield  {author} {\bibinfo {author} {\bibfnamefont {E.}~\bibnamefont
  {Ma}},\ }\href {\doibase 10.1142/S0217751X01004414} {\bibfield  {journal}
  {\bibinfo  {journal} {Int.J.Mod.Phys.}\ }\textbf {\bibinfo {volume} {A16}},\
  \bibinfo {pages} {3099} (\bibinfo {year} {2001})},\ \Eprint
  {http://arxiv.org/abs/hep-ph/0101355} {arXiv:hep-ph/0101355 [hep-ph]}
  \BibitemShut {NoStop}%
\bibitem [{\citenamefont {Jora}\ \emph {et~al.}(2013)\citenamefont {Jora},
  \citenamefont {Nasri},\ and\ \citenamefont {Schechter}}]{Jora:2013opa}%
  \BibitemOpen
  \bibfield  {author} {\bibinfo {author} {\bibfnamefont {R.}~\bibnamefont
  {Jora}}, \bibinfo {author} {\bibfnamefont {S.}~\bibnamefont {Nasri}}, \ and\
  \bibinfo {author} {\bibfnamefont {J.}~\bibnamefont {Schechter}},\ }\href
  {\doibase 10.1142/S0217751X1350036X} {\bibfield  {journal} {\bibinfo
  {journal} {Int.J.Mod.Phys.}\ }\textbf {\bibinfo {volume} {A28}},\ \bibinfo
  {pages} {1350036} (\bibinfo {year} {2013})},\ \Eprint
  {http://arxiv.org/abs/1302.6344} {arXiv:1302.6344 [hep-ph]} \BibitemShut
  {NoStop}%
\bibitem [{\citenamefont {Biswas}\ and\ \citenamefont
  {Lahiri}(2014)}]{Biswas:2014uba}%
  \BibitemOpen
  \bibfield  {author} {\bibinfo {author} {\bibfnamefont {A.}~\bibnamefont
  {Biswas}}\ and\ \bibinfo {author} {\bibfnamefont {A.}~\bibnamefont
  {Lahiri}},\ }\href@noop {} {\  (\bibinfo {year} {2014})},\ \Eprint
  {http://arxiv.org/abs/1412.6187} {arXiv:1412.6187 [hep-ph]} \BibitemShut
  {NoStop}%
\bibitem [{\citenamefont {Grzadkowski}\ and\ \citenamefont
  {Osland}(2010{\natexlab{a}})}]{Grzadkowski:2009iz}%
  \BibitemOpen
  \bibfield  {author} {\bibinfo {author} {\bibfnamefont {B.}~\bibnamefont
  {Grzadkowski}}\ and\ \bibinfo {author} {\bibfnamefont {P.}~\bibnamefont
  {Osland}},\ }\href {\doibase 10.1103/PhysRevD.82.125026} {\bibfield
  {journal} {\bibinfo  {journal} {Phys.Rev.}\ }\textbf {\bibinfo {volume}
  {D82}},\ \bibinfo {pages} {125026} (\bibinfo {year} {2010}{\natexlab{a}})},\
  \Eprint {http://arxiv.org/abs/0910.4068} {arXiv:0910.4068 [hep-ph]}
  \BibitemShut {NoStop}%
\bibitem [{\citenamefont {Grzadkowski}\ and\ \citenamefont
  {Osland}(2011)}]{Grzadkowski:2010dn}%
  \BibitemOpen
  \bibfield  {author} {\bibinfo {author} {\bibfnamefont {B.}~\bibnamefont
  {Grzadkowski}}\ and\ \bibinfo {author} {\bibfnamefont {P.}~\bibnamefont
  {Osland}},\ }\href {\doibase 10.1002/prop.201000098} {\bibfield  {journal}
  {\bibinfo  {journal} {Fortsch.Phys.}\ }\textbf {\bibinfo {volume} {59}},\
  \bibinfo {pages} {1041} (\bibinfo {year} {2011})},\ \Eprint
  {http://arxiv.org/abs/1012.0703} {arXiv:1012.0703 [hep-ph]} \BibitemShut
  {NoStop}%
\bibitem [{\citenamefont {Grzadkowski}\ and\ \citenamefont
  {Osland}(2010{\natexlab{b}})}]{Grzadkowski:2010se}%
  \BibitemOpen
  \bibfield  {author} {\bibinfo {author} {\bibfnamefont {B.}~\bibnamefont
  {Grzadkowski}}\ and\ \bibinfo {author} {\bibfnamefont {P.}~\bibnamefont
  {Osland}},\ }\href {\doibase 10.1088/1742-6596/259/1/012055} {\bibfield
  {journal} {\bibinfo  {journal} {J.Phys.Conf.Ser.}\ }\textbf {\bibinfo
  {volume} {259}},\ \bibinfo {pages} {012055} (\bibinfo {year}
  {2010}{\natexlab{b}})},\ \Eprint {http://arxiv.org/abs/1012.2201}
  {arXiv:1012.2201 [hep-ph]} \BibitemShut {NoStop}%
\bibitem [{\citenamefont {Chakraborty}\ and\ \citenamefont
  {Kundu}(2014)}]{Chakraborty:2014oma}%
  \BibitemOpen
  \bibfield  {author} {\bibinfo {author} {\bibfnamefont {I.}~\bibnamefont
  {Chakraborty}}\ and\ \bibinfo {author} {\bibfnamefont {A.}~\bibnamefont
  {Kundu}},\ }\href@noop {} {\  (\bibinfo {year} {2014})},\ \Eprint
  {http://arxiv.org/abs/1404.3038} {arXiv:1404.3038 [hep-ph]} \BibitemShut
  {NoStop}%
\bibitem [{\citenamefont {Delgado}\ \emph {et~al.}(2013)\citenamefont
  {Delgado}, \citenamefont {Nardini},\ and\ \citenamefont
  {Quiros}}]{Delgado:2013zfa}%
  \BibitemOpen
  \bibfield  {author} {\bibinfo {author} {\bibfnamefont {A.}~\bibnamefont
  {Delgado}}, \bibinfo {author} {\bibfnamefont {G.}~\bibnamefont {Nardini}}, \
  and\ \bibinfo {author} {\bibfnamefont {M.}~\bibnamefont {Quiros}},\ }\href
  {\doibase 10.1007/JHEP07(2013)054} {\bibfield  {journal} {\bibinfo  {journal}
  {JHEP}\ }\textbf {\bibinfo {volume} {1307}},\ \bibinfo {pages} {054}
  (\bibinfo {year} {2013})},\ \Eprint {http://arxiv.org/abs/1303.0800}
  {arXiv:1303.0800 [hep-ph]} \BibitemShut {NoStop}%
\bibitem [{\citenamefont {Carena}\ \emph {et~al.}(2014)\citenamefont {Carena},
  \citenamefont {Low}, \citenamefont {Shah},\ and\ \citenamefont
  {Wagner}}]{Carena:2013ooa}%
  \BibitemOpen
  \bibfield  {author} {\bibinfo {author} {\bibfnamefont {M.}~\bibnamefont
  {Carena}}, \bibinfo {author} {\bibfnamefont {I.}~\bibnamefont {Low}},
  \bibinfo {author} {\bibfnamefont {N.~R.}\ \bibnamefont {Shah}}, \ and\
  \bibinfo {author} {\bibfnamefont {C.~E.}\ \bibnamefont {Wagner}},\ }\href
  {\doibase 10.1007/JHEP04(2014)015} {\bibfield  {journal} {\bibinfo  {journal}
  {JHEP}\ }\textbf {\bibinfo {volume} {1404}},\ \bibinfo {pages} {015}
  (\bibinfo {year} {2014})},\ \Eprint {http://arxiv.org/abs/1310.2248}
  {arXiv:1310.2248 [hep-ph]} \BibitemShut {NoStop}%
\bibitem [{\citenamefont {Grzadkowski}\ \emph {et~al.}(2014)\citenamefont
  {Grzadkowski}, \citenamefont {Ogreid},\ and\ \citenamefont
  {Osland}}]{Grzadkowski:2014ada}%
  \BibitemOpen
  \bibfield  {author} {\bibinfo {author} {\bibfnamefont {B.}~\bibnamefont
  {Grzadkowski}}, \bibinfo {author} {\bibfnamefont {O.}~\bibnamefont {Ogreid}},
  \ and\ \bibinfo {author} {\bibfnamefont {P.}~\bibnamefont {Osland}},\
  }\href@noop {} {\  (\bibinfo {year} {2014})},\ \Eprint
  {http://arxiv.org/abs/1409.7265} {arXiv:1409.7265 [hep-ph]} \BibitemShut
  {NoStop}%
\bibitem [{\citenamefont {Deshpande}\ and\ \citenamefont
  {Ma}(1978)}]{Deshpande:1977rw}%
  \BibitemOpen
  \bibfield  {author} {\bibinfo {author} {\bibfnamefont {N.~G.}\ \bibnamefont
  {Deshpande}}\ and\ \bibinfo {author} {\bibfnamefont {E.}~\bibnamefont {Ma}},\
  }\href {\doibase 10.1103/PhysRevD.18.2574} {\bibfield  {journal} {\bibinfo
  {journal} {Phys.Rev.}\ }\textbf {\bibinfo {volume} {D18}},\ \bibinfo {pages}
  {2574} (\bibinfo {year} {1978})}\BibitemShut {NoStop}%
\bibitem [{\citenamefont {Ginzburg}\ and\ \citenamefont
  {Ivanov}(2005)}]{Ginzburg:2005dt}%
  \BibitemOpen
  \bibfield  {author} {\bibinfo {author} {\bibfnamefont {I.}~\bibnamefont
  {Ginzburg}}\ and\ \bibinfo {author} {\bibfnamefont {I.}~\bibnamefont
  {Ivanov}},\ }\href {\doibase 10.1103/PhysRevD.72.115010} {\bibfield
  {journal} {\bibinfo  {journal} {Phys.Rev.}\ }\textbf {\bibinfo {volume}
  {D72}},\ \bibinfo {pages} {115010} (\bibinfo {year} {2005})}\BibitemShut
  {NoStop}%
\bibitem [{\citenamefont {Barroso}\ \emph
  {et~al.}(2013{\natexlab{b}})\citenamefont {Barroso}, \citenamefont
  {Ferreira}, \citenamefont {Ivanov},\ and\ \citenamefont
  {Santos}}]{Barroso:2013awa}%
  \BibitemOpen
  \bibfield  {author} {\bibinfo {author} {\bibfnamefont {A.}~\bibnamefont
  {Barroso}}, \bibinfo {author} {\bibfnamefont {P.}~\bibnamefont {Ferreira}},
  \bibinfo {author} {\bibfnamefont {I.}~\bibnamefont {Ivanov}}, \ and\ \bibinfo
  {author} {\bibfnamefont {R.}~\bibnamefont {Santos}},\ }\href@noop {} {\
  (\bibinfo {year} {2013}{\natexlab{b}})},\ \Eprint
  {http://arxiv.org/abs/1303.5098} {arXiv:1303.5098 [hep-ph]} \BibitemShut
  {NoStop}%
\bibitem [{\citenamefont {Misiak}\ \emph {et~al.}(2015)\citenamefont {Misiak},
  \citenamefont {Asatrian}, \citenamefont {Boughezal}, \citenamefont {Czakon},
  \citenamefont {Ewerth} \emph {et~al.}}]{Misiak:2015xwa}%
  \BibitemOpen
  \bibfield  {author} {\bibinfo {author} {\bibfnamefont {M.}~\bibnamefont
  {Misiak}}, \bibinfo {author} {\bibfnamefont {H.}~\bibnamefont {Asatrian}},
  \bibinfo {author} {\bibfnamefont {R.}~\bibnamefont {Boughezal}}, \bibinfo
  {author} {\bibfnamefont {M.}~\bibnamefont {Czakon}}, \bibinfo {author}
  {\bibfnamefont {T.}~\bibnamefont {Ewerth}},  \emph {et~al.},\ }\href@noop {}
  {\  (\bibinfo {year} {2015})},\ \Eprint {http://arxiv.org/abs/1503.01789}
  {arXiv:1503.01789 [hep-ph]} \BibitemShut {NoStop}%
\bibitem [{ATL(2015)}]{ATLAS-CONF-2015-007}%
  \BibitemOpen
  \href@noop {} {}\bibinfo {howpublished} {{ATLAS conference note,
  ATLAS-CONF-2015-007}} (\bibinfo {year} {2015})\BibitemShut {NoStop}%
\bibitem [{\citenamefont {Aad}\ \emph {et~al.}(2015{\natexlab{b}})\citenamefont
  {Aad} \emph {et~al.}}]{Aad:2015gra}%
  \BibitemOpen
  \bibfield  {author} {\bibinfo {author} {\bibfnamefont {G.}~\bibnamefont
  {Aad}} \emph {et~al.} (\bibinfo {collaboration} {ATLAS}),\ }\href@noop {} {\
  (\bibinfo {year} {2015}{\natexlab{b}})},\ \Eprint
  {http://arxiv.org/abs/1503.05066} {arXiv:1503.05066 [hep-ex]} \BibitemShut
  {NoStop}%
\bibitem [{\citenamefont {Khachatryan}\ \emph
  {et~al.}(2014{\natexlab{a}})\citenamefont {Khachatryan} \emph
  {et~al.}}]{Khachatryan:2014jba}%
  \BibitemOpen
  \bibfield  {author} {\bibinfo {author} {\bibfnamefont {V.}~\bibnamefont
  {Khachatryan}} \emph {et~al.} (\bibinfo {collaboration} {CMS}),\ }\href@noop
  {} {\  (\bibinfo {year} {2014}{\natexlab{a}})},\ \Eprint
  {http://arxiv.org/abs/1412.8662} {arXiv:1412.8662 [hep-ex]} \BibitemShut
  {NoStop}%
\bibitem [{\citenamefont {Khachatryan}\ \emph
  {et~al.}(2015{\natexlab{a}})\citenamefont {Khachatryan} \emph
  {et~al.}}]{Khachatryan:2015ila}%
  \BibitemOpen
  \bibfield  {author} {\bibinfo {author} {\bibfnamefont {V.}~\bibnamefont
  {Khachatryan}} \emph {et~al.} (\bibinfo {collaboration} {CMS}),\ }\href@noop
  {} {\  (\bibinfo {year} {2015}{\natexlab{a}})},\ \Eprint
  {http://arxiv.org/abs/1502.02485} {arXiv:1502.02485 [hep-ex]} \BibitemShut
  {NoStop}%
\bibitem [{\citenamefont {Aad}\ \emph {et~al.}(2014{\natexlab{a}})\citenamefont
  {Aad} \emph {et~al.}}]{Aad:2014yja}%
  \BibitemOpen
  \bibfield  {author} {\bibinfo {author} {\bibfnamefont {G.}~\bibnamefont
  {Aad}} \emph {et~al.} (\bibinfo {collaboration} {ATLAS Collaboration}),\
  }\href@noop {} {\  (\bibinfo {year} {2014}{\natexlab{a}})},\ \Eprint
  {http://arxiv.org/abs/1406.5053} {arXiv:1406.5053 [hep-ex]} \BibitemShut
  {NoStop}%
\bibitem [{ATL(2013)}]{ATLAS-CONF-2013-067}%
  \BibitemOpen
  \href@noop {} {}\bibinfo {howpublished} {{ATLAS conference note,
  ATLAS-CONF-2013-067}} (\bibinfo {year} {2013})\BibitemShut {NoStop}%
\bibitem [{\citenamefont {Aad}\ \emph {et~al.}(2014{\natexlab{b}})\citenamefont
  {Aad} \emph {et~al.}}]{Aad:2014ioa}%
  \BibitemOpen
  \bibfield  {author} {\bibinfo {author} {\bibfnamefont {G.}~\bibnamefont
  {Aad}} \emph {et~al.} (\bibinfo {collaboration} {ATLAS Collaboration}),\
  }\href {\doibase 10.1103/PhysRevLett.113.171801} {\bibfield  {journal}
  {\bibinfo  {journal} {Phys.Rev.Lett.}\ }\textbf {\bibinfo {volume} {113}},\
  \bibinfo {pages} {171801} (\bibinfo {year} {2014}{\natexlab{b}})},\ \Eprint
  {http://arxiv.org/abs/1407.6583} {arXiv:1407.6583 [hep-ex]} \BibitemShut
  {NoStop}%
\bibitem [{ATL(2014)}]{ATLAS-CONF-2014-049}%
  \BibitemOpen
  \href@noop {} {}\bibinfo {howpublished} {{ATLAS conference note,
  ATLAS-CONF-2014-049}} (\bibinfo {year} {2014})\BibitemShut {NoStop}%
\bibitem [{\citenamefont {Aad}\ \emph {et~al.}(2015{\natexlab{c}})\citenamefont
  {Aad} \emph {et~al.}}]{Aad:2015wra}%
  \BibitemOpen
  \bibfield  {author} {\bibinfo {author} {\bibfnamefont {G.}~\bibnamefont
  {Aad}} \emph {et~al.} (\bibinfo {collaboration} {ATLAS Collaboration}),\
  }\href@noop {} {\  (\bibinfo {year} {2015}{\natexlab{c}})},\ \Eprint
  {http://arxiv.org/abs/1502.04478} {arXiv:1502.04478 [hep-ex]} \BibitemShut
  {NoStop}%
\bibitem [{\citenamefont {Khachatryan}\ \emph
  {et~al.}(2014{\natexlab{b}})\citenamefont {Khachatryan} \emph
  {et~al.}}]{Khachatryan:2014wca}%
  \BibitemOpen
  \bibfield  {author} {\bibinfo {author} {\bibfnamefont {V.}~\bibnamefont
  {Khachatryan}} \emph {et~al.} (\bibinfo {collaboration} {CMS
  Collaboration}),\ }\href {\doibase 10.1007/JHEP10(2014)160} {\bibfield
  {journal} {\bibinfo  {journal} {JHEP}\ }\textbf {\bibinfo {volume} {1410}},\
  \bibinfo {pages} {160} (\bibinfo {year} {2014}{\natexlab{b}})},\ \Eprint
  {http://arxiv.org/abs/1408.3316} {arXiv:1408.3316 [hep-ex]} \BibitemShut
  {NoStop}%
\bibitem [{\citenamefont {Khachatryan}\ \emph
  {et~al.}(2015{\natexlab{b}})\citenamefont {Khachatryan} \emph
  {et~al.}}]{Khachatryan:2015cwa}%
  \BibitemOpen
  \bibfield  {author} {\bibinfo {author} {\bibfnamefont {V.}~\bibnamefont
  {Khachatryan}} \emph {et~al.} (\bibinfo {collaboration} {CMS}),\ }\href@noop
  {} {\  (\bibinfo {year} {2015}{\natexlab{b}})},\ \Eprint
  {http://arxiv.org/abs/1504.00936} {arXiv:1504.00936 [hep-ex]} \BibitemShut
  {NoStop}%
\bibitem [{\citenamefont {Khachatryan}\ \emph
  {et~al.}(2015{\natexlab{c}})\citenamefont {Khachatryan} \emph
  {et~al.}}]{Khachatryan:2015qba}%
  \BibitemOpen
  \bibfield  {author} {\bibinfo {author} {\bibfnamefont {V.}~\bibnamefont
  {Khachatryan}} \emph {et~al.} (\bibinfo {collaboration} {CMS}),\ }\href@noop
  {} {\  (\bibinfo {year} {2015}{\natexlab{c}})},\ \Eprint
  {http://arxiv.org/abs/1506.02301} {arXiv:1506.02301 [hep-ex]} \BibitemShut
  {NoStop}%
\bibitem [{CMS(2013{\natexlab{a}})}]{CMS-PAS-HIG-13-032}%
  \BibitemOpen
  \href@noop {} {}\bibinfo {howpublished} {{CMS physics analysis summary,
  CMS-PAS-HIG-13-032}} (\bibinfo {year} {2013}{\natexlab{a}})\BibitemShut
  {NoStop}%
\bibitem [{CMS(2014{\natexlab{a}})}]{CMS-PAS-HIG-14-011}%
  \BibitemOpen
  \href@noop {} {}\bibinfo {howpublished} {{CMS physics analysis summary,
  CMS-PAS-HIG-14-011}} (\bibinfo {year} {2014}{\natexlab{a}})\BibitemShut
  {NoStop}%
\bibitem [{\citenamefont {Khachatryan}\ \emph
  {et~al.}(2015{\natexlab{d}})\citenamefont {Khachatryan} \emph
  {et~al.}}]{Khachatryan:2015yea}%
  \BibitemOpen
  \bibfield  {author} {\bibinfo {author} {\bibfnamefont {V.}~\bibnamefont
  {Khachatryan}} \emph {et~al.} (\bibinfo {collaboration} {CMS
  Collaboration}),\ }\href@noop {} {\  (\bibinfo {year}
  {2015}{\natexlab{d}})},\ \Eprint {http://arxiv.org/abs/1503.04114}
  {arXiv:1503.04114 [hep-ex]} \BibitemShut {NoStop}%
\bibitem [{\citenamefont {Aad}\ \emph {et~al.}(2015{\natexlab{d}})\citenamefont
  {Aad} \emph {et~al.}}]{Aad:2014kga}%
  \BibitemOpen
  \bibfield  {author} {\bibinfo {author} {\bibfnamefont {G.}~\bibnamefont
  {Aad}} \emph {et~al.} (\bibinfo {collaboration} {ATLAS}),\ }\href {\doibase
  10.1007/JHEP03(2015)088} {\bibfield  {journal} {\bibinfo  {journal} {JHEP}\
  }\textbf {\bibinfo {volume} {1503}},\ \bibinfo {pages} {088} (\bibinfo {year}
  {2015}{\natexlab{d}})},\ \Eprint {http://arxiv.org/abs/1412.6663}
  {arXiv:1412.6663 [hep-ex]} \BibitemShut {NoStop}%
\bibitem [{CMS(2013{\natexlab{b}})}]{CMS-PAS-HIG-13-026}%
  \BibitemOpen
  \href@noop {} {}\bibinfo {howpublished} {{CMS physics analysis summary,
  CMS-PAS-HIG-13-026}} (\bibinfo {year} {2013}{\natexlab{b}})\BibitemShut
  {NoStop}%
\bibitem [{CMS(2014{\natexlab{b}})}]{CMS-PAS-HIG-14-020}%
  \BibitemOpen
  \href@noop {} {}\bibinfo {howpublished} {{CMS physics analysis summary,
  CMS-PAS-HIG-14-020}} (\bibinfo {year} {2014}{\natexlab{b}})\BibitemShut
  {NoStop}%
\bibitem [{\citenamefont {Mahmoudi}\ and\ \citenamefont
  {Stal}(2010)}]{Mahmoudi:2009zx}%
  \BibitemOpen
  \bibfield  {author} {\bibinfo {author} {\bibfnamefont {F.}~\bibnamefont
  {Mahmoudi}}\ and\ \bibinfo {author} {\bibfnamefont {O.}~\bibnamefont
  {Stal}},\ }\href {\doibase 10.1103/PhysRevD.81.035016} {\bibfield  {journal}
  {\bibinfo  {journal} {Phys.Rev.}\ }\textbf {\bibinfo {volume} {D81}},\
  \bibinfo {pages} {035016} (\bibinfo {year} {2010})},\ \Eprint
  {http://arxiv.org/abs/0907.1791} {arXiv:0907.1791 [hep-ph]} \BibitemShut
  {NoStop}%
\bibitem [{\citenamefont {Crivellin}\ \emph {et~al.}(2012)\citenamefont
  {Crivellin}, \citenamefont {Greub},\ and\ \citenamefont
  {Kokulu}}]{Crivellin:2012ye}%
  \BibitemOpen
  \bibfield  {author} {\bibinfo {author} {\bibfnamefont {A.}~\bibnamefont
  {Crivellin}}, \bibinfo {author} {\bibfnamefont {C.}~\bibnamefont {Greub}}, \
  and\ \bibinfo {author} {\bibfnamefont {A.}~\bibnamefont {Kokulu}},\ }\href
  {\doibase 10.1103/PhysRevD.86.054014} {\bibfield  {journal} {\bibinfo
  {journal} {Phys.Rev.}\ }\textbf {\bibinfo {volume} {D86}},\ \bibinfo {pages}
  {054014} (\bibinfo {year} {2012})},\ \Eprint {http://arxiv.org/abs/1206.2634}
  {arXiv:1206.2634 [hep-ph]} \BibitemShut {NoStop}%
\bibitem [{\citenamefont {Eberhardt}\ \emph {et~al.}(2012)\citenamefont
  {Eberhardt}, \citenamefont {Herbert}, \citenamefont {Lacker}, \citenamefont
  {Lenz}, \citenamefont {Menzel} \emph {et~al.}}]{Eberhardt:2012gv}%
  \BibitemOpen
  \bibfield  {author} {\bibinfo {author} {\bibfnamefont {O.}~\bibnamefont
  {Eberhardt}}, \bibinfo {author} {\bibfnamefont {G.}~\bibnamefont {Herbert}},
  \bibinfo {author} {\bibfnamefont {H.}~\bibnamefont {Lacker}}, \bibinfo
  {author} {\bibfnamefont {A.}~\bibnamefont {Lenz}}, \bibinfo {author}
  {\bibfnamefont {A.}~\bibnamefont {Menzel}},  \emph {et~al.},\ }\href
  {\doibase 10.1103/PhysRevLett.109.241802} {\bibfield  {journal} {\bibinfo
  {journal} {Phys.Rev.Lett.}\ }\textbf {\bibinfo {volume} {109}},\ \bibinfo
  {pages} {241802} (\bibinfo {year} {2012})},\ \Eprint
  {http://arxiv.org/abs/1209.1101} {arXiv:1209.1101 [hep-ph]} \BibitemShut
  {NoStop}%
\bibitem [{\citenamefont {Lyonnet}\ \emph {et~al.}(2014)\citenamefont
  {Lyonnet}, \citenamefont {Schienbein}, \citenamefont {Staub},\ and\
  \citenamefont {Wingerter}}]{Lyonnet:2013dna}%
  \BibitemOpen
  \bibfield  {author} {\bibinfo {author} {\bibfnamefont {F.}~\bibnamefont
  {Lyonnet}}, \bibinfo {author} {\bibfnamefont {I.}~\bibnamefont {Schienbein}},
  \bibinfo {author} {\bibfnamefont {F.}~\bibnamefont {Staub}}, \ and\ \bibinfo
  {author} {\bibfnamefont {A.}~\bibnamefont {Wingerter}},\ }\href {\doibase
  10.1016/j.cpc.2013.12.002} {\bibfield  {journal} {\bibinfo  {journal}
  {Comput.Phys.Commun.}\ }\textbf {\bibinfo {volume} {185}},\ \bibinfo {pages}
  {1130} (\bibinfo {year} {2014})},\ \Eprint {http://arxiv.org/abs/1309.7030}
  {arXiv:1309.7030 [hep-ph]} \BibitemShut {NoStop}%
\bibitem [{\citenamefont {Bardin}\ \emph {et~al.}(1990)\citenamefont {Bardin},
  \citenamefont {Bilenky}, \citenamefont {Riemann}, \citenamefont {Sachwitz},\
  and\ \citenamefont {Vogt}}]{Bardin:1989tq}%
  \BibitemOpen
  \bibfield  {author} {\bibinfo {author} {\bibfnamefont {D.~Y.}\ \bibnamefont
  {Bardin}}, \bibinfo {author} {\bibfnamefont {M.~S.}\ \bibnamefont {Bilenky}},
  \bibinfo {author} {\bibfnamefont {T.}~\bibnamefont {Riemann}}, \bibinfo
  {author} {\bibfnamefont {M.}~\bibnamefont {Sachwitz}}, \ and\ \bibinfo
  {author} {\bibfnamefont {H.}~\bibnamefont {Vogt}},\ }\href {\doibase
  10.1016/0010-4655(90)90179-5} {\bibfield  {journal} {\bibinfo  {journal}
  {Comput.Phys.Commun.}\ }\textbf {\bibinfo {volume} {59}},\ \bibinfo {pages}
  {303} (\bibinfo {year} {1990})}\BibitemShut {NoStop}%
\bibitem [{\citenamefont {Bardin}\ \emph {et~al.}(2001)\citenamefont {Bardin},
  \citenamefont {Christova}, \citenamefont {Jack}, \citenamefont
  {Kalinovskaya}, \citenamefont {Olchevski} \emph {et~al.}}]{Bardin:1999yd}%
  \BibitemOpen
  \bibfield  {author} {\bibinfo {author} {\bibfnamefont {D.~Y.}\ \bibnamefont
  {Bardin}}, \bibinfo {author} {\bibfnamefont {P.}~\bibnamefont {Christova}},
  \bibinfo {author} {\bibfnamefont {M.}~\bibnamefont {Jack}}, \bibinfo {author}
  {\bibfnamefont {L.}~\bibnamefont {Kalinovskaya}}, \bibinfo {author}
  {\bibfnamefont {A.}~\bibnamefont {Olchevski}},  \emph {et~al.},\ }\href
  {\doibase 10.1016/S0010-4655(00)00152-1} {\bibfield  {journal} {\bibinfo
  {journal} {Comput.Phys.Commun.}\ }\textbf {\bibinfo {volume} {133}},\
  \bibinfo {pages} {229} (\bibinfo {year} {2001})},\ \Eprint
  {http://arxiv.org/abs/hep-ph/9908433} {arXiv:hep-ph/9908433 [hep-ph]}
  \BibitemShut {NoStop}%
\bibitem [{\citenamefont {Arbuzov}\ \emph {et~al.}(2006)\citenamefont
  {Arbuzov}, \citenamefont {Awramik}, \citenamefont {Czakon}, \citenamefont
  {Freitas}, \citenamefont {Grunewald} \emph {et~al.}}]{Arbuzov:2005ma}%
  \BibitemOpen
  \bibfield  {author} {\bibinfo {author} {\bibfnamefont {A.}~\bibnamefont
  {Arbuzov}}, \bibinfo {author} {\bibfnamefont {M.}~\bibnamefont {Awramik}},
  \bibinfo {author} {\bibfnamefont {M.}~\bibnamefont {Czakon}}, \bibinfo
  {author} {\bibfnamefont {A.}~\bibnamefont {Freitas}}, \bibinfo {author}
  {\bibfnamefont {M.}~\bibnamefont {Grunewald}},  \emph {et~al.},\ }\href
  {\doibase 10.1016/j.cpc.2005.12.009} {\bibfield  {journal} {\bibinfo
  {journal} {Comput.Phys.Commun.}\ }\textbf {\bibinfo {volume} {174}},\
  \bibinfo {pages} {728} (\bibinfo {year} {2006})},\ \Eprint
  {http://arxiv.org/abs/hep-ph/0507146} {arXiv:hep-ph/0507146 [hep-ph]}
  \BibitemShut {NoStop}%
\bibitem [{\citenamefont {Hahn}(2001)}]{Hahn:2000kx}%
  \BibitemOpen
  \bibfield  {author} {\bibinfo {author} {\bibfnamefont {T.}~\bibnamefont
  {Hahn}},\ }\href {\doibase 10.1016/S0010-4655(01)00290-9} {\bibfield
  {journal} {\bibinfo  {journal} {Comput.Phys.Commun.}\ }\textbf {\bibinfo
  {volume} {140}},\ \bibinfo {pages} {418} (\bibinfo {year} {2001})},\ \Eprint
  {http://arxiv.org/abs/hep-ph/0012260} {arXiv:hep-ph/0012260 [hep-ph]}
  \BibitemShut {NoStop}%
\bibitem [{\citenamefont {Hahn}\ and\ \citenamefont
  {Perez-Victoria}(1999)}]{Hahn:1998yk}%
  \BibitemOpen
  \bibfield  {author} {\bibinfo {author} {\bibfnamefont {T.}~\bibnamefont
  {Hahn}}\ and\ \bibinfo {author} {\bibfnamefont {M.}~\bibnamefont
  {Perez-Victoria}},\ }\href {\doibase 10.1016/S0010-4655(98)00173-8}
  {\bibfield  {journal} {\bibinfo  {journal} {Comput. Phys. Commun.}\ }\textbf
  {\bibinfo {volume} {118}},\ \bibinfo {pages} {153} (\bibinfo {year}
  {1999})},\ \Eprint {http://arxiv.org/abs/hep-ph/9807565}
  {arXiv:hep-ph/9807565} \BibitemShut {NoStop}%
\bibitem [{\citenamefont {Hahn}\ and\ \citenamefont
  {Rauch}(2006)}]{Hahn:2006qw}%
  \BibitemOpen
  \bibfield  {author} {\bibinfo {author} {\bibfnamefont {T.}~\bibnamefont
  {Hahn}}\ and\ \bibinfo {author} {\bibfnamefont {M.}~\bibnamefont {Rauch}},\
  }\href {\doibase 10.1016/j.nuclphysbps.2006.03.026} {\bibfield  {journal}
  {\bibinfo  {journal} {Nucl. Phys. Proc. Suppl.}\ }\textbf {\bibinfo {volume}
  {157}},\ \bibinfo {pages} {236} (\bibinfo {year} {2006})},\ \Eprint
  {http://arxiv.org/abs/hep-ph/0601248} {arXiv:hep-ph/0601248} \BibitemShut
  {NoStop}%
\bibitem [{\citenamefont {Djouadi}\ \emph {et~al.}(1998)\citenamefont
  {Djouadi}, \citenamefont {Kalinowski},\ and\ \citenamefont
  {Spira}}]{Djouadi:1997yw}%
  \BibitemOpen
  \bibfield  {author} {\bibinfo {author} {\bibfnamefont {A.}~\bibnamefont
  {Djouadi}}, \bibinfo {author} {\bibfnamefont {J.}~\bibnamefont {Kalinowski}},
  \ and\ \bibinfo {author} {\bibfnamefont {M.}~\bibnamefont {Spira}},\ }\href
  {\doibase 10.1016/S0010-4655(97)00123-9} {\bibfield  {journal} {\bibinfo
  {journal} {Comput.Phys.Commun.}\ }\textbf {\bibinfo {volume} {108}},\
  \bibinfo {pages} {56} (\bibinfo {year} {1998})},\ \Eprint
  {http://arxiv.org/abs/hep-ph/9704448} {arXiv:hep-ph/9704448 [hep-ph]}
  \BibitemShut {NoStop}%
\bibitem [{\citenamefont {Butterworth}\ \emph {et~al.}(2010)\citenamefont
  {Butterworth}, \citenamefont {Maltoni}, \citenamefont {Moortgat},
  \citenamefont {Richardson}, \citenamefont {Schumann} \emph
  {et~al.}}]{Butterworth:2010ym}%
  \BibitemOpen
  \bibfield  {author} {\bibinfo {author} {\bibfnamefont {J.}~\bibnamefont
  {Butterworth}}, \bibinfo {author} {\bibfnamefont {F.}~\bibnamefont
  {Maltoni}}, \bibinfo {author} {\bibfnamefont {F.}~\bibnamefont {Moortgat}},
  \bibinfo {author} {\bibfnamefont {P.}~\bibnamefont {Richardson}}, \bibinfo
  {author} {\bibfnamefont {S.}~\bibnamefont {Schumann}},  \emph {et~al.},\
  }\href@noop {} {\  (\bibinfo {year} {2010})},\ \Eprint
  {http://arxiv.org/abs/1003.1643} {arXiv:1003.1643 [hep-ph]} \BibitemShut
  {NoStop}%
\bibitem [{\citenamefont {Djouadi}\ \emph {et~al.}(2007)\citenamefont
  {Djouadi}, \citenamefont {Muhlleitner},\ and\ \citenamefont
  {Spira}}]{Djouadi:2006bz}%
  \BibitemOpen
  \bibfield  {author} {\bibinfo {author} {\bibfnamefont {A.}~\bibnamefont
  {Djouadi}}, \bibinfo {author} {\bibfnamefont {M.}~\bibnamefont
  {Muhlleitner}}, \ and\ \bibinfo {author} {\bibfnamefont {M.}~\bibnamefont
  {Spira}},\ }\href@noop {} {\bibfield  {journal} {\bibinfo  {journal} {Acta
  Phys.Polon.}\ }\textbf {\bibinfo {volume} {B38}},\ \bibinfo {pages} {635}
  (\bibinfo {year} {2007})},\ \Eprint {http://arxiv.org/abs/hep-ph/0609292}
  {arXiv:hep-ph/0609292 [hep-ph]} \BibitemShut {NoStop}%
\bibitem [{\citenamefont {Alloul}\ \emph {et~al.}(2014)\citenamefont {Alloul},
  \citenamefont {Christensen}, \citenamefont {Degrande}, \citenamefont {Duhr},\
  and\ \citenamefont {Fuks}}]{Alloul:2013bka}%
  \BibitemOpen
  \bibfield  {author} {\bibinfo {author} {\bibfnamefont {A.}~\bibnamefont
  {Alloul}}, \bibinfo {author} {\bibfnamefont {N.~D.}\ \bibnamefont
  {Christensen}}, \bibinfo {author} {\bibfnamefont {C.}~\bibnamefont
  {Degrande}}, \bibinfo {author} {\bibfnamefont {C.}~\bibnamefont {Duhr}}, \
  and\ \bibinfo {author} {\bibfnamefont {B.}~\bibnamefont {Fuks}},\ }\href
  {\doibase 10.1016/j.cpc.2014.04.012} {\bibfield  {journal} {\bibinfo
  {journal} {Comput.Phys.Commun.}\ }\textbf {\bibinfo {volume} {185}},\
  \bibinfo {pages} {2250} (\bibinfo {year} {2014})},\ \Eprint
  {http://arxiv.org/abs/1310.1921} {arXiv:1310.1921 [hep-ph]} \BibitemShut
  {NoStop}%
\bibitem [{\citenamefont {Alwall}\ \emph {et~al.}(2014)\citenamefont {Alwall},
  \citenamefont {Frederix}, \citenamefont {Frixione}, \citenamefont {Hirschi},
  \citenamefont {Maltoni} \emph {et~al.}}]{Alwall:2014hca}%
  \BibitemOpen
  \bibfield  {author} {\bibinfo {author} {\bibfnamefont {J.}~\bibnamefont
  {Alwall}}, \bibinfo {author} {\bibfnamefont {R.}~\bibnamefont {Frederix}},
  \bibinfo {author} {\bibfnamefont {S.}~\bibnamefont {Frixione}}, \bibinfo
  {author} {\bibfnamefont {V.}~\bibnamefont {Hirschi}}, \bibinfo {author}
  {\bibfnamefont {F.}~\bibnamefont {Maltoni}},  \emph {et~al.},\ }\href
  {\doibase 10.1007/JHEP07(2014)079} {\bibfield  {journal} {\bibinfo  {journal}
  {JHEP}\ }\textbf {\bibinfo {volume} {1407}},\ \bibinfo {pages} {079}
  (\bibinfo {year} {2014})},\ \Eprint {http://arxiv.org/abs/1405.0301}
  {arXiv:1405.0301 [hep-ph]} \BibitemShut {NoStop}%
\bibitem [{\citenamefont {Hocker}\ \emph {et~al.}(2001)\citenamefont {Hocker},
  \citenamefont {Lacker}, \citenamefont {Laplace},\ and\ \citenamefont
  {Le~Diberder}}]{Hocker:2001xe}%
  \BibitemOpen
  \bibfield  {author} {\bibinfo {author} {\bibfnamefont {A.}~\bibnamefont
  {Hocker}}, \bibinfo {author} {\bibfnamefont {H.}~\bibnamefont {Lacker}},
  \bibinfo {author} {\bibfnamefont {S.}~\bibnamefont {Laplace}}, \ and\
  \bibinfo {author} {\bibfnamefont {F.}~\bibnamefont {Le~Diberder}},\ }\href
  {\doibase 10.1007/s100520100729} {\bibfield  {journal} {\bibinfo  {journal}
  {Eur.Phys.J.}\ }\textbf {\bibinfo {volume} {C21}},\ \bibinfo {pages} {225}
  (\bibinfo {year} {2001})},\ \Eprint {http://arxiv.org/abs/hep-ph/0104062}
  {arXiv:hep-ph/0104062 [hep-ph]} \BibitemShut {NoStop}%
\bibitem [{\citenamefont {Wilks}(1938)}]{Wilks:1938dza}%
  \BibitemOpen
  \bibfield  {author} {\bibinfo {author} {\bibfnamefont {S.}~\bibnamefont
  {Wilks}},\ }\href {\doibase 10.1214/aoms/1177732360} {\bibfield  {journal}
  {\bibinfo  {journal} {Annals Math.Statist.}\ }\textbf {\bibinfo {volume}
  {9}},\ \bibinfo {pages} {60} (\bibinfo {year} {1938})}\BibitemShut {NoStop}%
\bibitem [{\citenamefont {Inoue}\ \emph {et~al.}(1980)\citenamefont {Inoue},
  \citenamefont {Kakuto},\ and\ \citenamefont {Nakano}}]{Inoue:1979nn}%
  \BibitemOpen
  \bibfield  {author} {\bibinfo {author} {\bibfnamefont {K.}~\bibnamefont
  {Inoue}}, \bibinfo {author} {\bibfnamefont {A.}~\bibnamefont {Kakuto}}, \
  and\ \bibinfo {author} {\bibfnamefont {Y.}~\bibnamefont {Nakano}},\ }\href
  {\doibase 10.1143/PTP.63.234} {\bibfield  {journal} {\bibinfo  {journal}
  {Prog.Theor.Phys.}\ }\textbf {\bibinfo {volume} {63}},\ \bibinfo {pages}
  {234} (\bibinfo {year} {1980})}\BibitemShut {NoStop}%
\bibitem [{\citenamefont {Haber}\ and\ \citenamefont
  {Hempfling}(1993)}]{Haber:1993an}%
  \BibitemOpen
  \bibfield  {author} {\bibinfo {author} {\bibfnamefont {H.~E.}\ \bibnamefont
  {Haber}}\ and\ \bibinfo {author} {\bibfnamefont {R.}~\bibnamefont
  {Hempfling}},\ }\href {\doibase 10.1103/PhysRevD.48.4280} {\bibfield
  {journal} {\bibinfo  {journal} {Phys.Rev.}\ }\textbf {\bibinfo {volume}
  {D48}},\ \bibinfo {pages} {4280} (\bibinfo {year} {1993})},\ \Eprint
  {http://arxiv.org/abs/hep-ph/9307201} {arXiv:hep-ph/9307201 [hep-ph]}
  \BibitemShut {NoStop}%
\bibitem [{\citenamefont {Grimus}\ and\ \citenamefont
  {Lavoura}(2005)}]{Grimus:2004yh}%
  \BibitemOpen
  \bibfield  {author} {\bibinfo {author} {\bibfnamefont {W.}~\bibnamefont
  {Grimus}}\ and\ \bibinfo {author} {\bibfnamefont {L.}~\bibnamefont
  {Lavoura}},\ }\href {\doibase 10.1140/epjc/s2004-02075-0} {\bibfield
  {journal} {\bibinfo  {journal} {Eur.Phys.J.}\ }\textbf {\bibinfo {volume}
  {C39}},\ \bibinfo {pages} {219} (\bibinfo {year} {2005})},\ \Eprint
  {http://arxiv.org/abs/hep-ph/0409231} {arXiv:hep-ph/0409231 [hep-ph]}
  \BibitemShut {NoStop}%
\bibitem [{\citenamefont {Cheung}\ \emph {et~al.}(2012)\citenamefont {Cheung},
  \citenamefont {Papucci},\ and\ \citenamefont {Zurek}}]{Cheung:2012nb}%
  \BibitemOpen
  \bibfield  {author} {\bibinfo {author} {\bibfnamefont {C.}~\bibnamefont
  {Cheung}}, \bibinfo {author} {\bibfnamefont {M.}~\bibnamefont {Papucci}}, \
  and\ \bibinfo {author} {\bibfnamefont {K.~M.}\ \bibnamefont {Zurek}},\ }\href
  {\doibase 10.1007/JHEP07(2012)105} {\bibfield  {journal} {\bibinfo  {journal}
  {JHEP}\ }\textbf {\bibinfo {volume} {1207}},\ \bibinfo {pages} {105}
  (\bibinfo {year} {2012})},\ \Eprint {http://arxiv.org/abs/1203.5106}
  {arXiv:1203.5106 [hep-ph]} \BibitemShut {NoStop}%
\bibitem [{\citenamefont {Dev}\ and\ \citenamefont
  {Pilaftsis}(2014)}]{Dev:2014yca}%
  \BibitemOpen
  \bibfield  {author} {\bibinfo {author} {\bibfnamefont {P.~B.}\ \bibnamefont
  {Dev}}\ and\ \bibinfo {author} {\bibfnamefont {A.}~\bibnamefont
  {Pilaftsis}},\ }\href {\doibase 10.1007/JHEP12(2014)024} {\bibfield
  {journal} {\bibinfo  {journal} {JHEP}\ }\textbf {\bibinfo {volume} {1412}},\
  \bibinfo {pages} {024} (\bibinfo {year} {2014})},\ \Eprint
  {http://arxiv.org/abs/1408.3405} {arXiv:1408.3405 [hep-ph]} \BibitemShut
  {NoStop}%
\bibitem [{\citenamefont {Costa}\ \emph {et~al.}(2014)\citenamefont {Costa},
  \citenamefont {Morais}, \citenamefont {Sampaio},\ and\ \citenamefont
  {Santos}}]{Costa:2014qga}%
  \BibitemOpen
  \bibfield  {author} {\bibinfo {author} {\bibfnamefont {R.}~\bibnamefont
  {Costa}}, \bibinfo {author} {\bibfnamefont {A.~P.}\ \bibnamefont {Morais}},
  \bibinfo {author} {\bibfnamefont {M.~O.~P.}\ \bibnamefont {Sampaio}}, \ and\
  \bibinfo {author} {\bibfnamefont {R.}~\bibnamefont {Santos}},\ }\href@noop {}
  {\  (\bibinfo {year} {2014})},\ \Eprint {http://arxiv.org/abs/1411.4048}
  {arXiv:1411.4048 [hep-ph]} \BibitemShut {NoStop}%
\bibitem [{\citenamefont {Huffel}\ and\ \citenamefont
  {Pocsik}(1981)}]{Huffel:1980sk}%
  \BibitemOpen
  \bibfield  {author} {\bibinfo {author} {\bibfnamefont {H.}~\bibnamefont
  {Huffel}}\ and\ \bibinfo {author} {\bibfnamefont {G.}~\bibnamefont
  {Pocsik}},\ }\href {\doibase 10.1007/BF01429824} {\bibfield  {journal}
  {\bibinfo  {journal} {Z.Phys.}\ }\textbf {\bibinfo {volume} {C8}},\ \bibinfo
  {pages} {13} (\bibinfo {year} {1981})}\BibitemShut {NoStop}%
\bibitem [{\citenamefont {Maalampi}\ \emph {et~al.}(1991)\citenamefont
  {Maalampi}, \citenamefont {Sirkka},\ and\ \citenamefont
  {Vilja}}]{Maalampi:1991fb}%
  \BibitemOpen
  \bibfield  {author} {\bibinfo {author} {\bibfnamefont {J.}~\bibnamefont
  {Maalampi}}, \bibinfo {author} {\bibfnamefont {J.}~\bibnamefont {Sirkka}}, \
  and\ \bibinfo {author} {\bibfnamefont {I.}~\bibnamefont {Vilja}},\ }\href
  {\doibase 10.1016/0370-2693(91)90068-2} {\bibfield  {journal} {\bibinfo
  {journal} {Phys.Lett.}\ }\textbf {\bibinfo {volume} {B265}},\ \bibinfo
  {pages} {371} (\bibinfo {year} {1991})}\BibitemShut {NoStop}%
\bibitem [{\citenamefont {Kanemura}\ \emph {et~al.}(1993)\citenamefont
  {Kanemura}, \citenamefont {Kubota},\ and\ \citenamefont
  {Takasugi}}]{Kanemura:1993hm}%
  \BibitemOpen
  \bibfield  {author} {\bibinfo {author} {\bibfnamefont {S.}~\bibnamefont
  {Kanemura}}, \bibinfo {author} {\bibfnamefont {T.}~\bibnamefont {Kubota}}, \
  and\ \bibinfo {author} {\bibfnamefont {E.}~\bibnamefont {Takasugi}},\ }\href
  {\doibase 10.1016/0370-2693(93)91205-2} {\bibfield  {journal} {\bibinfo
  {journal} {Phys.Lett.}\ }\textbf {\bibinfo {volume} {B313}},\ \bibinfo
  {pages} {155} (\bibinfo {year} {1993})},\ \Eprint
  {http://arxiv.org/abs/hep-ph/9303263} {arXiv:hep-ph/9303263 [hep-ph]}
  \BibitemShut {NoStop}%
\bibitem [{\citenamefont {Akeroyd}\ \emph {et~al.}(2000)\citenamefont
  {Akeroyd}, \citenamefont {Arhrib},\ and\ \citenamefont
  {Naimi}}]{Akeroyd:2000wc}%
  \BibitemOpen
  \bibfield  {author} {\bibinfo {author} {\bibfnamefont {A.~G.}\ \bibnamefont
  {Akeroyd}}, \bibinfo {author} {\bibfnamefont {A.}~\bibnamefont {Arhrib}}, \
  and\ \bibinfo {author} {\bibfnamefont {E.-M.}\ \bibnamefont {Naimi}},\ }\href
  {\doibase 10.1016/S0370-2693(00)00962-X} {\bibfield  {journal} {\bibinfo
  {journal} {Phys.Lett.}\ }\textbf {\bibinfo {volume} {B490}},\ \bibinfo
  {pages} {119} (\bibinfo {year} {2000})},\ \Eprint
  {http://arxiv.org/abs/hep-ph/0006035} {arXiv:hep-ph/0006035 [hep-ph]}
  \BibitemShut {NoStop}%
\bibitem [{\citenamefont {Eberhardt}(2014)}]{Eberhardt:2014kaa}%
  \BibitemOpen
  \bibfield  {author} {\bibinfo {author} {\bibfnamefont {O.}~\bibnamefont
  {Eberhardt}},\ }\href@noop {} {\bibfield  {journal} {\bibinfo  {journal}
  {{2014 Electroweak Interactions and Unified Theories, Proceedings of the 49th
  Rencontres de Moriond}}\ ,\ \bibinfo {pages} {523}} (\bibinfo {year}
  {2014})},\ \Eprint {http://arxiv.org/abs/1405.3181} {arXiv:1405.3181
  [hep-ph]} \BibitemShut {NoStop}%
\bibitem [{\citenamefont {Kolda}\ and\ \citenamefont
  {Murayama}(2000)}]{Kolda:2000wi}%
  \BibitemOpen
  \bibfield  {author} {\bibinfo {author} {\bibfnamefont {C.~F.}\ \bibnamefont
  {Kolda}}\ and\ \bibinfo {author} {\bibfnamefont {H.}~\bibnamefont
  {Murayama}},\ }\href {\doibase 10.1088/1126-6708/2000/07/035} {\bibfield
  {journal} {\bibinfo  {journal} {JHEP}\ }\textbf {\bibinfo {volume} {0007}},\
  \bibinfo {pages} {035} (\bibinfo {year} {2000})},\ \Eprint
  {http://arxiv.org/abs/hep-ph/0003170} {arXiv:hep-ph/0003170 [hep-ph]}
  \BibitemShut {NoStop}%
\bibitem [{\citenamefont {Einhorn}\ and\ \citenamefont
  {Jones}(1992)}]{Einhorn:1992um}%
  \BibitemOpen
  \bibfield  {author} {\bibinfo {author} {\bibfnamefont {M.}~\bibnamefont
  {Einhorn}}\ and\ \bibinfo {author} {\bibfnamefont {D.}~\bibnamefont
  {Jones}},\ }\href {\doibase 10.1103/PhysRevD.46.5206} {\bibfield  {journal}
  {\bibinfo  {journal} {Phys.Rev.}\ }\textbf {\bibinfo {volume} {D46}},\
  \bibinfo {pages} {5206} (\bibinfo {year} {1992})}\BibitemShut {NoStop}%
\end{thebibliography}%

\end{document}